\documentclass[10pt]{article}
\usepackage{eqsection,latexsym,epsf,cite,epsfig,amssymb,graphics}
\usepackage{hyperref}

\begin{document}

\title{\bf
$\theta$ dependence of $SU(N)$ gauge theories in the presence of a
topological term
}
\author{
  {\bf  Ettore Vicari} \\ 
        Dipartimento di Fisica, Universit\`a 
        di Pisa, and INFN, I-56127 Pisa, Italy \\ 
        E-mail: {\tt vicari@df.unipi.it} \\
\\
  {\bf  Haralambos Panagopoulos} \\ 
        Department of Physics, University of Cyprus,
        Lefkosia, CY-1678, Cyprus\\ 
        E-mail: {\tt haris@ucy.ac.cy} 
}

\maketitle
\thispagestyle{empty}   

\begin{abstract}
\normalsize
We review results concerning the $\theta$ dependence of 4D $SU(N)$ gauge
theories and QCD, where $\theta$ is the coefficient of the CP-violating
topological term in the Lagrangian. In particular, we discuss $\theta$
dependence in the large-$N$ limit.  

Most results have been obtained
within the lattice formulation of the theory via numerical
simulations. We review results at zero and finite
temperature. We show that the results support the scenario obtained by
general large-$N$ scaling arguments, and in particular the
Witten-Veneziano mechanism to explain the $U(1)_A$ problem. We also
compare with results obtained by other approaches, especially in the
large-$N$ limit, where the issue has been also addressed using, for
example, the AdS/CFT correspondence.  

We discuss issues related to
theta dependence in full QCD: the neutron electric dipole moment, the
dependence of the topological susceptibility on the quark masses, the
$U(1)_A$ symmetry breaking at finite temperature.  

We also review results in the 2D $CP^{N-1}$ model, which is an
interesting theoretical laboratory to study issues related to topology. 

Finally, we discuss the main features of the two-point correlation
function of the topological charge density.



\end{abstract}

\newcommand{\be}{\begin{equation}}
\newcommand{\ee}{\end{equation}}
\newcommand{\bea}{\begin{eqnarray}}
\newcommand{\eea}{\end{eqnarray}}
\newcommand{\<}{\langle}
\renewcommand{\>}{\rangle}

\def\spose#1{\hbox to 0pt{#1\hss}}
\def\ltapprox{\mathrel{\spose{\lower 3pt\hbox{$\mathchar"218$}}
 \raise 2.0pt\hbox{$\mathchar"13C$}}}
\def\gtapprox{\mathrel{\spose{\lower 3pt\hbox{$\mathchar"218$}}
 \raise 2.0pt\hbox{$\mathchar"13E$}}}

\newpage

\tableofcontents

\clearpage

\def\slashed{\mskip5.0mu{/}\mskip-13.0mu}

\section{Introduction}
\label{intro}


Four-dimensional $SU(N)$ gauge theories have a nontrivial dependence on the
parameter $\theta$ which appears in the Euclidean 
Lagrangian as~\footnote{
The Euclidean Lagrangian (\ref{lagrangian}) corresponds to
\begin{eqnarray}
{\cal L}_\theta  = -{1\over 4} F_{\mu\nu}^a(x)F_{\mu\nu}^a(x)
- \theta {g^2\over 64\pi^2} \epsilon^{\mu\nu\rho\sigma}
F_{\mu\nu}^a(x) F_{\rho\sigma}^a(x),
\nonumber
\end{eqnarray}
in Minkowski space-time.
}
\begin{equation}
{\cal L}_\theta  = {1\over 4} F_{\mu\nu}^a(x)F_{\mu\nu}^a(x)
- i \theta {g^2\over 64\pi^2} \epsilon_{\mu\nu\rho\sigma}
F_{\mu\nu}^a(x) F_{\rho\sigma}^a(x),
\label{lagrangian}
\end{equation}
where 
\begin{equation}
q(x)=\frac{g^2}{64\pi^2} \epsilon_{\mu\nu\rho\sigma} F_{\mu\nu}^a(x)
F_{\rho\sigma}^a(x)
\label{topchden}
\end{equation}
is the topological charge density.  $\theta$ is a parameter of the strong
interactions, which leads to the breaking of parity and time reversal.

The spacetime integral of the topological charge density assumes integer
values $Q = \int d^4x\, q(x)$ for any continuous classical field configuration
with a finite action.  Such configurations, satisfying the classical equations
of motion, were first constructed explicitly over 30 years ago, starting with
the instanton solution~\cite{BPST-75,Hooft-76}. Instantons have been widely
employed in a semiclassical approach, in studying tunneling phenomena, in
supersymmetric theories, in instanton gas and instanton liquid effective
descriptions of QCD, etc. (See
Refs.~\cite{Coleman,Shifman,Shuryak,SS-98,GPY-81,ZJ-book} for some books and
reviews describing the uses of instantons in various contexts.)

In the semiclassical picture, instantons give rise to barrier-penetration
processes between different classical $n$-vacua, which are obtained from the
perturbative vacuum through nontrivial gauge transformations carrying winding
number $n$.  The existence of such barrier-penetration processes implies that
the true vacua are linear combinations of $n$-vacua. These $\theta$-vacua are
given by
\begin{equation}
|\theta\rangle = \sum_n e^{in\theta}|n\rangle.  
\label{thva}
\end{equation}
In a functional integral formulation, $\theta$-vacua are described by adding
the $\theta$ term in the Lagrangian, as in Eq.~(\ref{lagrangian}).

The most plausible explanation of how the solution of the so-called $U(1)_A$
problem can be compatible with the $1/N$ expansion (performed keeping $g^2N$
fixed \cite{Hooft-74}) requires a nontrivial $\theta$ dependence of the
ground-state energy density $F(\theta)$~\cite{Witten-79,Veneziano-79},
\begin{equation}
\exp[ - V F(\theta) ] = 
\int [dA] \exp \left(  - \int d^4 x {\cal L}_\theta \right),
\label{vftheta}
\end{equation}
where $V$ is the space-time volume.

Some of the first approaches to study $\theta$ dependence have employed
effective Lagrangians, see e.g.
Refs.~\cite{Weinberg,DV-80,Witten-80,RST-80,AP-82}.  Attempts for a more
quantitative assessment of this problem have focused largely on the lattice
formulation of the theory, using Monte Carlo (MC) simulations.  However, the
complex nature of the $\theta$ term in the Euclidean QCD Lagrangian prohibits
a direct MC simulation at $\theta\ne 0$; instead, one can obtain information
on the $\theta$ dependence of physically relevant quantities, such as the
ground state energy and the spectrum, by computing their expansion around
$\theta=0$; the coefficients of such an expansion can be determined from
appropriate correlation functions with insertions of the topological charge
density at $\theta=0$.  The numerical evidence for $\theta$ dependence,
obtained through MC simulations of the lattice formulation, appears quite
robust.

The presence of the $\theta$ term has important phenomenological consequences,
since it violates parity and time reversal symmetry.  Experimental bounds on
the $\theta$ parameter in QCD are best obtained from the electric dipole
moment of the neutron~\cite{Baker,Harris-etal-99}, which leads to an
unnaturally small value for $\theta$, $|\theta| \ltapprox 10^{-10}$.  This
suggests the idea that there must be a mechanism responsible for suppressing
the value of $\theta$ in the context of QCD.

The $\theta$ dependence is particularly interesting in the large-$N$ limit
where the issue may also be addressed by other approaches, such as AdS/CFT
correspondence applied to nonsupersymmetric and non conformal theories, see
e.g. Refs.~\cite{AGMOO-00,C-00,DiVecchia-99,GK-07,GKN-07}.

A particularly useful class of models, in which $\theta$ dependence can be
studied by various approaches, consists of the two-dimensional (2D) $CP^{N-1}$
models. These models share a number of properties
with four-dimensional $SU(N)$ gauge theories, such as asymptotic freedom and
nontrivial topology. In addition, they afford us the possibility of a
systematic $1/N$ expansion which, combined with numerical simulations, leads
to a precise quantitative description of the models' properties. 

\bigskip
This review is organized as follows:

In Sec.~\ref{U1} we discuss the $U(1)_A$ problem, and its explanation in the
framework of the large-$N$ limit based on the Witten-Veneziano mechanism,
which requires a nonzero topological susceptibility $\chi\equiv \partial^2
F(\theta)/\partial \theta^2|_{\theta=0}$ in the pure four-dimensional
(4D) $SU(N)$ gauge theory, and
therefore a nontrivial $\theta$ dependence.

Section \ref{thetadep} is devoted to a discussion of the general features of
the $\theta$ dependence of the ground state energy $F(\theta)$ and the
spectrum of 4D $SU(N)$ gauge theories.  In particular, we consider their
expansion in powers of $\theta$ around $\theta=0$, which provides information
in the region of small $\theta$ values.  We discuss the $\theta$ dependence in
the large-$N$ limit, where standard scaling arguments indicate that the
relevant variable is $\bar\theta\equiv \theta/N$.  This is also interesting
because the large-$N$ limit has been addressed by other approaches, such as
the AdS/CFT correspondence.  Finally, we discuss the $\theta$ term in QCD and
chiral perturbation theory, where the $\theta$ term is closely connected with
complex masses of the quarks. Calculations within the effective chiral
Lagrangian lead to several important relations among the topological
susceptibility, and the quark mass and condensate.

In Sec.~\ref{lattform} we introduce the lattice formulation of 4D $SU(N)$
gauge theory and QCD.  In particular, we discuss issues related to the lattice
regularization of fermion Dirac operators.  Besides the standard Wilson and
staggered fermions, we consider Dirac operators satisfying the so-called
Ginsparg-Wilson relation which lead to lattice regularizations preserving
chiral symmetry.  We mention in particular the Neuberger overlap Dirac
operator. Aside from an exact chiral symmetry, Ginsparg-Wilson Dirac operators
satisfy an exact index theorem on the lattice, at finite lattice spacing, thus
leading to a natural definition of topological charge.

Section \ref{latttopsec} reviews the lattice methods which have been employed
to investigate the topological properties of 4D $SU(N)$ gauge theories and
QCD. We show how a topological charge density operator can be defined on the
lattice, and its renormalization properties. We discuss the problems arising
when correlators of topological charge density operators are integrated over
all space-time, thus including coincident points, as is required in the
computation of the topological susceptibility, and more generally of the
$\theta$ expansion of the ground state energy $F(\theta)$.  In particular, a
naive lattice definition of the topological susceptibility is affected by
power-divergent additive contributions, which become eventually dominant in
the continuum limit. We describe some methods which have been used to address
this problem when employing bosonic definitions, such as the so-called
geometric, cooling/smearing, and heating methods.  Then, we consider a
fermionic definition of topological charge density which exploits the lattice
index theorem of Dirac operators satisfying the Ginsparg-Wilson relation.
This definition circumvents the problem of renormalization arising in bosonic
approaches, leading to well-defined field-theoretical estimators for
integrated correlators of topological charge density operators, even though at
a much higher computational cost.  We finally discuss the large-$N$ solution
of the $U(1)_A$ problem within lattice QCD formulations with Ginsparg-Wilson
fermions, where one can derive, at finite lattice spacing, a relation
analogous to the large-$N$ Witten-Veneziano formula relating the $\eta'$ mass
to topological susceptibility.

In Sec.~\ref{MCres} we review the results obtained for the $\theta$ dependence
of the ground state energy and the spectrum in 4D $SU(N)$ gauge theories.
They mostly refer to the first few terms of the $\theta$ expansion around
$\theta=0$, in particular to $\chi\equiv \partial^2 F(\theta)/\partial
\theta^2|_{\theta=0}$, and have been obtained by MC simulations.  We review
results at zero and finite temperature.  The results for $N=3$ and larger
values of $N$ support the scenario obtained by general large-$N$ scaling
arguments, and in particular the Witten-Veneziano mechanism to explain the
$U(1)_A$ problem.  We also compare with results obtained by other approaches,
especially in the large-$N$ limit, where the issue has been also addressed
using, for example, the AdS/CFT correspondence.

In Sec.~\ref{fullQCD} we discuss issues related to the $\theta$ dependence in
full QCD, such as the neutron electric dipole moment, the dependence of the
topological susceptibility on the quark masses, and the $U(1)_A$ symmetry
breaking at finite temperature and across the deconfinement transition.

Section ~\ref{cpn} is dedicated to the 2D $CP^{N-1}$ models, and, in
particular, to their $\theta$ dependence in the presence of a topological term
analogous to the $\theta$ term of 4D $SU(N)$ gauge theories. We review
analytical results obtained in the large-$N$ limit, and numerical results by
MC simulations of their lattice formulation.  The $\theta$ dependence
of 2D $CP^{N-1}$ models appears analogous to that conjectured for the 4D
$SU(N)$ gauge theories, in particular when comparing their large-$N$ limits.

Section~\ref{qqcorr} is devoted to a discussion of the peculiar features of
the two-point correlation function $G(x)$ of the topological charge density;
in particular, it addresses the question of how a nonnegative
topological susceptibility can arise from 
the integral of the correlation function $G(x)$, i.e. $\chi = \int d^d x\,
G(x)\ge 0$, where $G(x)<0$ for $|x|>0$ due to reflection positivity.  Of
course, this requires an important positive contribution from the contact term
at $x=0$, which compensates the negative integral for $|x|>0$.  These issues
are discussed within a solvable model, the large-$N$ limit of the 2D
$CP^{N-1}$ model, in the continuum and on the lattice.

Finally, Sec.~\ref{qsampling} is devoted to the somewhat technical issue of
the slow dynamics of topological modes in MC simulations; this gives
rise to a dramatic critical slowing down, which poses serious limitations to
numerical studies in the continuum limit.  We discuss the implications for
simulations of full QCD, and the question of whether one can obtain the
interesting observables averaged over the full $\theta$ vacuum from
simulations trapped at a fixed topological sector.

\def\slashed{\mskip5.0mu{/}\mskip-13.0mu}

\section{$U(1)_A$ problem and large-$N$ limit} 
\label{U1}

\subsection{The $U(1)_A$ problem}
\label{ua1pro}
The axial $U(1)_A$ problem is a long standing issue in QCD, which can be
traced back to the early 1970's and before; it is intimately related to
another, as yet unresolved, puzzle: the ``Strong CP problem''.

The QCD Lagrangian describing $N_f$ quark flavors in Euclidean space is
\begin{equation}
{\cal L}= {1\over 4} F_{\mu\nu}^a(x)F_{\mu\nu}^a(x)
+ \sum_{f=1}^{N_f} \bar\psi_f\,( D + m_f)\,\psi_f \, .
\end{equation}
where $D$ is the Dirac operator $D\equiv \gamma_\mu (\partial_\mu + g A_\mu)$.
In the chiral limit of massless quarks, assuming $N_f>1$, the Lagrangian is
invariant under rotations in flavor space, performed independently for left-
and right-handed components, resulting in a global group of vector and axial
symmetries:
\begin{equation}
U(N_f)_L \otimes U(N_f)_R \simeq
SU(N_f)_V \otimes SU(N_f)_A \otimes U(1)_V \otimes U(1)_A\,.
\end{equation}
Let us assume $N_f=3$.  In the real world, $U(1)_V$ remains intact,
corresponding to baryon number conservation, and $SU(3)_V$ is only softly
broken by differences in the quark masses, leading to the approximate symmetry
of isospin and strangeness. The axial $SU(3)_A$ is spontaneously broken by the
dynamical formation of nonzero quark condensates, leading to the octet of the
lightest pseudoscalar mesons which, in the chiral limit, are presumed to be
the massless Goldstone bosons corresponding to the eight generators of
$SU(3)_A$. As for the $U(1)_A$ symmetry, if it were intact in the chiral
limit, one would expect all massless hadrons to have a partner of opposite
parity. Since this does not appear to be the case, e.g. there are no scalar
pions, one may next assume that $U(1)_A$\, too, is spontaneouly broken. In
this case, there should be a corresponding isosinglet pseudoscalar Goldstone
boson. However, as originally estimated by Weinberg~\cite{Weinberg-75} using
chiral perturbation theory, such a particle, away from the chiral limit,
should have a mass less than $\sqrt{3} m_\pi$\,; the closest candidates are
the mesons $\eta(549)$ (already assigned to the $0^-$ octet) and $\eta'(985)$,
whose masses are clearly beyond the Weinberg bound.

The resolution of the $U(1)_A$ puzzle was proposed by
't~Hooft~\cite{Hooft-76,Hooft-76b}, who showed that, as a result of instanton
effects in the QCD vacuum, $U(1)_A$ is not a true symmetry of the theory.
Under $U(1)_A$ transformations:
\begin{equation} 
\psi \to e^{i\alpha\gamma_5}\,\psi ,
\label{chiralpsi}
\end{equation} 
the action of massless fermions is invariant, however the path integral
measure over fermion fields is modified~\cite{Fujikawa-79}:
\begin{equation} 
[d\psi][d\bar\psi] \to 
\exp\left({-i\, \alpha \, g^2\,N_f \over 32\pi^2}\int d^4x \,
\epsilon_{\mu\nu\rho\sigma} F_{\mu\nu}^a F_{\rho\sigma}^a\right)\, 
[d\psi][d\bar\psi].
\label{fermionmeasure}
\end{equation}
This noninvariance of the measure gives rise to the well known axial anomaly
equation~\cite{Adler-69,BJ-69,Bardeen-69}, which reads
\begin{equation}
\partial_\mu j_5^\mu(x)= i 2 N_f q(x)
\label{anomaly}
\end{equation}
in the chiral limit, where $q(x)$ is the topological charge density defined in
Eq.~(\ref{topchden}).

The integrand in Eq.~(\ref{fermionmeasure}) is a total derivative; nevertheless
the existence of nontrivial topological configurations, such as instantons,
renders Eq.~(\ref{fermionmeasure}) nontrivial.  This is essentially related to
the boundary conditions. If the naive boundary conditions $A_\mu=0$ at spatial
infinity are used, the integral in Eq.~(\ref{fermionmeasure}) would not
contribute and $U(1)_A$ would appear to be a symmetry again.  't Hooft
showed~\cite{Hooft-76,Hooft-76b} that the correct choice of the boundary
conditions is that $A_\mu$ is a pure gauge field at spatial infinity, i.e.
either $A_\mu=0$ or a gauge transformation of it. With these boundary
conditions, there are gauge configurations for which $\int d^4 x \,q(x)\ne 0$.
Despite being superficially a total divergence, the anomaly term has nonzero
matrix elements at zero momentum.  Thus, the $U(1)_A$ charge is not conserved,
with the result that the theory contains neither a conserved $U(1)_A$ quantum
number, nor an extra Goldstone boson.

Comparing Eq.~(\ref{fermionmeasure}) with the $\theta$ term in the Lagrangian,
$-i\theta q(x)$, we see that a chiral rotation on massless fermions is
equivalent to a shift in $\theta$: $\theta \to \theta - 2\alpha N_f$\,.
Conversely, an initial $\theta$ term could be rotated away in the presence of
massless fermions; on the other hand, if the fermions have nonzero mass, as is
the case with all quark species, such a chiral rotation affects their mass
matrix. Allowing for complex quark masses, we may write the Lagrangian mass
term in the form
\begin{equation}
{\cal L}_m =  {1\over 2} \sum_f m_f \bar\psi_f (1+\gamma_5)\psi_f +
{1\over 2} \sum_f m_f^* \bar\psi_f (1-\gamma_5)\psi_f =
\sum_{f=1}^{N_f} \bar\psi_f\,({\rm Re} \,m_f + 
i\, {\rm Im} \,m_f\, \gamma_5)\,\psi_f \,.
\label{lmdef}
\end{equation}
P and T symmetries are broken for
${\rm Im}\,m_f \ne 0$. The transformation~(\ref{chiralpsi}) modifies $m_f$ as:
\begin{equation}
m_f \to e^{2 i \alpha}\, m_f \,.
\label{mfshift}
\end{equation} 
Thus, we may rotate $m_f$ to a real value by an appropriate choice of $\alpha$
(possibly different for each flavor). The shifted value of $\theta$ after
these rotations is still a source of P and T noninvariance; in particular, on
dimensional grounds, it leads to an electric dipole moment for the neutron of
order~\cite{Weinberg}
\begin{equation}
d_n \sim \theta {e m_f\over m_n^2} \sim
\theta {e m_\pi^2\over m_n^3}
 \approx 10^{-16} \theta \,e\,{\rm cm}.   
\label{dno}
\end{equation}
The experimental bound on $d_n$ is $|d_n| < 2.9 \times 10^{-26}e\,{\rm
  cm}$~\cite{Baker,Harris-etal-99}.  Supplementing this experimental bound
with the rough estimate (\ref{dno}), and also with further theoretical
calculations indicating $|d_n/\theta| \sim O(10^{-15}$-$10^{-16}) \,e\,{\rm
  cm}$ (which are reviewed in Sec.~\ref{neutron}), one then gets an
exceedingly small value for $\theta$: $|\theta|\ltapprox 10^{-10}$.  A similar
bound has also been obtained by measuring the electric dipole moment of the
$^{199}{\rm Hg}$ atom~\cite{RGJF-01}.  We shall return to this issue in
Sec.~\ref{neutron}.

A number of possible explanations have been proposed for the origin of this
small value of $\theta$, such as spontaneous breaking of CP, vanishing of the
mass of the lightest $u$-quark, and postulating the axion, which effectively
promotes $\theta$ into a dynamical variable (see,
e.g.,~\cite{SW-06,Peccei-06,Weinberg}, for some relevant recent reviews).  In
particular, in the latter theory, originally proposed by Peccei and
Quinn~\cite{PQ-77}, the dynamical field replacing the $\theta$ parameter could
relax to a minimum of the effective potential where the parity and time
reversal symmetries are recovered.  This would also require the existence of a
new particle~\cite{Weinberg-78,Wilczek-78}, the axion.  The various models
using this basic mechanism that have been proposed to solve the CP problem
have in common the existence of a further chiral symmetry $U(1)_{\rm PQ}$,
which is spontaneously broken at high energies, much higher than the QCD
scale, and is also broken by the anomaly.  This explanation still represents
an interesting open possibility, although the parameter regions related to the
axion have been much restricted, 
see Refs.~\cite{Weinberg,Peccei-06,PDG-06} and
references therein.  The vanishing of the mass $m_u$ of the lightest $u$-quark
has also been proposed as solution of the strong CP problem~\cite{KM-86}. This
possibility is disfavored by a standard current algebra analysis~\cite{GL-82},
which shows that the experimental data are consistent with a nonzero mass.
Nonzero quark masses are also strongly supported by calculations within
lattice gauge theory, see e.g.
Refs.~\cite{Aubin-etal-04,Aoki-etal-03,ALPHA-05,GHIPRSS-06,Becirevic-etal-06}.
For a recent review on quark masses see Ref.~\cite{PDG-06}.  Moreover, in
Refs.~\cite{Creutz-04-2,Creutz-07-3} it has been argued that the vanishing of
a single quark mass is renormalization scheme dependent due to nonperturbative
effects related to the anomaly, therefore the vanishing of $m_u$ should not be
relevant to a physical issue such as violation of the CP symmetry in strong
interactions.

Through the anomaly equation physically interesting matrix elements of
hadron phenomenology are related to the topological properties of the theory.
An example is given by the large mass of the $\eta'$, as we shall see
later.  Another example is related to the so-called proton spin problem, see
e.g.  Refs.~\cite{Veneziano-89,SV-92-2,NSV-99,Shore-06,Shore-07}.

\subsection{The large-$N$ limit}

The large-$N$ limit, where $N$ is the number of colors, is a very useful
framework to investigate the physics of strong interactions.  In the
context of particle physics the $1/N$ expansion was introduced by
't~Hooft~\cite{Hooft-74}.  He proposed to generalize the $SU(3)$ gauge
symmetry of QCD to $SU(N)$ and to expand in powers of $1/N$.  In fact, $N$ is
the only free parameter in QCD~\cite{Witten-79c}.  $N$ is an intrinsically
dimensionless parameter; the origin of $N$ dependence is basically
group-theoretical, and leads to well-defined field representations for all
integer values, hence it is not subject to any kind of renormalization.  The
't Hooft large-$N$ limit is given by
\begin{equation}
N \rightarrow \infty\qquad {\rm keeping} \quad g^2 N, N_f \quad {\rm fixed}.
\label{largenlimit}
\end{equation}
The dominant Feynman graphs in the large-$N$ limit can be classified by simply
counting powers of $N$~\cite{Hooft-74,Witten-79c,Coleman-82}. In general,
the leading large-$N$ contributions come from planar diagrams, each fermion
loop introduces a factor $1/N$, while each nonplanar crossing is suppressed by
$1/N^2$.  The Feynman diagrams contributing to connected correlation functions
of gauge invariant operators constructed using gauge fields are $O(N^2)$. They
are planar and do not contain fermion loops. The connected parts of
correlation functions of fermionic currents are $O(N)$. The corresponding
diagrams are planar, they do not have internal fermion loops, and all current
insertions are on a single fermion loop which bounds the graph.

The most interesting feature of the large-$N$ expansion is that the
phenomenology of QCD in the large-$N$ limit presents remarkable analogies with
that of the real world.  Assuming that the $N=\infty$ theory confines, so that
the physical spectrum consists of color singlets only, one can study
properties of hadrons by applying power counting in $N$, and
analyzing the intermediate states that contribute to the various $n$-point
correlation functions, as discussed e.g. in
Refs.~\cite{Witten-79c,Coleman-82}.  This leads to a scenario which is
qualitatively, and also semi-quantitatively, consistent with those of the real
world (see for example Refs.~\cite{Polyakov-87,Das-87,Manohar-98,RCV-98} for
reviews).

We also mention the Veneziano large-$N$ limit~\cite{Veneziano-76}
\begin{equation}
N, N_f \rightarrow \infty\qquad {\rm keeping} \quad g^2 N, N_f/N 
\quad {\rm fixed}.
\label{largenlimitv}
\end{equation}
This limit provides a better explanation of certain aspects of low-energy
phenomenology. However, the 't~Hooft limit is simpler, and has been studied in
more detail.

We should finally note that the large-$N$ solution of 4D $SU(N)$ gauge
theories is still unknown, and therefore it is not possible to perform a
systematic $1/N$ expansion.  The problems in determining the large-$N$
saddle point are essentially related to the matrix nature of the theory, see
for example Ref.~\cite{Polyakov-87} for a detailed discussion. Nevertheless,
the $1/N$ expansion of QCD represents a very useful framework, within which
several phenomenological issues can be discussed, and nontrivial predictions can
be inferred, see e.g.  Refs.~\cite{Polyakov-87,Manohar-98}.  We also mention
that the $1/N$ expansion has been instead successfully developed in vector
models, such as O($N$)-symmetric theories and $CP^{N-1}$ models (see e.g.
Refs.~\cite{Polyakov-87,ZJ-book,MZ-03}).

The conjectured large-$N$ scenario of QCD (discussed at length in Polyakov's
book~\cite{Polyakov-87}) can be investigated by performing lattice
calculations at relatively large $N$, studying their convergence for
$N\rightarrow\infty$ in the `quenched' case (i.e. with no dynamical fermions,
given that contributions of quark loops should be depressed by a factor
$1/N$).

\subsection{Solution of the $U(1)_A$ problem within the large-$N$ limit: 
The Witten-Veneziano formula}
\label{u1asolln}

In general a broken symmetry is best understood by studying the limit leading
to a theory where the symmetry is conserved.  In the case of $U(1)_A$
symmetry, such a limit should be that of a large number of colors:
$N\rightarrow \infty$.  An explanation of the explicit breaking of the
$U(1)_A$ symmetry based on large-$N$ arguments was originally proposed by
Witten~\cite{Witten-79} and then refined by Veneziano~\cite{Veneziano-79}.
According to Witten's argument the $U(1)_A$ problem should be solved at the
lowest nonplanar level, i.e. at the next-to-leading order of its $1/N$
expansion.  Let us sketch this argument.  
We introduce the topological susceptibility $\chi$ 
\begin{equation}
\chi = \left. {\partial^2 F(\theta)\over \partial \theta^2} \right|_{\theta=0}=
\int d^4 x \langle q(x) q(0) \rangle ,
\label{chidefu1}
\end{equation}
as defined in Euclidean space.  We recall that in Minkowski space-time the
topological susceptibility reads
\begin{equation}
\chi = -i \int d^4 x \langle 0 | T q(x) q(0)| 0 \rangle
\label{chimin}
\end{equation}
(more details on this definition and its relation with the corresponding
Euclidean quantity can be found in
Refs.~\cite{Crewther-79,Witten-79,Meggiolaro-98}).

While there is no $\theta$ dependence in perturbation theory, there may be a
nontrivial dependence at a nonperturbative level, and in particular within the
$1/N$ expansion.  Assume that the pure gauge theory without quarks is
$\theta$-dependent to leading order in $1/N$, and therefore that $\chi\neq 0$
at $N=\infty$.  This assumption leads to an apparent paradox: on the one hand,
when massless quarks are introduced the $\theta$ dependence must disappear and
therefore $\chi=0$, on the other hand quark loops give only nonleading
contributions of order $1/N$ to the physical processes of the pure gauge
theory~\cite{Hooft-74}.  This paradox can be solved by the presence of a
particle with mass squared $m_s^2$ of order $1/N$, having the same quantum
numbers of the topological charge density. Such a particle should be related
to the $\eta'$, i.e.  the lightest flavor-singlet pseudoscalar in nature.

The existence of a particle of mass squared $O(1/N)$ can be justified noting
that in the large-$N$ limit the singlet axial current should be conserved.
The anomaly is an $O(1/N)$ effect, because it arises from a diagram with one
quark loop, which is depressed by a factor $1/N$.  Therefore at $N=\infty$
there should be an axial singlet Goldstone boson.  At $O(1/N)$ the anomaly is
recovered and, as a consequence, the $N=\infty$ Goldstone boson gets a mass.
At order $1/N$, $m_{s}^2$ should receive an $O(1/N)$ contribution, in that the
mass squared of an approximate Goldstone boson is in general linear with
respect to the symmetry breaking parameter, which in this context is
represented by $1/N$.

So as a consequence of a nonzero large-$N$ limit of $\chi$ in the pure gauge
theory, we would have $m_s^2\sim 1/N$, while the other Goldstone bosons
associated with the nonsinglet axial symmetry remain massless at the chiral
limit.  A further development of these ideas results in the leading order
relationship
\begin{equation}
\chi={f_s^2 m_s^2\over 4N_f},
\label{wf}
\end{equation}
where $f_s$ is defined by 
\begin{equation}
\langle 0 | \; \partial_\mu j_5^\mu \;|s\rangle=\sqrt{N_f} m_s^2 f_s.  
\label{mj5}
\end{equation}
Notice that $f_s$ is of order $\sqrt{N}$ and in the large-$N$ limit $f_s
=f_{ns}=f_\pi$ (where $f_\pi$ is defined as in Ref.~\cite{PDG-06}).
By performing a more accurate analysis, based on the
large-$N$ limit (\ref{largenlimitv}) and using also the anomalous
flavor-singlet Ward-Takahashi identities, Veneziano~\cite{Veneziano-79}
refined the relationship (\ref{wf}) obtaining
\begin{equation}
{4N_f\over f_\pi^2} \chi=m_{\eta'}^2 + m_\eta^2 - 2m_K^2.
\label{vf}
\end{equation}

In Eqs.~(\ref{wf}) and (\ref{vf}), due to their large-$N$ based derivation,
$\chi$ should be considered that of the pure gauge theory.  This fact favors a
check by Monte Carlo simulation of the lattice formulation, in that pure gauge
simulations are much simpler than full QCD ones.  Substituting $N_f=3$ and the
experimental values in (\ref{vf}) (i.e.~\cite{PDG-06} $f_\pi\approx 131$ MeV,
$m_{\eta'}\approx 958$ MeV, $m_{\eta}\approx 547$ MeV, $m_{K}\approx 494$ MeV)
the prediction is $\chi\approx \left( 180 \;{\rm MeV}\right)^4$.  This
prediction has been substantially verified by lattice computations, as we
shall see below.

We mention that the Witten-Veneziano formula has been also derived within the
lattice formulation of QCD, when the so-called Ginsparg-Wilson lattice
fermions are considered~\cite{GRTV-02}, see also Sec.~\ref{ua1latt}.
Moreover, it has also been investigated within the AdS/CFT framework in
Refs.~\cite{Armoni-04,BHMM-04,HO-99,SS-05,Katz-07}.

Models based on instanton semiclassical pictures can hardly explain the
nontrivial large-$N$ behavior of the topological susceptibility,
$\chi=O(N^0)$, and therefore of the $\eta'$ mass, $m_{\eta'}^2\sim 1/N$.  The
naive dilute-instanton picture would suggest that the topological
susceptibility and $m_{\eta'}$ vanish exponentially as $e^{-c N}$, $c$ being
some constant, because the instanton weight behaves as $\exp(-8\pi/g^2)$, and
the 't~Hooft large-$N$ limit must be taken keeping $g^2 N$ fixed, thus leading
to an apparent exponential suppression.  Actually, since a nontrivial
$N$-dependent prefactor~\cite{Hooft-76b} appears in the weight of the
instanton distribution, this argument strictly applies only to small
instantons, leaving open the possibility to get a consistent scenario by
appropriately taking into account instanton contributions at the QCD scale.
The consistency of the semiclassical instanton calculation and $1/N$ expansion
of QCD has been discussed in Refs.~\cite{Schafer-02,Schafer-04}; in particular
it was argued that the instanton liquid model, where large (overlapping)
instantons get suppressed by ad-hoc short-range repulsive instanton
interactions~\cite{Shuryak-94}, is not necessarily incompatible with the
large-$N$ scenario.  This issue has been also
discussed~\cite{Jevicki-79,David-84} within the 2D $CP^{N-1}$ models,
see also Sec.~\ref{thetadepcpnln}.

\section{$\theta$ dependence of 4D $SU(N)$ gauge theories}
\label{thetadep}

\subsection{General features}
\label{generalfeat}

In order to study the $\theta$ dependence of the ground-state energy in 4D
$SU(N)$ gauge theories, it is convenient to introduce the dimensionless {\it
  scaling} function
\begin{equation}
f(\theta) = {F(\theta)-F(0)\over \sigma^2},
\label{scge}
\end{equation}
where $F(\theta)$ is the ground-state energy defined in Eq.~(\ref{vftheta}),
and $\sigma$ is the string tension at $\theta=0$, which can be obtained from
the area law of Wilson loops. Of course, one may use any other energy scale to
define a dimensionless function related to the ground state energy, for
example the lowest glueball mass.

As a consequence of the topological nature of the $\theta$ term when applied
to continuous field configurations, which leads to the quantization of
topological charge, the ground-state energy $F(\theta)$, and therefore
$f(\theta)$, is expected to be periodic in $\theta$, with periodicity $2\pi$.

The $\theta$ dependence can be studied in the region of small $\theta$ values
by expanding $f(\theta)$ around $\theta=0$.  Of course, we are assuming that
the theory is not singular at $\theta=0$, and in particular that CP is not
spontaneously broken at $\theta=0$~\cite{VW-84}. For this purpose, the scaling
function $f(\theta)$ may be parametrized as
\begin{eqnarray}
f(\theta)={1\over 2} C \theta^2 s(\theta),\label{ftheta}
\end{eqnarray}
where $s(\theta)$ is a dimensionless function of $\theta$ such that $s(0)=1$.
$C$ is the ratio $\chi/\sigma^2$ and $\chi$ is the topological susceptibility
at $\theta=0$, 
\begin{equation}
\chi = \int d^4 x \langle q(x)q(0) \rangle = {\langle Q^2 \rangle\over V},
\label{chidef}
\end{equation}
where 
\begin{equation}
Q=\int d^4 x \,q(x)
\label{topoch}
\end{equation}
is the topological charge.
The function $s(\theta)$ can be expanded around $\theta=0$ as
\begin{eqnarray}
s(\theta) = 1 + b_2 \theta^2 + b_4 \theta^4 + \cdots,
\label{stheta}
\end{eqnarray}
where only even powers of $\theta$ appear.  The coefficients of the expansion
of $f(\theta)$ are related to the zero-momentum $n$-point connected
correlation functions of the topological charge density, and therefore to the
moments of the probability distribution $P(Q)$ of the topological charge $Q$.
If all $b_{2n}$ were vanishing, leading to $s(\theta)=1$, then the
corresponding distribution $P(Q)$ would be Gaussian, i.e.
\begin{equation}
P(Q)={1\over \sqrt{2\pi\langle Q^2\rangle}}\,
{\rm exp}\left( -{Q^2\over 2\langle Q^2\rangle}
\right).
\end{equation}
Therefore the coefficients $b_{2n}$ of the expansion of $s(\theta)$
parametrize the deviations from a simple Gaussian behavior.  For example, the
coefficients of the first few nontrivial terms are given by
\begin{eqnarray}
&&b_2 = - {\chi_4 \over 12 \chi},\label{b2chi4} \\
&& \chi_4 = {1\over V} \left[ 
\langle Q^4 \rangle - 3  \langle Q^2 \rangle^2 \right]_{\theta=0}, \label{chi4}
\end{eqnarray}
and
\begin{eqnarray}
&&b_4 = {\chi_6\over 360 \chi},\label{b4chi6} \\
&& \chi_6 = {1\over V} \left[ 
\langle Q^6 \rangle  -  15 \langle Q^2 \rangle \langle Q^4 \rangle  +
30 \langle Q^2 \rangle^3 \right]_{\theta=0},  \label{chi6}
\end{eqnarray}
etc. It has been recently shown by L\"uscher~\cite{Luscher-04} (see also
\cite{GRT-04}) that correlation functions involving multiple zero-momentum
insertions of the topological charge density can be defined in a nonambiguous,
regularization-independent way, and therefore the expansion coefficients
$b_{2n}$ are well defined renormalization-group invariant quantities.

Besides the $\theta$ dependence of the ground-state energy, one may also
consider the $\theta$ dependence of the glueball spectrum and of the string
tension $\sigma(\theta)$, from the area law of Wilson loops. By analogy with
the ground state energy, $\sigma(\theta)$ may be expanded around $\theta=0$
as:
\begin{equation}
\sigma(\theta) = \sigma \left( 1 + s_2 \theta^2 + ... \right),
\label{sigmaex}
\end{equation}
where $\sigma$ is the string tension at $\theta=0$.  
Similarly for the lowest glueball state:
\begin{equation}
M(\theta) = M\left( 1 + g_2 \theta^2 + ... \right),
\label{gmex}
\end{equation} 
where $M$ is the $0^{++}$ glueball mass at $\theta=0$.  Of course, at
$\theta\ne 0$, the lightest glueball state does not have a definite parity
anymore, but it becomes a mixed state of $0^{++}$ and $0^{-+}$ glueballs.  The
coefficients of the above expansions can be computed from appropriate
correlators at $\theta=0$, see also Sec.~\ref{thspres}.

Another interesting related issue is the possible spontaneous breaking
of CP at $\theta=\pi$.  $SU(N)$ gauge theories, as well as QCD, in the
presence of the $\theta$ term are invariant under CP not only at
$\theta=0$, but also at $\theta=\pi$.  This is because $\theta=\pi$
goes to $\theta= -\pi$ under a CP transformation, but physics should
be unchanged for $\theta\to\theta+2\pi$, so that the values
$\theta=\pm \pi$ are equivalent. However, CP can be spontaneosuly
broken at $\theta=\pi$, with the appearance of two CP violating
degenerate vacua~\cite{Dashen-71}.  According to the Vafa-Witten
theorem,~\cite{VW-84,AG-99} which has been argued to hold under quite
general conditions (the argument makes positivity assumptions on the
path integral), this possibility is excluded at $\theta=0$. The
spontaneous breaking of CP at $\theta=\pi$ is expected to be generally
related to a first order transition when varying $\theta$.  These
issues have been investigated in
Refs.~\cite{DV-80,Witten-80,Creutz-95,EHNS-97,Smilga-99,Tytgat-00,LW-02,ALS-02,KN-05,ADD-05}.
They have also been addressed within softly broken supersymmetric QCD,
see e.g. Refs.~\cite{EHS-97,Konishi-97}, arriving at analogous
conclusions for the dependence on $\theta$ around $\theta=\pi$.  The
$\theta$ dependence in softly broken supersymmetric QCD-like theories
has been also discussed in
Refs.~\cite{KT-98,Ferrari-97,ShVa-88,KS-97,DK-96}.

\subsection{$\theta$ dependence in the large-$N$ limit}
\label{largeNsec}

Large-$N$ scaling arguments~\cite{Witten-79b,Witten-80,Witten-98} applied to
the Lagrangian (\ref{lagrangian}) indicate that the relevant scaling variable
in the large-$N$ limit is
\begin{equation}
\bar\theta\equiv {\theta\over N}\,.
\label{thetabar}
\end{equation}
This can easily be derived by recalling that the Lagrangian is $O(N^2)$ and
the large-$N$ limit is performed keeping $g^2N$ fixed.  Accordingly, the
ground-state energy is expected to behave as
\begin{equation} 
f(\theta) \equiv {F(\theta)-F(0)\over \sigma^2}=
N^2 \bar{f}(\bar{\theta}),
\label{fthetabarln} 
\end{equation}
where the function $\bar{f}(x)$ has a nontrivial limit as $N\to \infty$.  

In the dilute instanton gas approximation, the $\theta$ dependence has the
form
\begin{equation}
e^{-8\pi^2/g^2}e^{i\theta} = 
\left( e^{-8\pi^2/(g^2N)}e^{i\theta/N} \right)^N 
\label{gasinst}
\end{equation}
in the one-instanton sector.  This is exponentially suppressed in $N$, thus it
might suggest that the $\theta$ dependence is exponentially small in $N$. This
conclusion is incorrect, essentially because the instanton gas approximation
is not valid due to infrared divergences.

The large-$N$ behavior of the coefficients $b_{2n}$ of the expansion
(\ref{stheta}) can then be easily derived:
\begin{eqnarray} 
\bar{f}(\bar{\theta}) = 
{1\over 2} C_\infty \bar{\theta}^2 ( 1 + \bar{b}_2 \bar{\theta}^2 + 
 \bar{b}_4 \bar{\theta}^4 + \cdots), 
\label{lnexp}
\end{eqnarray}
where $C_\infty$ is the large-$N$ limit of the ratio $C=\chi/\sigma^2$.
Comparing with Eq.~(\ref{ftheta}), one obtains
\begin{eqnarray}
&&C=C_\infty + {c_2\over N^2} + ... , \label{largeNC}\\
&&b_{2j}={\bar{b}_{2j}\over N^{2j}}+...\label{largeNco}
\end{eqnarray}
We recall that a nonzero value of $C_{\infty}$ is essential to provide an
explanation to the $U(1)_A$ problem in the large-$N$ limit, and can be related
to the $\eta'$ mass \cite{Witten-79} through the relation
\begin{equation}
\chi_\infty = {f_{\eta'}^2 m_{\eta'}^2\over 4 N_f} + O(1/N).
\label{wittenformula}
\end{equation}
We note in passing that the quantity $b_2$ also lends itself to a physical
interpretation, being related to the $\eta^\prime$-$\eta^\prime$ elastic
scattering amplitude \cite{Veneziano-79}.

The large-$N$ scaling behavior is apparently incompatible with the periodicity
condition $f(\theta) = f(\theta+2\pi)$, which is a consequence of the
quantization of the topological charge, as indicated by semiclassical
arguments based on its geometrical meaning for continuous field
configurations.  Indeed a regular function of $\bar\theta=\theta/N$ cannot be
invariant for $\theta \to \theta+2\pi$, unless it is constant.  A plausible way
out has been proposed by Witten in Ref.~\cite{Witten-80}: the ground-state
energy $F(\theta)$ in the large-$N$ limit is a multibranched function because
of many candidate vacuum states which all become stable, although not
degenerate, for $N=\infty$. Such behavior is exhibited by some two-dimensional
models~\cite{Coleman-76,Witten-79}.  This scenario leads to the ground-state
energy~\cite{Witten-80}
\begin{equation}
F(\theta) = N^2 \, {\rm Min}_k\, H\left( {\theta+2\pi k\over N}\right),
\label{conj1}
\end{equation}
where $H$ is an unspecified function; $F(\theta)$ is then periodic in
$\theta$, but not regular everywhere, since at some value of $\theta$ there is
a jump between two different branches.  This issue, and in particular the
consistency between the $\theta/N$ dependence in the large-$N$ limit and the
$2\pi$ periodicity in $\theta$, has been discussed in
Refs.~\cite{HZ-98,HHI-96,HHI-95,Ohta-81,Witten-80,Ferrari-02}.

The conjecture (\ref{conj1}) was then refined in Ref.~\cite{Witten-98}.
$F(\theta)$ must have its absolute mininum at $\theta=0$ essentially because
at $\theta=0$ the integrand of the Euclidean path integral is real and
positive. If the vacuum is unique at $\theta=0$, then the mininum must occur
for $k=0$. Moreover, the large-$N$ solution of the $U(1)_A$ problem requires
$d^2 F/d\theta^2|_{\theta = 0} > 0$ in the large-$N$ limit and, therefore, we
can write $F(\theta)-F(0) \propto \theta^2$\,; higher orders do not contribute
to leading order in $1/N$.  The simplest expression combining this behavior
with periodicity is then given by
\begin{equation}
F(\theta) - F(0) = {\cal A} \, {\rm Min}_k \, (\theta+2\pi k)^2 + O\left(
1/N\right).
\label{conj2}
\end{equation} 
In particular, for sufficiently small values of $\theta$, i.e. $|\theta|<\pi$,
\begin{equation}
F(\theta) - F(0) = {\cal A} \, \theta^2  + O\left( 1/N^2\right).
\label{conj2b}
\end{equation} 
This issue has been discussed within a field-theoretical framework in
Refs.~\cite{Gabadadze-99,Shifman-99}.  The same result has been derived using
the conjectured correspondence between large-$N$ gauge theories and
supergravity/string theory on some particular compactified
spacetimes~\cite{Witten-98}, and also by calculations based on the fivebrane
of M theory to study the $\theta$ dependence of softly broken supersymmetric
$SU(N)$ gauge theories~\cite{OP-98}.  As we shall see, this scenario is 
supported by the results of Monte Carlo simulations of the lattice formulation
of 4D $SU(N)$ gauge theories, presented in Sec.~\ref{MCres}.

Concerning the spectrum of the theory, the large-$N$ scaling arguments, which
indicate $\bar{\theta}\equiv \theta/N$ as the relevant Lagrangian parameter in
the large-$N$ limit, imply that the coefficients of the $\theta$ expansion in
Eqs.~(\ref{sigmaex}) and (\ref{gmex}) are suppressed, in particular $s_2$ and
$g_2$ should decrease as $1/N^{2}$.  This is suggestive of a scenario in which
the $\theta$ dependence of the spectrum disappears in the large-$N$ limit, at
least for sufficiently small values of $\theta$ around $\theta=0$.  The only
effect of the $\theta$ term on the lowest spin-zero glueball state is that it
becomes a mixed state of $0^{++}$ and $0^{-+}$ glueballs, as a consequence of
the fact that the $\theta$ term breaks parity, but the mass of the state does
not change. This conclusion has been also reached by an analysis of the
$\theta$ dependence of the glueball spectrum using AdS/CFT~\cite{GI-04}. This
scenario is also supported by the results of lattice Monte Carlo
simulations reported in Sec.~\ref{thspres}.

\subsection{$\theta$ term in QCD and effective chiral Lagrangians}
\label{chiralpert}

An important approach to the study of $\theta$ dependence in QCD is provided
by chiral perturbation theory, see e.g. Refs.~\cite{GL-85,GL-84},
also~\cite{Weinberg} and references therein.  The low-energy physics of
Goldstone bosons in QCD can be described systematically using an effective
theory: In the case of $N_f$ light quarks, the chiral Lagrangian of the
Goldstone boson $SU(N_f)$ matrix field $U$ in the lowest-order approximation
is given by
\begin{equation}
{\cal L}_{\rm ch} =
{1\over 4} F^2 \,{\rm Tr} [\partial_\mu U^\dagger \partial_\mu U]
- {1\over 2} \Sigma\,
{\rm Tr} [M e^{i\theta/N_f} U^\dagger + M^\dagger e^{-i\theta/N_f} U ],
\label{chirallagr}
\end{equation}
where the coupling $F$ is related to the pion decay constant,
$F=f_\pi/\sqrt{2}\approx 92$ MeV at lowest order of chiral
perturbation theory, and
$\Sigma$ is related to the quark condensate in the massless theory,
\begin{equation}
\Sigma = - {\langle \bar{\psi} \psi \rangle \over N_f} .
\label{sigmadef}
\end{equation}
$M$ is the real diagonal quark mass matrix, $M={\rm diag}(m_u,m_d,...)$.  The
kinetic term proportional to $F^2$ is invariant under chiral $SU(N_f)_L\otimes
SU(N_f)_R$ transformations, while the mass term breaks chiral symmetry.

Let us consider the simplest case of $N_f$ degenerate flavors of mass $m$. In
this case the chiral symmetry is broken to $SU(N_f)_V$. From
Eq.~(\ref{chirallagr}), one can derive an expression for the ground-state
energy by minimizing the action, i.e.~\cite{Witten-80}
\begin{equation}
F(\theta) - F(0) = N_f m \Sigma [1 - {\rm cos}(\theta/N_f)],
\label{etheta}
\end{equation}
which is valid for $|\theta|\le \pi$.  When it is extended periodically to
other values of $\theta$, it shows cusps at
$\theta=2\pi(k+1/2)$~\cite{Witten-80}.  From Eq.~(\ref{etheta}) one can derive
the topological susceptibility
\begin{equation}
\chi = {\partial^2 F(\theta)\over \partial \theta^2}\Bigg|_{\theta{=}0} = 
{m \Sigma \over N_f}.
\label{chichi}
\end{equation}
An analogous expression can also be determined
for different quark masses: 
\begin{equation}
\chi = \Sigma \,\Bigl( \sum_f {1\over m_f} \Bigr)^{-1}.
\label{chichi2}
\end{equation}

Given that, by virtue of the anomaly, the flavor-singlet pseudoscalar particle
$\eta'$ is heavy, it is not a Goldstone boson and, therefore, its physics
cannot be described using chiral perturbation theory in the form of
Eq.~(\ref{chirallagr}). However, in the large-$N$ limit the anomaly disappears
and the $\eta'$ becomes a Goldstone boson, as discussed above; thus, it may be
studied in chiral perturbation theory. To this end, the chiral Lagrangian is
modified as~\cite{Witten-80,DV-80,RST-80,NA-81,Leutwyler-00}
\begin{equation}
{\cal L}_{\rm ch} = 
{1\over 4} F^2 {\rm Tr} [\partial_\mu U^\dagger \partial_\mu U]
- {1\over 2 } \Sigma\,
{\rm Tr} [M e^{i\theta/N_f} U^\dagger + M^\dagger e^{-i\theta/N_f} U ]
- {1\over 2} \chi \, (\ln {\rm det} U)^2 ,
\label{chirallagrN}
\end{equation}
where $U$ is now a $U(N_f)$ matrix, whose determinant is given by
$\exp(i\sqrt{2N_f}\eta'/F)$.  The last term in this chiral Lagrangian gives
the $\eta'$ mass $m_{\eta'}^2=4N_f\chi/f_\pi^2$.  At large $N$ the pion decay
constant $f_\pi$ is $O(\sqrt{N})$, while the topological susceptibility $\chi$ of the
quenched theory is $O(1)$. Therefore, at large $N$, $m_{\eta'}^2$ is
suppressed by $1/N$, consistently with the $\eta'$ transforming into a
Goldstone boson.  Since $\theta$ is related to the determinant of the quark
mass matrix, by appropriately redefining the field associated with the $\eta'$
particle, one can rewrite Eq.~(\ref{chirallagrN}) as
\begin{equation}
{\cal L}_{\rm ch} = 
{1\over 4} F^2 {\rm Tr} [\partial_\mu U^\dagger \partial_\mu U]
- {1\over 2} \Sigma\, 
{\rm Tr} [M U^\dagger + M^\dagger U ]
+ {1\over 2} \chi \, (i\ln {\rm det} U - \theta)^2 .
\label{chirallagrN2}
\end{equation}
This theory can be used to obtain some information on the $\theta$ dependence
of the spectrum at large $N$, see e.g. Ref.~\cite{BCNW-03}.  Of course, the
$\theta$ dependence must disappear at leading order, and this is encoded in
the above chiral Lagrangian because its last term is suppressed, as a
consequence of the fact that the r.h.s. of the anomaly equation
(\ref{anomaly}) vanishes.

\section{Lattice formulation of the theory}
\label{lattform}

\begin{figure}
\centerline{\psfig{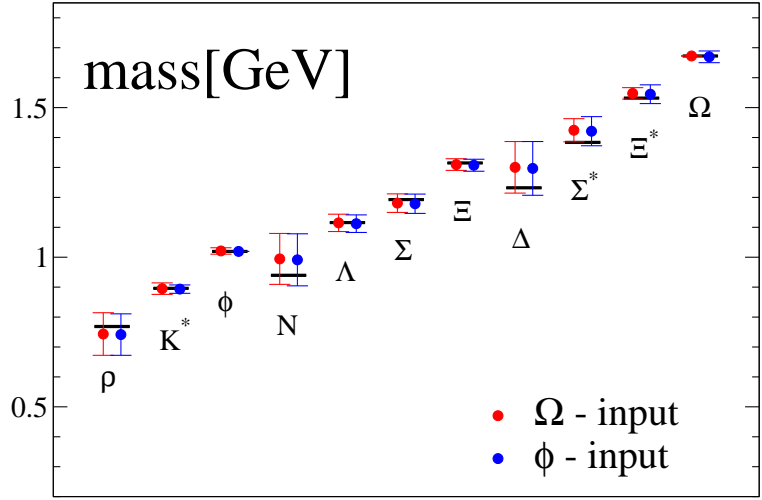}}
\caption{
  Lattice results for the spectrum compared with the experimental values, from
  Monte Carlo simulations of QCD with $N_f=2+1$ dynamical $O(a)$ improved
  Wilson quarks, taken from Ref.~\protect\cite{PACS-CS-07}.  In each pair of
  data, the left (right) point has been expressed in GeV using as input the
  mass of $\Omega\ (\phi)$.  }
\label{spectrum}
\end{figure}

A nonperturbative formulation of QCD is essential for understanding the
low-energy hadronic physics. In this respect the Euclidean lattice formulation
originally proposed by Wilson represented an important breakthrough.  Indeed,
it provided an elegant nonperturbative formulation of QCD, emerging from the
critical continuum limit of a statistical four-dimensional system.  This
opened the road to the use of the powerful numerical techniques of statistical
mechanics, and in particular of Monte Carlo simulations.  During the last few
decades, the results of this approach and their agreement with experiments
have been remarkable (see e.g. the recent
Refs.~\cite{Davies-etal-04,UTfit-06,PACS-CS-07} and Ref.~\cite{PDG-06} for
reviews). For example, the hadron 
masses computed from lattice QCD with $N_f=2+1$ flavors of dynamical quarks
agree quite well with the observed spectrum in nature, as shown in
Fig.~\ref{spectrum}~\cite{PACS-CS-07}.

In statistical mechanics, continuum field theory may be seen as a tool in the
study of critical phenomena~\cite{Wilson-71,WF-72,WK-74}, since the real world
is represented by the lattice models, which for example may be associated to
crystalline structures of solids. Lattices play an important role at short
distance, but become irrelevant in the critical region, where they can be just
seen as regulators of the field theory which describes the critical behavior.
Conversely, in studying fundamental interactions one adopts the opposite point
of view: The real world is represented by a continuum field theory, and a
lattice may be introduced as a regulator for this theory, with the lattice
spacing $a$ as cutoff.  The lattice system has no physical meaning and within
a large class of universality its choice is not unique, but it can be studied
at any temperature/coupling by exploiting the lattice techniques of
statistical mechanics, so that one can get information about its critical
region, which should be described by the initial continuum theory.  This idea
provides a nonperturbartive formulation of a quantum field theory, from the
critical behavior of statistical models.

\subsection{Lattice formulation of  $SU(N)$ gauge theories}
\label{sunlattf}

A lattice formulation of 4D $SU(N)$ gauge theories, preserving local gauge
invariance, is given by the Wilson action~\cite{Wilson-74} defined on a
four-dimensional hypercubic lattice and constructed using link variables
$U_\mu(x)\in SU(N)$, defined for each link between neighboring sites $x$ and
$x+a \hat\mu$ (here, $a$ is the lattice spacing, and $\hat\mu = \hat 1,\hat
2,\hat 3,\hat 4$ stands for a unit vector along one of the spacetime
directions).  The Wilson action is
\begin{equation}
S_L = - {\beta\over N} a^4 \sum_{x,\mu>\nu} {\rm Re} {\rm Tr}\, 
\Pi_{\mu\nu}(x),\qquad \beta={2N\over g_0^2},
\label{wilsonac} 
\end{equation}
where  $g_0$ is the bare coupling and $\Pi_{\mu\nu}$ is the product of
link variables along a $1\times 1$ plaquette of the lattice,
\begin{equation}
\Pi_{\mu\nu}(x) = U_\mu(x)U_\nu(x+a\hat\mu)U^\dagger_\mu(x+a\hat\nu)
U^\dagger_\nu(x).
\label{plaquette}
\end{equation}
Formally, in the $a\rightarrow 0$ limit, performed by setting
$U_\mu=\exp(iaA_\mu)$, where $A_\mu\equiv t^a A_\mu^a$ ($t^a$ are the group
generators, normalized so that $2{\rm Tr}\,t^at^b=\delta^{ab}$),
and expanding in powers of the lattice spacing $a$, we recover the continuum
action
\begin{equation}
S = \int d^4 x {1\over 2g_0^2}{\rm Tr}\,F_{\mu\nu}F_{\mu\nu}.
\end{equation}

The statistical theory develops a mass gap, and therefore a length scale
$\xi$.  The continuum theory is nonperturbatively defined in the critical
limit $g_0^2\to 0$, when $\xi\to \infty$ in units of the lattice spacing.
According to standard renormalization-group arguments, a physical length scale
$\xi$ satisfies the equation
\begin{equation}
{d\over da} \xi = \left( {\partial \over \partial a} 
+ \beta_L(g_0) {\partial \over \partial g_0}\right) \xi = O(a^2),
\label{rgequa}
\end{equation}
where $\beta_L(g_0)$ is the lattice $\beta$-function, defined by
\begin{equation}
\beta_L(g_0)=a{d\over da}g_0, 
\end{equation}
and the derivative is performed keeping the physical quantities fixed.  A
weak coupling expansion gives
\begin{equation}
\beta_L(g_0)= -b_0 g_0^3 - b_1 g_0^5 + ..., 
\end{equation} 
where 
\begin{equation}
b_0={11\over 3}{N\over 16\pi^2}, \qquad
b_1={34\over 3}\left({N\over 16\pi^2}\right)^2. 
\end{equation}
By solving Eq.~(\ref{rgequa}) we obtain the behavior of the length scale as a
function of the bare coupling $g_0$
\begin{equation}
\xi = a \exp \int^{g_0} {dg\over \beta_L(g)}.
\end{equation}
Replacing the perturbative expansion of the $\beta$-function we obtain the
asymptotic behavior
\begin{equation}
\xi 
\sim a\left( b_0g_0^2\right)^{-b_1/2b_0^2}\exp \left( {1\over 2b_0 g_0^2}\right)
\left[ 1 + O(g_0^2)\right].
\label{asysc}
\end{equation}

Dimensionless and renormalization-group invariant physical quantities $R$,
such as the ratio $C\equiv \chi/\sigma^2$ introduced in the preceding section,
approach their continuum limit $R^*$, and therefore their physical value, with
$O(a^2)$ scaling corrections, apart from logarithms of $a$.  Therefore, in the
limit $g_0\to 0$, we have
\begin{equation}
R(g_0^2) - R^* \sim {a^2\over \xi^2}\,.
\label{scal}
\end{equation}
The calculation of dimensionless and renormalization-group invariant
quantities is then favored in that the scaling regime (\ref{scal}) should
set in much before that of the asymptotic scaling formula (\ref{asysc}), whose
corrections get suppressed only logarithmically, i.e. they are $O(1/\ln\xi)$.

The relevant continuum limit must be taken at infinite volume, therefore it
must be reached keeping $1 \ll \xi/a \ll L$ where $L$ is the size of the
lattice.  At finite $\xi$, in order to get infinite volume results, one may
perform calculations at large but finite volume (finite size scaling theory
allows one to control finite size effects in the scaling region, see e.g.
Ref.~\cite{Cardy-88}), thus working with a finite, although large, number of
variables.  One can then perform numerical Monte Carlo simulations in order to
obtain nonperturbative information close to criticality; extrapolations
controlled by field theory (using Eqs. (\ref{asysc}) and (\ref{scal})) lead to
the desired physical numbers of the continuum theory.

\subsection{Fermions on the lattice}
\label{fermions}

The major difficulty in introducing fermions on the lattice is that one
usually ends up with having more fermions in the continuum limit than
intended. This is dubbed the ``fermion doubling'' problem; for example, the
naive discretization of the Dirac operator leads to sixteen fermions in the
continuum limit.  The no-go theorem~\cite{NN-81} of Nielsen and Ninomiya
states that this is unavoidable under a few plausible assumptions.  See, for
example, Refs.~\cite{Creutz-book,MM-94} for general discussions of this issue.
In the following we briefly discuss the various proposals that have been
considered so far, up to the various versions of so-called Ginsparg-Wilson
fermions, and in particular the overlap fermions, which have circumvented the
no-go theorem, leading to a lattice regularization of fermions coupled
vectorially to a gauge field which preserves chiral symmetry.  Recent reviews
discussing chiral symmetry on the lattice are found in
Refs.~\cite{Neuberger-01,Creutz-01,CW-04,Hasenfratz-04,Zinn-Justin-05,Niedermayer-98}.

\subsubsection{Wilson fermions}
\label{wferm}

In Wilson's formulation~\cite{Wilson-74} of lattice QCD, the desired number of
fermions in the continuum limit is achieved by adding an appropriate
irrelevant term to the naive lattice discretization of the Dirac operator.
The Wilson action for QCD is given by
\begin{eqnarray}
S_L =  -{\beta\over N} \sum_{x,\mu,\nu} {\rm Re} {\rm Tr}\, \Pi_{\mu\nu}(x)
+ \sum_{x,f} \bar\psi_f(x) (D_{\rm W} + m_{0,f}) \psi_f(x).
\label{latact}
\end{eqnarray}
$D_W$ is the Wilson Dirac operator
\begin{equation}
D_{\rm W} = {1\over 2} \left[ \gamma_\mu \left( \nabla_\mu^*+\nabla_\mu\right)
 - r a\nabla_\mu^*\nabla_\mu \right],\label{wdiracop} 
\end{equation}
where
\begin{eqnarray}
&&\nabla_\mu\psi(x) = {1\over a} \left[ U(x,\mu) \psi(x + a\hat{\mu})
- \psi(x)\right],\\
&&\nabla_\mu^*\psi(x) = {1\over a} 
\left[ \psi(x) - U(x-a\hat\mu,\mu)^\dagger \psi(x - a\hat{\mu})\right],
\nonumber
\end{eqnarray}
$r$ is the Wilson parameter and $f$ is a flavor index.  The $r$ term breaks
chiral symmetry, but it becomes an irrelevant operator in the continuum limit,
where chiral symmetry should be recovered. Therefore, the price paid to have
the correct number of fermions in the continuum limit is an explicit breaking
of the chiral symmetry, which causes $O(a)$ lattice artefacts and other
unwanted effects at finite lattice spacing.  For example, the quark masses are
affected by an additive renormalization, which shifts the chiral limit to a
negative value of $m_{0,f}$.  Improved versions of the Wilson action are
obtained by implementing the Symanzik improvement
procedure~\cite{Symanzik-83}, using for example the so-called clover
action~\cite{LSSWW-97,SW-85}, where an appropriate ($\beta$ dependent) tuning
of an additional parameter in the action can achieve the cancellation of the
$O(a)$ scaling corrections, leaving only $O(a^2)$ corrections.

An alternative discretization of QCD is given by staggered
fermions~\cite{Kogut-75,Susskind-77}.  This is for example described in
textbooks such as \cite{MM-94,Rothe-book}.  Staggered fermions achieve a
partial reduction in the degrees of freedom of naive fermions by a so-called
spin diagonalization.  Moreover, they exhibit a residual lattice symmetry
which protects the quark mass from additive renormalization, and scaling
corrections are $O(a^2)$.  The major drawback of staggered fermions is that
fermion doubling is only partially reduced, from sixteen to four species, but
this does not reproduce QCD. Monte Carlo simulations with fewer that four
fermion species have been performed by taking fractional powers of the
fermionic determinant in the functional integral. While the results obtained
so far appear quite promising, see e.g. Ref.~\cite{Davies-etal-04}, the full
correctness of the corresponding continuum limit is still debated, see e.g.
Refs.~\cite{Creutz-07,Kronfeld-07}.

\subsubsection{Ginsparg-Wilson fermions}
\label{gwferm}

The conceptual problems related to chiral symmetry can be overcome if the
lattice Dirac operator $D$ satisfies the so-called Ginsparg-Wilson (GW)
relation~\cite{GW-82} which, in its simplest form, may be cast as
\begin{equation}
\gamma_5 D + D\gamma_5 = a D\gamma_5 D.
\label{GWrel}
\end{equation}
The GW relation implies the existence of an exact chiral symmetry of the lattice
action under the transformation~\cite{Luscher-98}
\begin{eqnarray}
&&\delta\psi(x) = \varepsilon \gamma_5 \left( 1 - {1\over 2}a D\right)\psi(x),
\label{latchsym} \\
&&\delta\bar{\psi}(x) = \varepsilon \bar{\psi}(x)\left(1-{1\over 2}a D\right)\gamma_5.
\nonumber
\end{eqnarray}
This symmetry protects the quark masses from additive
renormalizations~\cite{Neuberger-98,Hasenfratz-98}, and lattice effects are
O($a^2$) in the continuum limit.  The associated chiral Ward identities ensure
the nonrenormalization of vector and flavor nonsinglet axial vector
currents, and the absence of mixing among operators in different chiral
representations.  Thus, the pattern of lattice renormalization becomes greatly
simplified with respect to lattice fermions violating chiral
symmetry~\cite{Hasenfratz-98,AFPV-00,Capitani-03}.  Lattice gauge theories
with GW fermions have been proved to be renormalizable to all orders of
perturbation theory~\cite{RR-99}.

Lattice Dirac operators satisfying Eq.~(\ref{GWrel}) are not affected by the
Nielsen-Ninomiya theorem~\cite{NN-81}, thus they need not suffer from fermion
doubling.  Note however that, although the GW relation implies some
generalized form of chirality on the lattice, it does not guarantee the
absence of doublers.  Solutions of the GW relation without doublers have been
found, such as domain wall fermions~\cite{Kaplan-92,Shamir-93,FS-95}, overlap
formulations~\cite{NN-93,NN-93-b,NN-95,Neuberger-98,Neuberger-98-2,Neuberger-99,
Neuberger-01}, and the so-called classical fixed-point Dirac
operator~\cite{HLN-98,BW-96,HN-94}.  A lattice formulation of QCD satisfying
the Ginsparg-Wilson relation overcomes the complications of the standard
approaches based on Wilson or staggered fermions, where chiral symmetry is
violated at the scale of the lattice spacing.  However, the numerical
implementation of these chirality preserving Dirac operators is significantly
more CPU intensive than the standard approaches.

The GW relation and the associated symmetry (\ref{latchsym}) imply relations
at finite lattice spacing which are substantially equivalent to those holding
in the low-energy phenomenology associated with chiral symmetry (see e.g.
Refs.~\cite{Chandrasekharan-99,KY-99-2,Niedermayer-98}).  A natural candidate
for the chiral condensate at finite lattice spacing is given by
\begin{equation}
\Sigma_L = - \langle \bar{\psi} \left( 1 - {1\over 2}a D\right) \psi \rangle,
\label{lattcond}
\end{equation}
which can be easily obtained by applying a nonsinglet transformation to the
quantity $\sum_x \langle \bar{\psi} t_a \gamma_5 \psi(x)\rangle$.  $\Sigma_L$
is the order parameter which is expected to be nonzero in the thermodynamic
limit,
\begin{equation}
\lim_{m\to 0 } \lim_{V\to\infty} \Sigma_L \neq 0,
\label{lattcond2}
\end{equation}
leading to the spontaneous breaking of chiral symmetry.  As shown in
Ref.~\cite{Chandrasekharan-99}, massless pions emerge when the lattice chiral
symmetry is broken, due to the nonzero condensate (\ref{lattcond2}). This can
be inferred by considering the zero-momentum pion correlation function
\begin{equation}
G_{ab} = \langle  \sum_x \langle \bar{\psi}(0) t_a \gamma_5 \psi(0)\, 
\bar{\psi}(x) t_b \gamma_5 \psi(x) \rangle
\label{gab}
\end{equation}
in the presence of a mass term $m$. Using the GW relation, in the chiral limit
one arrives at the relation~\cite{Chandrasekharan-99}
\begin{equation}
\lim_{m\to 0 } G_{ab} = \delta_{ab} {1 \over a^4 m} 
\langle \bar{\psi} \left( 1 - {1\over 2}a D\right) \psi \rangle .
\label{gab2}
\end{equation}
Thus, a nonzero condensate implies that the r.h.s.  is singular in the chiral
limit. Since $G_{ab}\sim m_\pi^{-2}$, this leads to the well known relation
$m_\pi^2 \sim m$.

The axial anomaly then arises from the noninvariance of the fermion integral
measure \cite{Luscher-98} under flavor-singlet chiral transformations
(\ref{latchsym})
\begin{equation}
\delta [d\bar{\psi} d\psi] 
= \varepsilon a {\rm Tr}(\gamma_5 D) [d\bar{\psi} d\psi] 
= - 2 \varepsilon N_f (n_+ - n_-) [d\bar{\psi} d\psi] ,
\label{measure}
\end{equation}
where $n_\pm$ are the number of zero modes of $D$ which are also eigenstates
of $\gamma_5$ with eigenvalues $\pm 1$, see also
Refs.~\cite{Luscher-99,Chiu-99,Fujikawa-99,Suzuki-99,Adams-02}. This implies
that the flavor-singlet chiral transformations present an anomaly on the
lattice like that in the continuum, and provide a natural fermionic definition
$q_i(x)$ of topological charge density as~\cite{HLN-98}
\begin{equation}
q_i(x) = {1\over 2} {\rm Tr} [\gamma_5 D(x,x)],
\label{qxgw}
\end{equation}
and an index theorem on the lattice, i.e. 
\begin{equation}
Q_i=\sum_x q_i(x) = {\rm index}(D).
\label{indtheo}
\end{equation}
We will return to this point in Sec.~\ref{fermdefGW}.  Considerations on the
flavor-singlet pion correlator~\cite{Chandrasekharan-99} lead also to the
relation (where all flavors are assumed degenerate)
\begin{equation}
{1\over V} \langle (n_+ - n_-)^2 \rangle 
= {m \Sigma_L \over N_f}  
\label{qms}
\end{equation}
to leading order in the chiral limit. This is analogous to the continuum
relation derived in Sec.~\ref{chiralpert}, cf. Eq.~(\ref{chichi}).

The same result can be inferred~\cite{Niedermayer-98} by introducing a
$\theta$ parameter as in the continuum.  In the case of $N_f$
degenerate quarks, we can write the lattice action as
\begin{equation}
S_\theta = 
S_{\rm gauge} + \sum_x \left[ \bar\psi D \psi + 
\mu \bar\psi_R \psi_L +  \mu^* \bar\psi_L \psi_R \right] - i \theta \sum_x q(x),
\end{equation}
where $\mu$ is a complex mass and $q(x)$ is given by Eq.~(\ref{qxgw}).  When
transforming the fermion fields as $\psi_L\to e^{-i\alpha} \psi_L$, one can
restore the invariance of the action by changing $\mu\to e^{i\alpha} \mu$ and
the change of the measure can be cancelled by $\theta\to \theta-N_f \alpha$.
Therefore, the ground-state energy should only depend on the combination
\begin{equation}
\mu e^{i\theta/N_f}
\label{depmth}
\end{equation}
as in the continuum.  By repeating the arguments of Ref.~\cite{LS-92}, leading
to Eq.~(\ref{etheta}), with the same natural assumptions, the ground-state
energy density can be written as
\begin{equation}
F(\theta) = - N_f \Sigma_L {\rm Re} \left( \mu e^{i\theta/N_f}\right) 
+ O(|\mu|^2).
\label{fffres}
\end{equation}
Then, by taking the second derivative with to respect to $\theta$ for real
$\mu=m$, one obtains Eq.~(\ref{qms}).

Let us briefly mention the known examples of Dirac operators that satisfy the
GW relation. One of them is the ``fixed-point'' Dirac operator, which is based
on the idea of a classically perfect action obtained from the fixed point of
appropriate renormalization-group transformations as determined by classical
saddle point equations~\cite{HN-94}.  At the classical level these actions do
not have cutoff effects; however, $O(a^2)$ cutoff effects reappear at the
quantum level.  This idea has been applied to QCD in
Refs.~\cite{DHHN-95,BW-96,DHZ-96}.  The corresponding Dirac operator satisfies
the GW relation~\cite{Hasenfratz-98}.  Domain wall fermions are based on the
idea~\cite{Kaplan-92} that massless 4D lattice fermions arise naturally as
zero modes localized on a domain wall embedded in a 5D space-time.  This
allows to construct a lattice vector-like gauge theory preserving chiral
symmetry.  The five-dimensional fermion action to represent quarks in lattice
QCD was proposed in Refs.~\cite{Shamir-93,FS-95}.  Finally, there is the
overlap Dirac operator proposed by Neuberger~\cite{Neuberger-98}, which is
described below.

\subsubsection{The Neuberger overlap Dirac operator}
\label{overlapop}

The simplest lattice formulation satisfying the GW relation is Neuberger's
overlap Dirac operator~\cite{Neuberger-98,Neuberger-98-2,Neuberger-99}; it
has been derived from the overlap formulation of chiral fermions on the
lattice~\cite{NN-95}, and can be shown to be an effective Dirac operator
which describes the massless chiral mode of the domain wall fermion. It is
given by
\begin{eqnarray}
D_{\rm N} &=& {\rho \over a} \left[  1 + X (X^\dagger X)^{-1/2} \right],
\label{Nop}\\
X &=& D_{\rm W} - {1\over a}\rho,
\label{Xdef}
\end{eqnarray}
where $D_{\rm W}$ is the Wilson-Dirac operator (\ref{wdiracop}), and $\rho$ is
a free real parameter whose value must lie within a certain range, in order to
guarantee the correct pole structure of the fermion propagator; the
perturbative limits of that range are $0< \rho < 2$. The Fourier
representation of $D_{\rm N}$ shows that no fermion doublers appear.  Several
perturbative studies have been performed with this fermion lattice
formulation, see e.g.~\cite{KY-99,AFPV-00,APV-00,CG-00,HPRSS-04,CP-07}.

As shown in Refs.~\cite{Horvath-98,HBM-01}, under some general assumptions
which include the absence of doublers, GW fermions cannot be ultralocal, i.e.
any lattice variable must be coupled to an infinite number of other variables.
Indeed, the Neuberger-Dirac operator $D_{\rm N}$ is not strictly local, and
locality should be recovered only in a more general sense, i.e.  allowing an
exponential decay of the kernel of $D_{\rm N}$ at a rate which scales with the
lattice spacing and not with the physical quantities.  In Ref.~\cite{HJL-99}
the locality of $D_{\rm N}$ has been proved for sufficiently smooth gauge
fields. Moreover numerical evidence has been presented for typical gauge
fields in present-day simulations.

Unlike the Wilson-Dirac operator $D_{\rm W}$, $D_N$ is not analytic in the
link variables when the operator $X$ in Eq.~(\ref{Nop}) has a zero eigenvalue.
However, such a lack of analyticity is expected to be harmless in the
continuum limit~\cite{Neuberger-99,HJL-99}. The nonanalyticity of general GW
fermions, and in particular of overlap fermions, is strictly related to the
existence of the lattice index theorem.  On the one hand, the lattice
configuration space is simply connected when using conventional actions. On
the other hand, the lattice index, corresponding to an integer winding number,
must be singular as one passes from one sector to another~\cite{Creutz-02}.
The locations of these singularities actually depend on the particular GW
Dirac operator.  One way to circumvent these nonanalyticities is by putting
constraints on the roughness of the gauge fields~\cite{HJL-99,Neuberger-00}.
Note that this would also imply that transitions among different topological
sectors become suppressed.  Actions implementing this idea have been
considered in Refs.~\cite{BJNNSS-06,FHHOO-06}.  As noted in
Ref.~\cite{Creutz-04}, such constraints would lead to actions without a
positive transfer matrix.

In conclusion, the Neuberger overlap Dirac operator provides a lattice
regularization of massless QCD, i.e.  of chiral fermions coupled vectorially
to a gauge field, without the need of fine tuning.  We should also say that
the numerical implementation of dynamical overlap fermions is still a great
challenge today. Indeed their Monte Carlo simulations are considerably slower
than simulations of Wilson or staggered fermions, see e.g.
Refs.~\cite{NNV-95,Neuberger-98-3,Neuberger-99,FKS-04,DS-05-2,DS-05,EFKS-06,
Aoki-etal-07,DS-07}.
A review of Monte Carlo algorithms for simulations with overlap fermions is
reported in Ref.~\cite{Schaefer-06}.

\section{Topology from the lattice}
\label{latttopsec}

The topological winding number, which classifies continuum 4D $SU(N)$
gauge field configurations~\cite{Hooft-76b}, relies on certain
smoothness assumptions.  This winding number is uniquely defined for
smooth fields, however the path integral requires integration over all
configurations, some of which may not be sufficiently smooth. In
particular, the lattice regularization makes the topology strictly
trivial, because the configuration space of lattice fields is simply
connected. It is only in the continuum limit that physical topological
properties are expected to be recovered.

\subsection{The topological charge density on 
the lattice and its renormalization}
\label{latttop}

We first recall that, in pure gauge theories, the topological charge
density $q(x)$ is a renormalization-group invariant composite
operator, i.e. its anomalous dimension is zero, when it is defined in
an appropriate renormalization scheme, such as minimal subtraction
$\overline{\rm MS}$~\cite{ET-82}.  One can straightforwardly define
lattice operators corresponding to the topological charge density, by
constructing a local function $q_L(x)$ of the lattice fields which has
the topological charge density $q(x)$ as its classical continuum
limit:
\begin{equation}
q_L(x)\longrightarrow a^4 q(x) + O(a^{6}).
\label{limitq}
\end{equation}
$q_L(x)$ is not unique, indeed infinitely many choices differing by
$O(a^6)$ terms can be conceived.  Various lattice versions $q_L(x)$
have been introduced in
Refs.~\cite{DFRV-81,DFRV-82,BGO-82,FR-83,MZG-83,CDPV-96}.  A simple
example is given by the twisted double plaquette
operator~\cite{DFRV-81}
\begin{equation}
q_{L}(x) = - {1\over 2^4\times 32 \pi^2} \sum^{\pm 4}_{\mu\nu\rho\sigma=\pm 1}
\epsilon_{\mu\nu\rho\sigma} {\rm Tr} \left[ \Pi_{\mu\nu}\Pi_{\rho\sigma}\right],
\label{qL}
\end{equation}
where $\Pi_{\mu\nu}$ is the $1\times 1$ plaquette operator defined in
Eq.~(\ref{plaquette}).  One can easily check that, classically,
Eq.~(\ref{limitq}) is satisfied.  The classical continuum limit of
lattice operators must be in general corrected by lattice
renormalizations at the quantum level, see e.g.
Ref.~\cite{BMMRT-85}. This also holds in the case of the topological
charge density, thus we have~\cite{CDP-88}
\begin{equation}
q_{L}(x)\longrightarrow a^4 Z_L(g_0^2) q(x) + O(a^{6}),
\label{renorm}
\end{equation}
where $Z_L(g_0^2)$ is a finite function of the bare coupling $g_0^2$
going to one in the limit $g_0^2\rightarrow 0$, which can be proven
using the fact that the continuum operator $q(x)$ has zero anomalous
dimension.  The function $Z_L(g_0)$ can be computed in perturbation
theory~\cite{CDP-88}; some perturbative calculations are reported in
Refs.~\cite{CDP-88,CDPV-90,DPV-90,AV-91,CDPV-96,SP-05}.  For example,
for the operator (\ref{qL}) a one loop calculation~\cite{CDP-88} gives
\footnote{The lattice topological charge density is renormalized
in such a way that its renormalized correlation functions coincide
with those obtained in the $\overline{\rm MS}$ renormalization scheme.}
$Z_L(g_0^2)= 1 - 0.908 g_0^2 + O(g_0^4)$ for $SU(3)$.  This straightforward
approach allows us to compute general $d$-dimensional Euclidean correlation
functions of the topological charge density operator,
\begin{equation}
G_q(x_1,x_2,...,x_n) \equiv \langle q(x_1) q(x_2) ... q(x_n) \rangle, 
\label{gqdef}
\end{equation}
by exploiting the field-theoretical relation
\begin{equation}
G_q^{(n)}(x_1,x_2,...x_n)=
a^{-nd} \, Z_L(g_0)^{-n} \langle q_L(x_1) q_L(x_2) ... q_L(x_n) \rangle + O(a).
\label{corrdef}
\end{equation}
Analogous relations can be written down for correlation functions with
insertions of other operators at different space-time positions.
However, it is important to stress that this holds only when $x_i\neq
x_j$ for any $i\neq j$. Indeed, the coincidence of two spacetime
arguments gives rise to peculiar contact terms, which make the
relation between lattice and continuum correlation functions
complicated, as we shall see later.

The above relations become more intricate in the presence of fermions,
essentially because the topological charge density is not
renormalization-group invariant anymore; instead, it presents a mixing
with fermionic operators at the quantum level.  Unlike pure gauge
theory, in full QCD the topological charge density mixes under
renormalization with $\partial_\mu j_\mu^5$.  The nonrenormalization
property of the anomaly in the ${\overline {\rm MS}}$ scheme means
that the anomaly equation should assume exactly the same form in terms
of bare and renormalized quantities~\cite{AB-69}. However, the
renormalization of $\partial_\mu j^5_\mu(x)$ and $q(x)$ is
nontrivial. In the chiral limit the Euclidean anomaly equation reads
\begin{equation}
\partial_\mu j^5_\mu(x)=i2N_f q(x).
\label{anomeq}
\end{equation}
Renormalized operators in ${\overline {\rm MS}}$ are obtained
by~\cite{Bardeen-69,AB-69,ET-82,SV-92,Larin-93,Testa-98}
\begin{equation}
\left( \matrix{  i2N_f\,q(x)\cr \partial_\mu j_\mu^5(x)\cr}\right)_R
\;=\
\left( \matrix{ 1&z-1\cr 0&z\cr}\right)\;
\left( \matrix{  i2N_f\,q(x)\cr \partial_\mu j_\mu^5(x)\cr}\right)_B
\label{cren}
\end{equation}
where
\begin{equation}
z= 1 + {g^4\over 16\pi^4} {3c_{_F}\over 8}  N_f  {1\over \epsilon}+
O\left( g^6\right),
\label{ztwol}
\end{equation}
$\epsilon = 2 - d/2$ and $c_{_F}=(N^2-1)/(2N)$.  The structure of the
renormalization matrix above assures the stability of the anomaly
equation under renormalization. As a consequence of Eq.~(\ref{cren}),
the topological charge density has an anomalous dimension, and
therefore the relation between the generic matrix elements of any
lattice version $q_L$ of the topological charge density and its
continuum counterpart cannot be as simple as in
Eq.~(\ref{corrdef}). Using general renormalization-group arguments,
one can show that~\cite{ADPV-95}
\begin{equation}
\langle a| i2N_f q_{_L} | b \rangle = Y(g_0^2)\,\langle a| R | b \rangle \;,
\label{ql}
\end{equation}
where $a,b$ are generic states, $Y(g_0^2)$ is a finite function of $g_0^2$,
which can be computed in perturbation theory, and
\begin{equation}
\langle a| R |b\rangle \;\equiv\;
\langle a| \partial_\mu j_\mu^5(x)_{{\overline {\rm MS}}} | b \rangle 
\exp \int^0_{g(\mu)}
{\bar{\gamma}(\tilde{g})\over \beta_{{\overline {\rm MS}}}(\tilde{g})} {\rm d} \tilde{g}
\label{tt}
\end{equation}
is a renormalization-group invariant quantity; $\partial_\mu
j_\mu^5(x)_{{\overline {\rm MS}}}$ indicates the operator $\partial_\mu
j_\mu^5(x)$ renormalized in the ${\overline {\rm MS}}$ scheme, and the
function $\bar{\gamma}(g)$ is related to the anomalous dimension of the
continuum operators $q(x)$, $\partial_\mu j^5_\mu(x)$ in the ${\overline {\rm
    MS}}$ scheme:
\begin{equation}
\bar{\gamma}(g)=
\mu {{\rm d} \over {\rm d}\mu} \ln z =
-{1\over 16\pi^4} {3c_{_F}\over 2} \, N_f \,g^4\;+\;O\left(g^6\right)\;.
\end{equation}
In the usual case of the operator (\ref{qL}), one obtains
\begin{equation}
Y(g_0^2)= 1-0.908 g_0^2 + {3N_f \over \pi^2 (33 - 2 N_f)} g_0^2  + O(g_0^4)
\end{equation}
for $N=3$, which leads to $Y(g_0^2)= 1-0.887 g_0^2 + O(g_0^4)$ for $N_f=2$.

As an example of physically relevant matrix elements of the topological charge
density, we mention the one over proton states which enters the so-called
proton spin problem~\cite{Veneziano-89,SV-92-2,NSV-99,Shore-06,Shore-07},
because it can be related to on-shell nucleon matrix elements of the singlet
axial current $j_\mu^5$ through the anomaly.  Indeed, the on-shell nucleon
matrix element of the topological charge density
\begin{equation}
\langle\, \vec{p},e |\,q\,| \vec{p}\,', e'\,\rangle\;=\;
MB(k^2)\,\bar{u}(\vec{p},e)i\gamma_5 u(\vec{p}\,',e')
\label{qp}
\end{equation}
can be related to the on-shell nucleon matrix element of the singlet axial
current $j_\mu^5$,
\begin{equation}
\langle\, \vec{p},e |\,j_\mu^5\,| \vec{p}\,', e'\,\rangle\;=\;
\bar{u}(\vec{p},e)\left[ G_1(k^2)\gamma_\mu \gamma_5\,-\,
G_2(k^2)k_\mu\gamma_5\right] u(\vec{p}\,',e'),
\label{jmatrix}
\end{equation}
where $e,e'$ label the helicity states and $k$ is the momentum transfer.  In a
naive wave function picture $G_1(0)$ can be interpreted as the fraction of the
nucleon spin carried by the quarks.  Using the axial anomaly equation one can
easily show that $B(0) =G_1(0)/N_f$.  Lattice studies of this issue have been
reported in Refs.~\cite{FKOU-95,SOOB-03,AliKhan-05,ALNT-06,Edwards-etal-06}.

\subsection{Nonrenormalizability of the $\theta$ term in full QCD}
\label{thetanor}

Although the topological charge density operator renormalizes
nontrivially, according to Eq.~(\ref{cren}), the relevant $\theta$
parameter in the Lagrangian of full QCD does not get
renormalized. This can be proved within the ${\overline {\rm MS}}$
scheme, exploiting the nonrenormalizability of the composite operator
\begin{equation}
O_f = {1\over 2}\left[ m_f \bar{\psi}_f(x)(1+\gamma_5)\psi_f(x) + {\rm h.c.}\right]
\label{mfdef}
\end{equation}
entering Eq.~(\ref{lmdef}).  Indeed, performing a chiral
transformation (\ref{chiralpsi}) on the Lagrangian (which allows us to
arbitrarily change $m_f \to e^{2 i \alpha_f}\, m_f$ and
correspondingly shift $\theta\to \theta - 2\sum_f \alpha_f$), the
topological term can be absorbed by the complex mass term; the fact
that the phase of the mass does not renormalize implies the
nonrenormalizability of $\theta$, which is defined by the condition
${\rm Im}\,m_f=0$ in order to fix the arbitrariness due to the chiral
transformation (\ref{chiralpsi}).  Consequently, the expansion of the
ground state energy (expressed in physical units) in powers of
$\theta$ will have coefficients which are renormalization group
invariant.

Alternatively, the nonrenormalizability of the parameter $\theta$ can
be also shown using the fact that the anomaly equation takes exactly
the same form in terms of bare and renormalized quantities,
\begin{eqnarray}
\partial_\mu j^5_\mu(x) = i 2 p(x) + i2N_f q(x),
\label{anomeqm}
\end{eqnarray}
where $j_\mu^5(x) = \sum_{f=1}^{N_f} \bar{\psi}_f(x)\gamma_\mu\gamma_5\psi_f(x)$
and $p(x) = \sum_{f=1}^{N_f} m_f \bar{\psi}_f(x)\gamma_5\psi_f(x)$.

Notice that the massless limit appears singular in this respect, just
because the $\theta$ term can be completely eliminated by a chiral
transformation without any physical effect.

The above results imply that in taking the correct continuum limit $a\to 0$
of lattice QCD, with real fermion masses and $\theta$ term, one must
keep the parameter $\theta$ fixed (apart
from a finite lattice renormalization, just as in pure SU($N$) gauge
theories).

\subsection{The topological susceptibility and 
problems from power-divergent additive contributions}
\label{chilren}

Serious problems arise when the quantity that we want to study involves also
correlation functions (\ref{gqdef}) with coincident points, i.e.  in the limit
$|x_i-x_j|\to 0$. This is for example required to compute the $\theta$
dependence of the ground-state energy around $\theta=0$.  In particular this
limit enters the definition of topological susceptibility
\begin{equation}
\chi = \int d^4x \langle q(0) q(x) \rangle .
\label{chic}
\end{equation} 
The lattice counterpart $\langle q_L(x) q_L(y) \rangle$ does not reconstruct
correctly the singular behavior for $x\to y$, which, as we shall see, is
essential to determine the physical value of the topological susceptibility,
see in particular Sec.~\ref{qqcorr}.  The relation of the zero-momentum
correlation of two lattice operators $q_{L}(x)$,
\begin{equation}
\chi_{L}= \sum_x \langle q_{L}(x)q_{L}(0) \rangle = {1\over V} \langle Q_L^2 \rangle,
\qquad Q_L = \sum_x q_L(x),
\label{childef}
\end{equation} 
with the continuum $\chi$ is affected by the presence of an unphysical
background term, which becomes dominant in the continuum limit. In general, we
expect
\begin{equation}
\chi_{L}(g_0)= a^4 Z_L(g_0)^2\chi+B.
\label{chileq}
\end{equation}
Heuristic arguments based on the operator product expansion
would lead to~\cite{CDPV-90} 
\begin{equation}
B(g_0)= P(g_0) \langle I \rangle  + a^4 A(g_0) \langle T \rangle + O(a^{5}).  
\label{back} 
\end{equation}
where $I$ is the identity operator and $T$ is the trace of the energy-momentum
tensor.  Therefore, the background term is a power divergent additive
contribution, which eventually becomes dominant in the continuum limit. The
so-called perturbative tail $P(g_0)$ can be computed in perturbation theory.
For some definitions of $q_L$ it has been computed to high order, see e.g.
Refs.~\cite{DFRV-81,ACFP-94,SP-05}.  However, the perturbative series is only
asymptotic, and it is expected to have nonperturbative contributions due to
renormalons which cannot be disentangled from the $O(a^4)$ physical signal,
see e.g.  Refs.~\cite{DMO-95,Beneke-99,DS-01}.  Therefore, perturbation theory
does not help to disentangle the interesting $O(a^4)$ physical signal in
Eq.~(\ref{chileq}) from the {\rm renormalization} effects.  The computation of
$\chi$ through Eq.~(\ref{chileq}) requires a well defined nonperturbative
subtraction of these terms, and therefore a rigorous field-theoretical
prescription to unambiguously extract the quantity which has the desired
continuum limit from the lattice data. In this respect, the main problem 
may be related to the fact that even in the continuum the integrand $\langle
q(x) q(0) \rangle$ in the definition of $\chi$ is singular for $x\to 0$, and
it is not completely clear what prescription provides the correct physical
quantity after integration.  In Sec.~\ref{fermdefGW} we will report the
solution of the problem as outlined in Ref.~\cite{Luscher-04}.  The singular
behavior of $\langle q(x) q(0) \rangle$ for $x\to 0 $ is discussed in some
detail in Sec.~\ref{qqcorr}.

\subsection{Bosonic definitions exploiting various methods}
\label{bosmeth}

Geometrical, smoothing (such as cooling and smearing), off-equilibrium (such
as heating) techniques have been used to address the problems caused by
power--divergent additive contributions and multiplicative renormalizations in
definitions of the topological susceptibility based on discretized versions of
the topological charge density operator $q(x)$.

\subsubsection{Geometrical method}

The so-called geometrical method~\cite{BL-81,Luscher-82,PS-86} meets the
demands that the topological charge on the lattice have the classical correct
continuum limit and that it take integer values for every lattice
configuration in a finite volume with periodic boundary conditions.  In 4D
$SU(N)$ gauge theories this can be achieved by performing an interpolation of
the lattice field, from which the principal fibre bundle is reconstructed. One
can unambiguously assign a topological charge to a configuration, provided it
is sufficiently smooth and satisfies certain bounds~\cite{Luscher-82}; for
example, in the case of $SU(2)$ gauge theory, plaquettes must
satisfy\cite{Luscher-82}
\begin{equation}
{\rm Tr} [1 - \Pi_{\mu\nu}]< 0.03.  
\end{equation}
However, the required constraints are far from the values of the plaquettes
in actual simulations, thus leaving some ambiguities in
the assignment of topological charge to each Monte Carlo configuration.
Due to their global topological stability, geometrical definitions are not
affected by perturbative renormalizations.  Since on the lattice each
configuration can be continuously deformed into any other, integer valued
geometrical definitions cannot have an analytical functional dependence on the
lattice field.

Geometrical definitions may be plagued by topological defects on the scale of
the lattice spacing, termed dislocations~\cite{Berg-81,Luscher-82-2}, whose
nonphysical contribution may either survive in the continuum limit (as
could happen in the $SU(2)$ gauge theory with Wilson action~\cite{PT-89}), or
push the scaling region for the topological susceptibility to large $\beta$
values.  These warnings have been substantially confirmed by the available
results for $SU(2)$ and $SU(3)$ gauge theories, and two-dimensional $CP^{N-1}$
models.  For example, the results for the
$SU(2)$~\cite{KLSW-87,GKLSW-88,KLSSW-88,PT-89} and the
$SU(3)$~\cite{GKLSW-87,GKLSW-88} gauge theories (the latter obtained for
$\beta\lesssim 6.0$) gave significantly larger values than those of other
methods. A better behavior has been observed when measuring the geometrical
charge on blocked configurations~\cite{PT-89}.  The problem was reanalyzed in
Ref.~\cite{RRV-97} by exploiting the fact that the geometrical definitions can
be obtained as limits of sequences of standard analytic operators. The results
indicate that, with increasing $N$, the unphysical contribution should get
suppressed, and that at least at large $N$ the geometrical charge should be
free from dislocations.  This scenario has been checked in two-dimensional
$CP^{N-1}$ models by Monte Carlo simulations~\cite{CRV-92-2}.

Monte Carlo results using this method can be found in
Refs.~\cite{GKLSW-87,KLSW-87,GKLSW-88,KLSSW-88,PT-89} for the 4D
$SU(2)$ and $SU(3)$ lattice gauge theories, and in
Refs.~\cite{Berg-81,JW-92,Wolff-92,CRV-92,CRV-92-2,v-93,DMV-04} for the 2D
$CP^{N-1}$ models.

\subsubsection{Smoothing methods}

In this class of methods the topological properties are read from the
configurations generated by Monte Carlo simulations after applying
appropriate smoothing procedures which eliminate the short-ranged
renormalization effects and provide an (approximately) integer value for the
topological charge. Various smoothing procedures have been proposed and
employed in numerical works.

In the cooling method, see e.g.
Refs.~\cite{Berg-81,IY-83,IIY-84,Teper-85,ILMSS-86}, $\chi$ is measured on an
ensemble of configurations which are ``cooled''; that is, link variables are
gradually changed in a way as to minimize the action locally. For example,
cooling can be achieved by setting $\beta=\infty$ in a standard Monte Carlo
updating procedure, such as Metropolis or heat bath.  By virtue of the
locality of this procedure, it is ideally expected to eliminate the short
ranged fluctuations responsible for the renormalization effects, without
modifying the global topological content of a configuration. The latter may
then be extracted from the corresponding cooled configuration, finding an
(almost) integer value of $Q_L=\sum_x q_{L}(x)$, where $q_L(x)$ may be any
lattice discretization of the topological charge density, such as the one in
Eq.~(\ref{qL}).  Different implementations may differ in the speed of the
minimization procedure and in the way one reads the value of the topological
charge $Q$ from the cooled configuration.  The topological susceptibility is
obtained by measuring $\chi_L=\langle Q^2 \rangle/V$ on the cooled
configurations, typically when the result becomes stable with respect to the
number of cooling steps.

One possible realization of cooling is by means of a smearing procedure,
analogous to that proposed in Ref.~\cite{APE-87}, in which the original links
of a configuration are replaced by {\em smeared} links, constructed as
follows:
\begin{eqnarray}
V_\mu^{(0)}(x)&\equiv& U_\mu(x),\nonumber \\
\widehat{V}_\mu^{(i)}(x)&=& (1-c)V_\mu^{(i-1)}(x)+{c\over 6} 
\sum_{\pm\nu, \nu\neq\mu}
V_\nu^{(i-1)}(x) 
V_\mu ^{(i-1)}(x+\nu) V_\nu^{(i-1)}(x+\mu)^\dagger ,\nonumber \\
V_\mu^{(i)}(x)&=& {\widehat{V}_\mu^{(i)}(x)\over 
\left[{1\over N}{\rm Tr}\,\widehat{V}^{(i)}_\mu(x)^\dagger
 \widehat{V}_\mu^{(i)}(x)\right]^{1/2} }\,,
\label{imppr}
\end{eqnarray}
where $V_{-\nu}^{(i)}(x) = V^{(i)}_\nu(x-\nu)^\dagger$, and $c$ is a free
parameter, which can be tuned to optimize the properties of $q_{_L}^{(i)}(x)$.
This approach lends itself to a {\em Heisenberg picture}
description~\cite{CDPV-96}: starting from a simple definition of $q_L$, such
as Eq.~(\ref{qL}), one constructs sequences of operators by replacing the link
variables with the smeared link variables.  One may consider
\begin{equation}
q_{L}^{(i)}(x)= - {1\over 2^9\pi^2}
\sum^{\pm 4}_{\mu\nu\rho\sigma=\pm 1}
\epsilon_{\mu\nu\rho\sigma} {\rm Tr}
\left[ \Pi^{(i)}_{\mu\nu}\,\Pi^{(i)}_{\rho\sigma}\right],
\label{impop}
\end{equation}
where $\Pi^{(i)}_{\mu\nu}$ is the product of smeared links $V^{(i)}_\mu(x)$
around a $1\times 1$ plaquette.  All these operators have the correct
classical continuum limit, i.e. $q_{L}^{(i)}(x)\rightarrow a^4 q(x)$
for $a\rightarrow 0$.  As shown in Ref.~\cite{CDPV-96}, the 
renormalization effects get suppressed with increasing number of iterations.
For example, the multiplicative renormalization of the operator with one
smearing step is given by $Z_L(g_0^2)= 1 - 0.247 g_0^2 + O(g_0^4)$ (for the
optimal value $c=0.6774$), as compared to $Z_L(g_0^2)= 1 - 0.908 g_0^2 +
O(g_0^4)$ for the operator (\ref{qL}).  However, since the size of
$q_L^{(i)}(x)$ increases with the number of iterations $i$, one must
keep a fixed maximum value of $i$ while approaching the continuum limit,
otherwise the operator cannot be strictly considered as a local operator
anymore.

\begin{figure*}[tb]
\hspace{0cm}
\vspace{0cm}
\centerline{\psfig{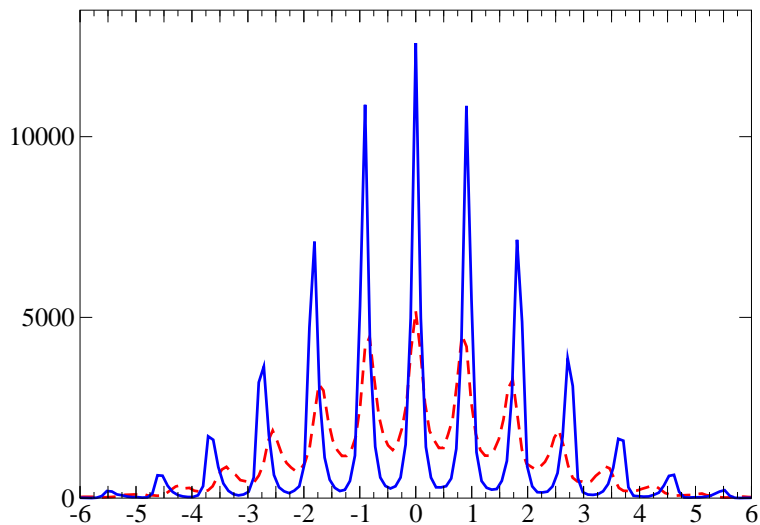}}
\vspace{2mm}
\caption{ Histograms of the lattice topological charge as measured by cooling,
  using the standard twisted double plaquette operator in
  Eq.~(\protect\ref{qL}), at $g_0^2=1$ ($\beta=6$) and on a $16^4$ lattice for
  the $SU(3)$ gauge theory, after $n=4$ (dashed line) and $n=12$ (full line)
  cooling steps.  The data clearly tend to cluster around integer values as
  the number of cooling steps is increased.  }
\label{histcool}
\end{figure*}

\begin{figure*}[tb]
\hspace{0cm}
\vspace{0cm}
\centerline{\psfig{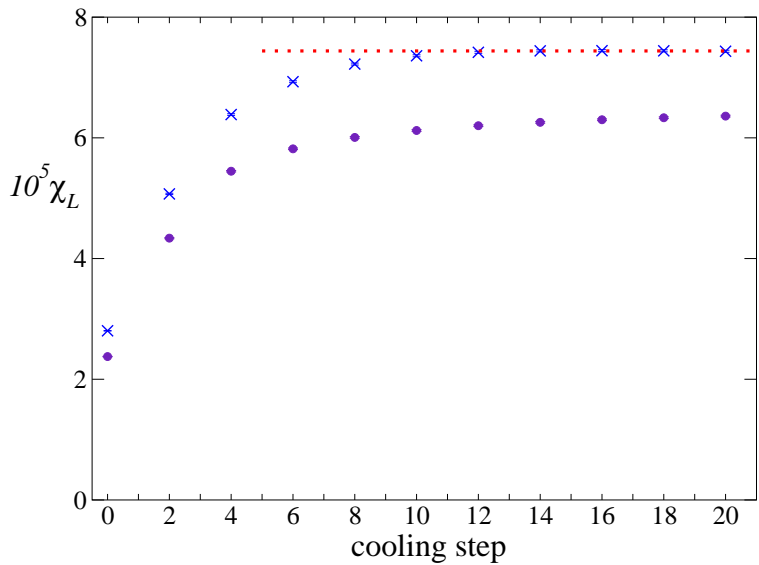}}
\vspace{2mm}
\caption{ 
  The lattice topological susceptibility $\chi_L\equiv \langle Q_L^2\rangle/V$
  versus cooling step at $g_0^2=1$ ($\beta=6$) for the $SU(3)$ gauge
  theory, on a lattice of size $16^3\times 36$. The data indicated by full
  circles are obtained using the operator in
  Eq.~(\protect\ref{qL}).  The data indicated by crosses have been obtained
  using the procedure of Ref.~\cite{DPV-02}, which reads integer values of $Q$
  from the cooled configurations.  The latter set of data show better
  convergence;  the dotted line indicates the plateau where the value of
  of $\chi_L$ is taken.  }
\label{coolingvsstep}
\end{figure*}

A delicate point in cooling is to check for eventual losses of topological
charge; such losses would not occur in the continuum, given the global
stability of the topological charge.  On the lattice, where instantons are
only {\em quasi} stable, the charge is bound to vanish eventually, after
protracted cooling: Given that lattice formulations are in general not scale
invariant, an instanton can actually decrease its action by shrinking in size,
and eventually disappear.  In this respect, some improvements have been
proposed, see e.g.
Refs.~\cite{GGSV-94,DGS-97,vB-98,GPS-99,Bruckmann-etal-04}, by selecting
procedures which make lattice instanton-like configurations more stable under
smoothing.  This is achieved, for example, by modifying the action in the
minimization procedure such that instantons tend to increase under
cooling~\cite{GGSV-94}.  Ref.~\cite{HK-01} proposes another smoothing
procedure based on hypercubic blocking. Other proposals can be found in
Refs.~\cite{DHK-97,HN-98,MP-04,ML-08,ILMKSW-08}.  Moreover, topological
structures have been also investigated using filtering methods constructed by
keeping the lowest eigenstates of the overlap Dirac
operator~\cite{Ho-etal-03,Ho-etal-03-2} or Laplacian bosonic
operators~\cite{BI-05}, see also Ref.~\cite{BGIMSS-07} for comparisons of the
results of these methods.

Figs.~\ref{histcool} and \ref{coolingvsstep} show typical results obtained
using the cooling technique at $g_0^2=1$ ($\beta=6$) for the $SU(3)$ gauge
theory.  The histograms of Fig.~\ref{histcool} show how the values $Q_L$ of
the topological charge of the cooled configurations, as obtained using the
lattice operator (\ref{qL}), tend to cluster around integer values when
increasing the number of cooling steps.  In Fig.~\ref{coolingvsstep} we show
the lattice topological susceptibility versus the number of cooling steps.  We
show data for two definitions from the same sample of cooled configurations,
i.e. results using the lattice operator (\ref{qL}), and results obtained by
taking the integer value closest to $Q_L$ as estimator of $Q$, as outlined in
Ref.~\cite{DPV-02}.  The latter procedure shows a better convergence, while
the first one requires a protracted cooling to reach a plateau with respect to
the cooling steps. Anyway the difference between the results of the two
procedures is expected to vanish as $O(a^2)$ in the continuum limit.

The basis for determining physical topological properties by these
smoothing methods is clearly heuristic, and their systematic errors
are not under robust theoretical control.  But, as we shall see, after
several numerical checks and comparison with other approaches, their
results regarding zero-momentum topological properties have been seen
to be quite reliable. 

Monte Carlo results using methods based on these smoothing procedures can be
found in
Refs.~\cite{IY-83,Teper-85,ILMSS-86,Teper-86,Teper-86-2,MT-87,PV-88,PT-89,
CDMPV-89,CDPV-90,GGSV-94,GMM-95,MS-95,DGS-97,DHK-98,DGHS-98,vB-98,GPS-99,
S-00,BGIMSS-07} ($SU(2)$ gauge theories),
\cite{IIY-84,HTW-87,Teper-88,APE-90,CGHN-94,DGHS-98,HN-98,ST-98,Negele-98,
Hasenfratz-00,S-00,Teper-00,ZBBLWZ-02,DFHK-07} ($SU(3)$ gauge theories),
\cite{LT-01,CTW-02,DPV-02,DPV-04,LTW-04,LTW-05} ($SU(N)$ gauge theories),
\cite{Berg-81,IIY-84,CRV-92,CRV-92-2,ACDP-00} ($CP^{N-1}$ / $O(3)$ models).

\subsubsection{Off-equilibrium methods to determine the renormalization effects}
\label{heatingmeth}

\begin{figure*}[tb]
\hspace{0cm}
\vspace{0cm}
\centerline{\psfig{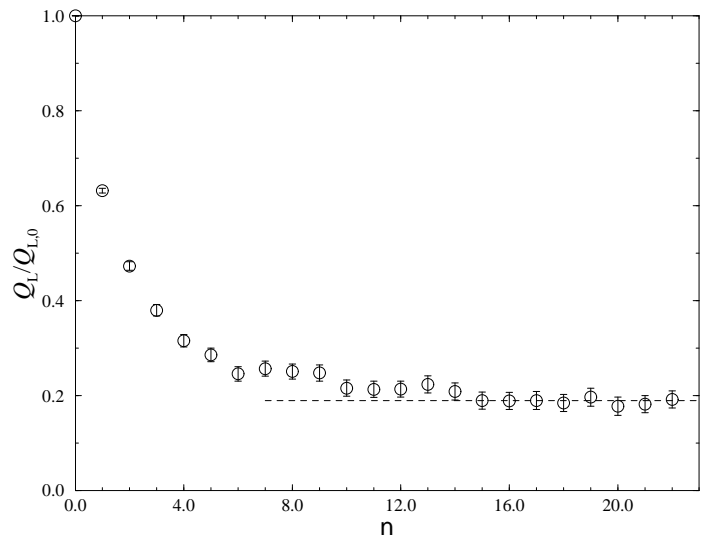}}
\vspace{2mm}
\caption{ This figure shows the ratio $Q_{L,n}/Q_{L,0}$ versus $n$, where
  $Q_{L,n}$ is the lattice topological charge as measured by the operator
  (\ref{qL}) after $n$ steps of a local heat-bath algorithm, when {\em
    heating} an instanton-like configuration of charge $Q_{L,0}\approx 1$ and
  averaged over many independent trajectories, using the Wilson action at
  $g_0^2=1$ ($\beta=6$) for the $SU(3)$ gauge theory.  The observed plateau
  provides an estimate of the multiplicative renormalization $Z$. From
  Ref.~\cite{ACDDPV-94}. }
\label{heatingz}
\end{figure*}

\begin{figure*}[tb]
\hspace{0cm}
\vspace{0cm}
\centerline{\psfig{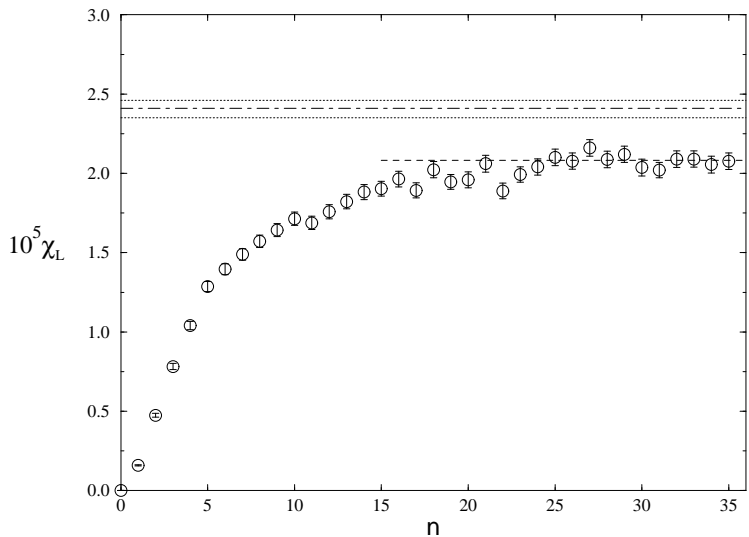}}
\vspace{2mm}
\caption{ This figure shows $\chi_{L,n}$ versus $n$, where $\chi_{L,n}$ is the
  lattice topological susceptibility as measured by the operator (\ref{qL})
  after $n$ steps of a local heat-bath algorithm at $g_0^2=1$ ($\beta=6$) for
  the $SU(3)$ gauge theory, starting from a flat configuration and averaged
  over many independent trajectories.  The dot-dashed line indicates the
  equilibrium value of $\chi_{L}$ (the dotted lines indicate the error).  The
  dashed line shows the estimate of the background $B$ obtained by averaging
  data on the plateau. From Ref.~\cite{v-95}. }
\label{heatingb}
\end{figure*}

The renormalization effects in the relation (\ref{chileq}) between the lattice
quantity $\chi_L$ and the topological susceptibility $\chi$, i.e. the
multiplicative renormalization $Z$ and the background contribution $B$, can be
estimated by exploiting off-equilibrium Monte Carlo simulations.  When using a
standard local algorithm, for instance Metropolis or heat-bath, quantities
like the topological charge, involving changes of global properties of the
configurations, show a slow dynamics, much slower than the dynamics of
quasi-Gaussian modes determining other features of the theory, see e.g.
Refs.~\cite{DPV-02,LTW-04b} and also Sec.~\ref{qsampling}.  The so-called
heating method~\cite{DV-92} relies on the possibility of somehow separating
the various contributions to Eq.~(\ref{chileq}) by means of off-equilibrium
simulations which exploit the slow dynamic behavior of the topological modes
during Monte Carlo simulations, and in particular during the thermalization
process, giving the opportunity to estimate $Z$ and $B$ in Eq.~(\ref{chileq}).
Similar considerations were also reported in Ref.~\cite{Teper-89}.

The main idea behind this method is to perform Monte Carlo simulations where
the lattice modes responsible for the additive and multiplicative
renormalizations in Eq.~(\ref{chileq}) are quasi-thermalized, while the
topological modes are instead frozen and essentially determined by the chosen
starting configuration.  In particular, one can estimate the multiplicative
renormalization $Z$ by starting from an instanton-like lattice configuration
with $Q_L\approx 1$, measuring how $Q_L$ changes under thermalization (i.e. MC
steps of local algorithms, such as Metropolis or heat-bath) and repeating this
procedure many times to eventually obtain averages over many different
trajectories. If a window exists in which the modes responsible for the
renormalizations are already approximately thermalized, while the original
topological modes remain substantially unchanged, then we expect to observe a
plateau in the data of $Q_L$ with respect to the number of heating steps,
which provides an estimate of the multiplicative renormalization
$Z$~\cite{DV-92}.  Fig.~\ref{heatingz} shows MC data in the case of the
operator (\ref{qL}) using the Wilson action at $g_0^2=1$ ($\beta=6$) for the
$SU(3)$ gauge theory: a plateau is clearly observed suggesting
$Z(g_0^2=1)\approx 0.2$. On the other hand, starting from flat configurations
where $Q_L=0$ and measuring how the lattice susceptibility $\chi_L= Q_L^2/V$
changes under thermalization, one can obtain an approximate estimate of the
background term $B$ in Eq.~(\ref{chileq}).  The data of $\chi_L$ are expected
to present a plateau with respect to the number of heating steps, which
provides an estimate of $B$.  Here one is assuming that the dependence of $B$
on the global topological charge $Q$ is negligible.  As an example, in
Fig.~\ref{heatingb} we show again results in the case of the operator
(\ref{qL}) using the Wilson action at $g_0^2=1$ for the $SU(3)$ gauge theory.
Finally, once $Z$ and $B$ are estimated, $\chi$ can be extracted from
Eq.~(\ref{chileq}), by supplementing the calculation with a standard Monte
Carlo simulation at equilibrium to determine $\chi_L$.  As shown in
Fig.~\ref{heatingb}, in the case $\chi_L$ is computed using the operator
(\ref{qL}), the background $B$ gives a large contribution to $\chi_L$ at
$g_0^2=1$ already, and this clearly affects the precision of the estimate of
$\chi$. Better results are obtained by considering improved smeared
operators~\cite{CDPV-96}, whose renormalization effects are smaller.

Like the cooling method, this method relies on heuristic arguments, and their
systematic errors are not under robust theoretical control. However, several
numerical checks and comparisons with other techniques have been reported,
showing that good estimates of $\chi$ are obtained by the heating method.

Monte Carlo results using this method can be found in
Refs.~\cite{ACDGV-93,v-93,ACDDPV-94,CDPV-96,ABF-96,ADD-97,ADD-97-2,ADDK-98,
Delia-03,ADD-05,ADDP-06}
for the 4D $SU(2)$ and $SU(3)$ lattice gauge theories, and in
Refs.~\cite{DV-92,CRV-92,FP-93,RRV-97} for the 2D $CP^{N-1}$ models.

\subsection{Fermionic definitions by the index theorem of
Ginsparg-Wilson fermions}
\label{fermdefGW}

Fermionic methods to determine the topological charge density, and its
correlations, are essentially based on the anomalous flavor-singlet axial
Ward-Takahashi identities.  The realization of the axial anomaly on the
lattice has been largely discussed in the literature, see e.g.
Refs.~\cite{KS-80,BRTY-84,SV-87,SV-87-2,SV-87-3,SV-88,NN-93,NN-95,Ungarelli-95,
Hernandez-98,Hasenfratz-98,Niedermayer-98,Neuberger-99,Luscher-99,Chiu-99,
Chiu-99-2,KY-99-2,RR-00,FR-01,Adams-02,CH-02,FNS-02,GS-03,AB-04}.

In the continuum theory and for the massless Dirac operator the Atiyah-Singer
index theorem~\cite{AS-71} states that
\begin{equation}
{1\over 32\pi^2}  \int d^4 x \epsilon_{\mu\nu\rho\sigma} 
{\rm Tr} F_{\mu\nu} F_{\rho\sigma} = {\rm index}(D) = n_+-n_- \,,
\label{attheo}
\end{equation}
where $n_\pm$ are the number of zero modes of the Dirac operator $D$ which are
also eigenstates of $\gamma_5$ with eigenvalues $\pm 1$.  Any acceptable
discretization of the Dirac operator may be used to determine the topological
charge for sufficiently small values of $a$, by determining its almost zero
modes. In this respect Dirac operators satisfying the GW relation are ideal,
because their lattice chiral symmetry allows exact zero modes.  As discussed
in Sec.~\ref{gwferm}, Dirac operators satisfing the GW relation preserve an
exact chiral symmetry at finite lattice spacing, cf. Eq.~(\ref{latchsym}). The
corresponding Jacobian gives rise to the axial anomaly~\cite{Luscher-98}, and
therefore to a natural definition of the topological charge
density~\cite{HLN-98}
\begin{equation}
q_i(x) = {1\over 2}  {\rm Tr} [\gamma_5 D(x,x)],
\label{qfl}
\end{equation}
where $D(x,x)$ represents the kernel of the lattice Dirac operator in position
space and the trace is taken over the Dirac and color indices. The associated
topological charge is equal to the index of the Dirac operator, i.e.
\begin{equation}
Q_i = \sum_x q_i(x) = n_+ - n_-
\label{index}
\end{equation} 
(as above, $n_\pm$ are the number of zero modes with
positive and negative chirality). Then, the topological susceptibility can be
defined as
\begin{equation}
\chi = \sum_x \langle q_i(0) q_i(x) \rangle = {\langle Q_i^2 \rangle\over V}\,.
\label{chidefl}
\end{equation}
Analogously, one may define zero-momentum $2n$-point correlation functions of
$q_i(x)$, which are needed for the computation of the $\theta$ expansion
around $\theta=0$.

\begin{figure*}[tb]
\hspace{0cm}
\vspace{0cm}
\centerline{\psfig{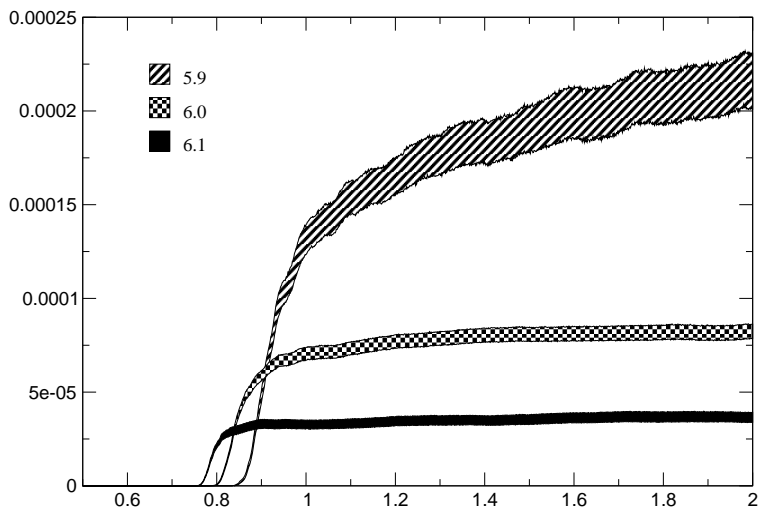}}
\vspace{2mm}
\caption{
  The topological susceptibility from the overlap method at
  $\beta=5.9,\,6.0,\,6.1$ for the Wilson lattice formulation of the 4D $SU(3)$
  gauge theory, as a function of the parameter $\rho$ of the overlap Dirac
  operator.  Taken from Ref.~\cite{DP-04}.  }
\label{DP04}
\end{figure*}

The index of the overlap Dirac operator, and in general of all Dirac operators
satisfying the GW relation, provides a well--defined estimator for the
topological charge~\cite{Neuberger-01,GRTV-02,GRT-04,Luscher-04}, which can
also be used in pure gauge theories. Indeed, as shown by
L\"uscher~\cite{Luscher-04}, the corresponding zero-momentum $2n$-point
correlation functions, and in particular the topological susceptibility, can
be expressed in terms of appropriate zero-momentum correlation functions of
scalar and pseudoscalar quark operators which are free of short-range
singularities, such as
\begin{equation}
\chi = m_1...m_s \, a^{4s-4} \sum_{x_1,...x_{s-1}} 
\langle P_{r1}(x_1) S_{12}(x_2)...S_{r-1,r}(x_r)\,
P_{s,r+1}(x_{r+1}) S_{r+1,r+2}(x_{r+2})...S_{s-1,s}(0)\rangle_c
\label{chips}
\end{equation}
where $1\le r< s \le N_f$, and
\begin{eqnarray}
&&P_{ab}(x)= \bar{\psi}_a(x) \gamma_5 \left( 1 -{1\over 2} a D\right) \psi_b(x), 
\label{psdef}\\
&&S_{ab}(x)= \bar{\psi}_a(x) \left( 1 -{1\over 2} a D\right) \psi_b(x), 
\nonumber
\end{eqnarray}
where $a,b$ are flavor indices.  Since the normalization of the correlation
functions is fixed by the nonsinglet chiral Ward identities, these expressions
provide an unambiguous definition of the topological susceptibility, and in
general of the zero-momentum $2n$-point correlation functions.  By a further
refining of this argument~\cite{Luscher-04}, these results can be shown to
also hold in the case of the pure gauge theory.  Therefore, also in pure
$SU(N)$ gauge theories the index of the overlap Dirac operator provides a well
defined fermionic definition of topological charge and the moments of its
distribution at $\theta=0$, which determine the coefficients of the $\theta$
expansion.  This method circumvents the problem of renormalization arising in
bosonic approaches, albeit at a much higher computational cost than bosonic
methods.

We also remark that in the case of the overlap Dirac operator, an explicit
ambiguity arises from the fact that when one varies the parameter $\rho$,
cf. Eq.~(\ref{Nop}), within its theoretically acceptable range, an eigenvalue
of the operator $X = D_{\rm W} - (\rho/a)$ may occasionally change sign, thus
changing the value of the operator index~\cite{EHN-99,DP-04}. This gives rise
to a corresponding ambiguity in the value of the topological charge to be
assigned to the underlying gauge field.  However, the effects of this
ambiguity are expected to disappear in the continuum limit, and therefore
lattice topological susceptibilities defined using different values of $\rho$
are expected to only differ by $O(a^2)$. The dependence of $\rho$ has been
numerically studied in Ref.~\cite{DP-04}. Some results for $\chi$ are shown in
Fig.~\ref{DP04} as a function of $\rho$, where one can clearly observe that
the dependence of $\rho$ gets reduced with increasing $\beta$, thus
approaching the continuum limit.

We should also mention the warning of Refs.~\cite{Creutz-04,Creutz-07-3} on
the possibility of a residual nonperturbative ambiguity in defining the
topological susceptibility in pure gauge theories.

More detailed discussions of the fermionic methods and Monte Carlo results can
be found in Refs.~\cite{SV-87,EHN-98,EHN-99-2,Edwards-02} (using ultralocal
fermion actions), and
Refs.~\cite{NN-93,NV-97,Neuberger-98-2,Niedermayer-98,HLN-98,EHN-99,
Chandrasekharan-99,Suzuki-99,Fujikawa-99,Chiu-99,EHKN-00,Neuberger-01,EH-01,
DH-01,GRTV-02,GHS-02,Adams-02,MILC-02,Edwards-02,G-02,HHJNH-02,ZBBLWZ-02,CH-03,
GLWW-03,GHLWW-04,DP-04,GRT-04,DGP-05,EFKS-06,BS-06,GPT-07}
(using Ginsparg-Wilson fermions).

We finally mention a recent proposal~\cite{AFHO-07} for the calculation of the
topological susceptibility, by computing correlation functions at fixed
topological sector.  Expectation values of observables at fixed global
topological charge $Q$ differ by $O(1/V)$ terms ($V$ is the space-time volume)
from those at $\theta=0$, obtained by sampling the topological sectors.  Under
some assumptions, these differences can be determined by a saddle-point
expansion~\cite{BCNW-03}.  In particular, the large-distance behavior of the
two-point correlation function of the topological charge density $q(x)$ at
fixed topology $Q$ is given by~\cite{AFHO-07}
\begin{equation}
\langle q(0) q(x) \rangle_Q = 
- {\chi \over V} + {1\over V^2}
\left( Q^2 + {\chi_4\over 2 \chi} \right)  + O(V^{-3})
+ O(e^{-m_{\eta'}|x|})
\label{chifv}
\end{equation}
with $\chi_4$ defined in Eq.~(\ref{chi4}).  Ref.~\cite{AFHO-07} proposes to
extract $\chi$ from the above formula, or from an analogous behavior of the
correlation function of flavor singlet pseudo-scalar densities, by performing
MC simulations at fixed topological sector, using overlap fermions and a
modified lattice action which suppresses zero eigenvalues, and therefore
prevents any change of the topological charge. Results for two-flavor QCD are
presented in Ref.~\cite{Aoki-etal-07}.

\subsection{The $U(1)_A$ problem on the lattice with Ginsparg-Wilson fermions} 
\label{ua1latt}

The large-$N$ solution of the $U(1)_A$ problem and the Witten-Veneziano
formula for the $\eta'$ mass has also been investigated on the lattice using
Ginsparg-Wilson fermions. We report the main steps of its derivation as
outlined in Ref.~\cite{GRTV-02}, which is based on the lattice topological
charge density $q_i(x) = {1\over 2} {\rm Tr} \left[ \gamma_5 D(x,x) \right]$
defined from a Dirac operator satisfying the GW relation (\ref{GWrel}), such
as the Neuberger overlap Dirac operator. For more details see the original
reference \cite{GRTV-02}.

In the presence of $N_f$ massless flavors, using the results of
Sec.~\ref{gwferm} and in particular the symmetry (\ref{latchsym}), one can
write the lattice anomalous flavor-singlet Ward-Takahashi indentity
\begin{equation}
\nabla_\mu \langle A_\mu P \rangle  = 2N_f 
\langle q_i(x) P \rangle + \langle \delta P \rangle ,
\label{lwti}
\end{equation}
where $A_\mu$ is the singlet axial current, $P$ is a generic product of local
operators, and $\delta P$ the corresponding variation under the symmetry
(\ref{latchsym}).  The operators $q_i(x)$ and the divergence $A_\mu(x)$ are
not RG invariant operators, as already discussed in Sec.~\ref{latttop} for
their continuum counterparts.  They can be renormalized by analogy to the
continuum case, leading to renormalized operators $q^{(r)}(x)$ and
$A^{(r)}_\mu$,
\begin{equation}
q_r(x) = q_i(x) - {z\over 2 N_f} \nabla_\mu A_\mu(x),
\qquad
A_{\mu,r} = (1-z)A_\mu 
\end{equation}
where $z$ diverges logarithmically at two-loop order, and vanishes
when $u\equiv N_f/N \to
0$, cf. Eq.~(\ref{ztwol}).  This allows us to write a renormalized
Ward-Takahashi identity analogous to (\ref{lwti}).

A formula for the $\eta'$ mass in the limit $u\to 0$ can be then obtained by
assuming that in this limit the $\eta'$ mass vanishes as $O(u)$.  For
vanishing quark mass, one obtains
\begin{equation}
{\rm lim}_{p\to 0} \, {\rm lim}_{u\to 0} \,
{1\over 2 N_f} \int d^4 x e^{-ipx} \nabla_\mu \langle
A_{\mu,r} q_r(0)\rangle 
= {f_{\pi}^2\over 4 N_f} m_{\eta'}^2|_{u=0}
\label{llc}
\end{equation}
because the integral is determined by the $\eta'$ pole.  Here, $f_{\eta'} =
f_\pi$ has been used, and $f_\pi$ is defined as in Sec.~\ref{u1asolln}, cf.
Eqs.~(\ref{wf}-\ref{vf}).  On the other hand, using the renormalized
Ward-Takahashi identity analogous to Eq.~(\ref{lwti}), one also arrives at
\begin{equation}
{\rm lim}_{p\to 0}\, {\rm lim}_{u\to 0}\, 
{1\over 2 N_f} \int d^4 x e^{-ipx} \nabla_\mu \langle
A_{\mu,r} q_r(0)\rangle 
= {\rm lim}_{p\to 0} \,{\rm lim}_{u\to 0}\, 
\int d^4x e^{ipx} \langle q_i(x) q_i(0) \rangle .
\label{llc2}
\end{equation}
Finally, Eqs.~(\ref{llc}) and  (\ref{llc2}) lead to
\begin{equation}
\chi = \int d^4x \langle q_i(x) q_i(0) \rangle 
= {f_\pi^2\over 4 N_f} m_{\eta'}^2|_{u=0}
\label{lwvf}
\end{equation}
which provides a lattice version of the Witten-Veneziano formula.

\section{Results from lattice Monte Carlo simulations}
\label{MCres}

The $\theta$ dependence of 4D $SU(N)$ gauge theories has been investigated by
Monte Carlo simulations of their lattice formulation, as described in
Sec.~\ref{sunlattf}.  The lattice action corresponding to the Lagrangian
(\ref{lagrangian}) cannot be directly simulated for $\theta\ne 0$, by virtue
of the complex nature of the $\theta$ term. On the other hand, the
coefficients in the expansion of the scaling ground-state energy $f(\theta)$,
cf. Eq.~(\ref{scge}), around $\theta=0$, i.e. $C$ and $b_{2i}$, can be
accessed by determining the moments of the topological charge distribution at
$\theta=0$.  They are dimensionless renormalization-group invariant
quantities, which approach a constant in the continuum limit, with $O(a^2)$
scaling corrections.

\subsection{The topological susceptibility in the 4D $SU(3)$ gauge theory}
\label{su3restop}

\begin{table*}
  \caption{
    MC results for the topological susceptibility of the 4D
    $SU(3)$ pure gauge theory. We report the type of lattice action
    (Wilson action~\cite{Wilson-74} (WA),
    improved action introduced in Ref.~\cite{Iwasaki-83} (IA),
    tadpole improved L\"uscher-Weisz action~\cite{LW-85,CMP-83,ADLHM-95} (TILW), 
    one-loop Symanzik improved (1lSI),
    classically perfect action~\cite{NRW-01} (FP),
    doubly blocked Wilson action~\cite{Takaishi-96} (DBW2)),
    the method employed to determine the topological susceptibility,
    the scaling quantities $C\equiv \chi/\sigma^2$ and 
    $\chi r_0^4$ (where $r_0$ is the length scale defined in~\cite{Sommer-94}),
    and the value of $\chi^{1/4}$ in MeV, which is often
    derived by using the typical value $\sqrt{\sigma}=440$ MeV
    (some authors prefer $\sqrt{\sigma}=420$ MeV) or
    $r_0=0.5$ fm.
    The data marked by an asterisk have been derived by us 
    (the original references do not report them),
    using also results for $\sigma$ and $r_0$ in the literature, such 
    as the estimate~\cite{NRW-01} $\sigma^{1/2}r_0=1.193(10)$
    (thus $\sigma^{2}r_0^4=2.03(7)$), and values for $\sigma a^2$ at
    various $\beta$-values~\cite{LT-01}.
  }
\label{su3chi}
\hspace*{-1cm}    
\tabcolsep 4pt        
\begin{center}
\begin{tabular}{rlllll}
\hline
\multicolumn{1}{c}{Ref. $_{\rm year}$}& 
\multicolumn{1}{c}{Lattice action}& 
\multicolumn{1}{c}{Method}& 
\multicolumn{1}{c}{$C\equiv \chi/\sigma^2$}&
\multicolumn{1}{c}{$\chi r_0^4$}& 
\multicolumn{1}{c}{$\chi^{1/4}$(MeV)} \\
\hline  
\cite{DFHK-07} $_{2007}$ & WA $\beta=[5.9,6.3]$ 
& HYP smearing & 0.0259(10)$^*$ & 0.0524(13) & 193(9) \\

\cite{BS-06}  $_{2006}$ & WA $\beta=5.85$  
& overlap HF & 0.035(2)$^*$  & 0.071(3) & 190(3)$^*$ \\

\cite{BBOS-05}  $_{2006}$ & DBW2 $\beta=0.87$  
& APE smearing & 0.028(2)$^*$  & 0.056(4) & 179(4)$^*$ \\

\cite{DGP-05}  $_{2005}$ & WA $\beta=[5.8,6.2]$  
& overlap  &  0.029(2)$^*$ & 0.059(3) & 191(5) \\

\cite{ADD-05}  $_{2005}$ & WA $\beta=6.0$  
& heating & 0.0263(8)$^*$ & 0.053(3)$^*$ & $173.4(1.7)^{+1.1}_{-0.2}$ \\

\cite{DS-05}  $_{2005}$ & TILW $\beta=7.2$  
& overlap  &  0.027(4)$^*$ & 0.055(7) & 191 \\

\cite{DP-04}  $_{2004}$ & WA $\beta=[5.9,6.1]$  
& overlap  &  $0.025^{+2}_{-10}$ & 0.055(10) & 188(17) \\

\cite{GLWW-03}  $_{2003}$ & WA $\beta=[5.84,6.14]$  
& overlap  &  0.029(3)$^*$ & 0.059(5) & 195(4)$^*$ \\

\cite{Bernard-etal-03} $_{2003}$ & 1lSI $\beta=8.0,\,8.4$ 
& Boulder/HYP  & 0.030(3)$^*$ & 0.061(5) & 183(4)$^*$ \\

\cite{Kovacs-03}  $_{2003}$ & WA $\beta=5.94$  
& APE smearing  & 0.030(2)$^*$  & 0.061(3) & 183(3)$^*$ \\

\cite{Kovacs-03}  $_{2003}$ & WA $\beta=5.94$  
& overlap & 0.031(2)$^*$  & 0.063(3) & 185(3)$^*$ \\

\cite{CH-03}  $_{2003}$ & WA $\beta=6.0$  
& overlap &  0.026(3)$^*$ & 0.052(6)$^*$ & 175(6)\\

\cite{CTW-02}  $_{2002}$ & WA $\beta=5.9$  
& overlap &  0.0211(53) & 0.043(11)$^*$ & 168(11)$^*$ \\

\cite{DPV-02}  $_{2002}$ & WA $\beta=[5.9,6.2]$  
& cooling & 0.0282(12) & 0.057(3)$^*$ & 180(2) \\

\cite{GHS-02}  $_{2002}$ & TILW $\beta=[8.1,8.6]$ 
& overlap &  & & 191(5) \\

\cite{HHJNH-02} $_{2002}$ & FP $\beta=[3.0,3.2]$ & overlap & 0.030(4)$^*$ 
& 0.0612(75) & 196(6) \\

\cite{MILC-02} $_{2002}$ & WA $\beta=5.9$ & overlap & 
0.043(6)$^*$ & 0.087(13)$^*$ & 213(7) \\

\cite{LT-01}  $_{2001}$ & WA $\beta=[5.7,6.2]$  & cooling & 
0.035(3) & 0.072(7)$^*$ & 191(4)$^*$ \\

\cite{CPPACS-01}  $_{2001}$ &IA $\beta=[1.8,2.1]$  & cooling & 
0.0333(27) & 0.0570(43)  & $197^{+13}_{-16}$ \\

\cite{Hasenfratz-00} $_{2000}$ & WA $\beta=6.0$ &  
APE smearing & & & 193(4) \\

\cite{BDET-00} $_{2000}$ & WA $\beta=5.7$ & fermionic & 
0.0263(24)$^*$ & 0.053(5)$^*$   & 188(5) \\

\cite{EHN-99-3} $_{1999}$ & WA $\beta=[5.7,6.0]$ & 
overlap &  0.038(8)$^*$ & 0.077(16)$^*$ & 194(10) \\

\cite{HN-98} $_{1998}$ & WA $\beta=[5.85,6.1]$  & APE smearing & 
0.036(4) & 0.074(9)$^*$  & 203(5) \\

\cite{DGHS-98} $_{1998}$ & WA $\beta=[5.85,6.0]$  & improved cooling & 
  & & 183(10) \\

\cite{ST-98} $_{1998}$ & WA $\beta=[6.0,6.4]$  & cooling & 
0.032(9) &  0.065(19)$^*$  & 187(22) \\

\cite{ADD-97}  $_{1997}$ & WA $\beta=[5.9,6.1]$  & heating & 
0.028(4)$^*$ & 0.057(8)$^*$ & 170(7) \\

\cite{v-95}  $_{1995}$ & WA $\beta=6.0$  & heating & 
0.04(1) & 0.08(2)$^*$ & 190(12)  \\

\cite{GKLSW-88}  $_{1988}$ & WA $\beta=6.0$  & geometrical & 
0.047(5)$^*$ & 0.095(11)$^*$ & 205(5)$^*$ \\

\cite{Teper-88}  $_{1988}$ & WA $\beta=[5.6,6.2]$  & cooling & 
0.033(3) & 0.067(7)$^*$ & 179(4) \\
\hline
\end{tabular}
\end{center}
\end{table*}

Many studies have been devoted to the determination of the topological
susceptibility of the 4D $SU(3)$ pure gauge theory.  Geometrical, smoothing
and off-equilibrium techniques have been used to address the problems caused
by power--divergent additive contributions and multiplicative renormalizations
in definitions of the topological susceptibility based on {\em bosonic}
discretized versions of the topological charge density operator $q(x)$.  As
discussed in the preceding section, these methods have their drawbacks, since
their systematic errors are not under robust theoretical control.  Substantial
progress has been achieved after the introduction of a fermionic definition
through the index of the overlap Dirac operator~\cite{Neuberger-98}, which
provides a well--defined estimator for the topological
charge~\cite{Luscher-04}, even in the case of pure gauge theories.  This
method circumvents the problem of renormalization arising in bosonic
approaches, albeit at a much higher computational cost.

\begin{figure}
\centerline{\psfig{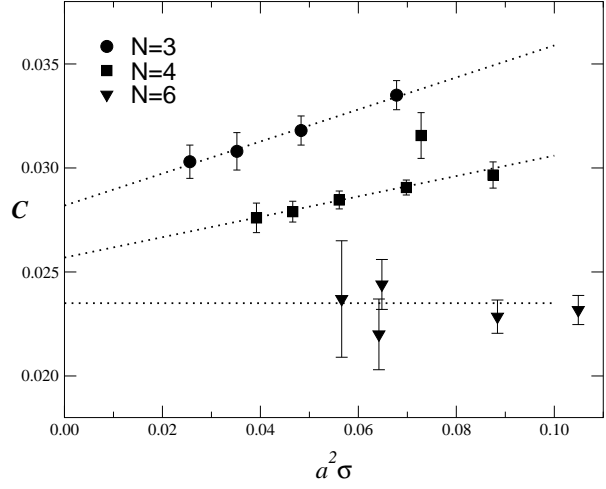}}
\caption{
   The scaling ratio $C\equiv\chi/\sigma^2$ versus $a^2\sigma$, where $\sigma$
  is the string tension and $a^2\sigma$ is the expected order of the scaling
  corrections, for $SU(3)$, $SU(4)$ and $SU(6)$ gauge theories,
  obtained using a cooling technique. From Ref.~\cite{DPV-02}.  }
\label{chiDPV}
\end{figure}

\begin{figure}
\vspace{0.5cm}
\centerline{\psfig{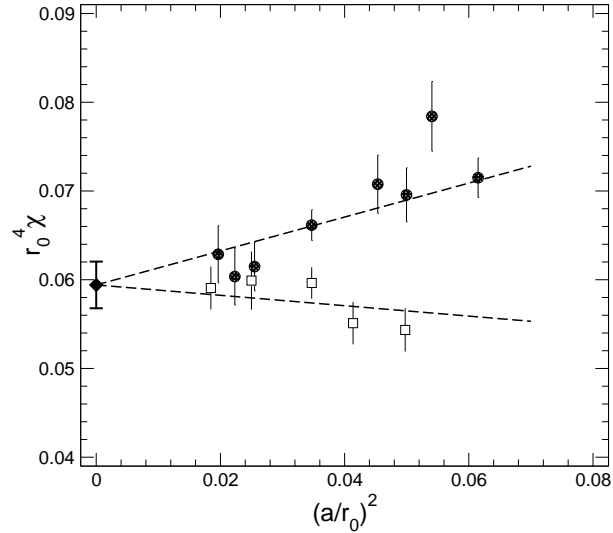}}
\caption{
Extrapolation to the continuum limit of the scaling quantity $\chi
r_0^4$ ($N=3$).
Data obtained by the overlap method for two different values of the
arbitrary parameter $\rho$ entering the definition of the overlap
definition. From Ref.~\cite{DGP-05}.}
\label{DGP05}
\end{figure}

\begin{figure}
\centerline{\psfig{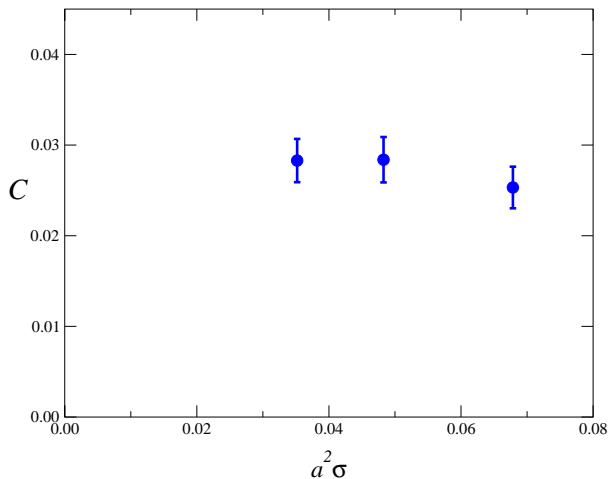}}
\caption{
  The scaling ratio $C\equiv\chi/\sigma^2$ versus $a^2\sigma$ for the $SU(3)$
  gauge theory, obtained by using the heating method to determine the lattice
  renormalizations in Eq.~(\ref{chileq}) (the data for $\chi$ at
  $\beta=5.9,\,6.0,\,6.1$ are taken from Ref.~\cite{ADD-97} and correspond to
  the 2-smeared operator, while the data for the string tension at the same
  values of $\beta$ come from Ref.~\cite{LT-01}).  They suggest the continuum
  extrapolation $C=0.028(4)$ (a linear fit to $C+b\sigma$ would give
  $C=0.032(5)$).  }
\label{chithm}
\end{figure}

In Table~\ref{su3chi} we report several results for the topological
susceptibility as obtained by various methods. We list the scaling quantities
$C\equiv \chi/\sigma^2$ and $\chi r_0^4$ (where $r_0$ is the length scale
defined in~\cite{Sommer-94}, which can be related to the string tension
$\sigma$ through the estimate~\cite{NRW-01} $\sigma^{1/2}r_0=1.193(10)$), and
the value $\chi^{1/4}$ in MeV as reported in the corresponding paper.  In
order to check consistency among the various results, the comparison of the
scaling quantities $C$ and $\chi r_0^4$ is more significant than the value of
$\chi^{1/4}$ in MeV, which may depend on the choice of the scale (typical
values used for the string tension are $\sqrt{\sigma}=420$-440 MeV, and
$r_0=0.5$ fm for the Sommer scale, but other choices have been also used in
some cases). In the most accurate numerical works the continuum limit of the
scaling quantities $\chi/\sigma^2$ or $\chi r_0^4$ is obtained by taking into
account the expected $O(a^2)$ corrections.  These scaling corrections are
clearly exemplified in Figs.~\ref{chiDPV} and \ref{DGP05}, where some MC
results for the scaling ratio $\chi/\sigma^2$ are plotted versus $a^2\sigma$
(data from Ref.~\cite{DPV-02}, using the cooling method), and for the scaling
quantity $\chi r_0^4$ versus $a^2/r_0^2$ (data from Ref.~\cite{DGP-05}, using
the overlap method).

The recent studies based on the overlap definition, see Table~\ref{su3chi},
have led to a quite precise estimate of the topological susceptibility of the
pure $SU(3)$ gauge theory. The most accurate estimate using this definition
is~\cite{DGP-05} $\chi r_0^4=0.059(3)$, corresponding to $C=0.029(2)$.  It is
important to note that the results obtained by the (less computer-power
demanding) bosonic methods are substantially consistent, showing their
effectiveness although they are supported by a weaker theoretical ground.  For
example, we mention the results: $C=0.0282(12)$ obtained using
cooling~\cite{DPV-02}, and $C=0.0259(10)$~\cite{DFHK-07} using the HYP
smoothing method~\cite{HK-01}.  Fig.~\ref{chithm} shows results for the
scaling ratio $C\equiv \chi/\sigma^2$ as obtained from the heating method (the
data for $\chi$ at $\beta=5.9,\,6.0,\,6.1$ are taken from Ref.~\cite{ADD-97}
and correspond to the 2-smeared operator, while the data for the string
tension at the same values of $\beta$ come from Ref.~\cite{LT-01}); they
suggest the continuum extrapolation $C=0.028(4)$. We also mention that
comparisons of determinations of $\chi$ by cooling and fermionic overlap
methods on the same sample of configurations have been done in
Refs.~\cite{ZBBLWZ-02,CTW-02}, observing a strong correlation between them at
the typical values of $\beta$ where MC simulations are actually performed.
Ref.~\cite{BS-06} used an overlap hypercubic fermion to determine $\chi$,
which was a variant of the standard overlap operator, replacing the Wilson
operator $D_{\rm W}$ in Eq.~(\ref{Xdef}) with another hypercubic kernel 
which improves the locality properties of the overlap Dirac
operator.

The values of $\chi^{1/4}$ can be compared with the Witten-Veneziano formulae,
cf. Sec.~\ref{u1asolln}. The r.h.s. of Eq.~(\ref{wf}), gives $\chi^{1/4}
\approx 190 \, {\rm MeV}$ using the actual values of $f_\pi$, $m_{\eta '}$,
and $N_f=3$; using the formula refined by Veneziano \cite{Veneziano-79}, cf.
Eq.~(\ref{vf}), for which $m_{\eta'}^2 \rightarrow m_{\eta'}^2 + m_\eta^2 - 2
m_K^2$, one obtains $\chi^{1/4}\approx 180 \, {\rm MeV}$. The agreement of the
results reported in Table~\ref{su3chi} is remarkable.  The results obtained in
recent years suggest the estimates
\begin{equation}
C\equiv \chi/\sigma^2 = 0.028(2), \qquad \chi\,r_0^4 = 0.057(5),
\label{sumest}
\end{equation}
where the error is intended to take into account the spread and the
typical uncertainty of the estimates of $\chi$.
Then, using the quite standard values $\sqrt{\sigma}=440$ MeV and $r_0
= 0.5$ fm, we obtain respectively:
\begin{equation}
\chi^{1/4} = 180(3) \;{\rm MeV}, \qquad \chi^{1/4} = 193(4) \;{\rm MeV}.
\label{chimex}
\end{equation}
The two estimates are in reasonable agreement with the Witten and Veneziano
formulae, cf. Eqs.~(\ref{wf}) and (\ref{vf}); the difference between
the estimates is
of course related to the uncertainty in the physical scale for $\sigma$ and
for $r_0$.

We also mention that the topological susceptibility of a pure $SU(3)$ gauge
theory has been also studied by QCD spectral sum rule
methods~\cite{Narison-84,Narison-91,Narison-06}, leading to the estimate
$\chi^{1/4}\approx 120 \, {\rm MeV}$, and by the field-correlator
method~\cite{CK-07}, obtaining $\chi^{1/4}=196(7)\,{\rm MeV}$.

Results for the 4D $SU(2)$ gauge theory can be found in
Refs.~\cite{Woit-82,BRSW-84,MT-87,KLSW-87,KLSSW-88,PT-89,CDPV-90,DPV-90,
ACDGV-93,DHZ-96-2,ADD-97-2,DGS-97,NV-97,DHK-97,EHN-98-2,DHK-98,EHN-99-3,
GPS-99,Teper-00,LT-01,BEF-01,LTW-05,GMPZ-05}.

Topological structures, such as instanton-like configurations, have been
investigated on the lattice in
Refs.~\cite{BGIMSS-07,BI-05,IIY-84,HN-98,FP-98,GGSV-94,CGHN-94,
  MS-95,vB-98,DHZ-96,EHN-98-2,DHK-98,ML-08,Hasenfratz-00,S-00,DH-01,HIMT-01,
  CTW-02,DH-02,EH-01,GGLRS-01,Ho-etal-03,Ho-etal-03-2,HAZCDDLLMTT-05,
  LTW-05,BGM-06,IKKSSW-07}.  In particular,
Refs.~\cite{BGIMSS-07,BI-05,Ho-etal-03,Ho-etal-03-2,HAZCDDLLMTT-05,HIMT-01,
  IKKSSW-07,DH-02,EH-01,BGM-06,EHN-98} investigated topological structures beyond
instanton-like configurations.  Some reviews of these results can be found in
Refs.~\cite{Thacker-06,Bruckmann-07}.  Issues concerning the relation of the
topological structures to the chiral structure of the lowest fermionic
eigenstates have been discussed in
Refs.~\cite{DH-01,HIMT-01,HLNSS-02,DH-02,EH-01,GGLRS-01,GGRSS-01,GGRSS-01-2,
  Horvath-etal-02,Blum-etal-02,Edwards-02,CTW-02,GMPZ-05,Bruckmann-07,ILMKSW-08}.

\subsection{The topological susceptibility for $N>3$ and 
in the large-$N$ limit}
\label{largentop}

\begin{figure}
\centerline{\psfig{width=8truecm,angle=0,file=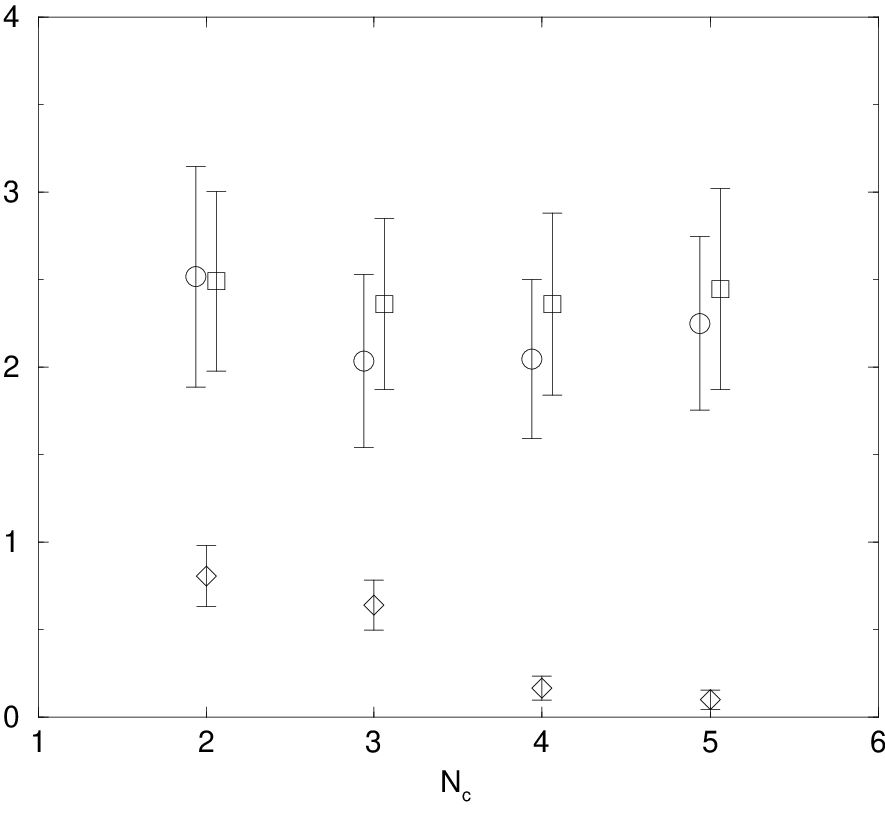}}
\caption{
Comparison between overlap and cooling determinations
of topological quantities on  the same sample of configurations,
for $N=2,3,4,5$, at the same value $a\sqrt{\sigma}\approx 0.261$
(corresponding to $\beta=5.9$ at $N=3$ using the Wilson action). 
In particular, the figure shows data for
the quantities 
$\langle Q_f^2 \rangle - \langle Q_f \rangle^2$ (circles), 
$\langle Q_g^2 \rangle - \langle Q_g \rangle^2$ (squares), 
and $\langle (Q_f-Q_g)^2 \rangle$ (diamonds)
where $Q_f$ and $Q_g$ are the topological charges as determined by
the overlap and cooling methods respectively. From Ref.~\cite{CTW-02}.
}
\label{overcoolcomp}
\end{figure}

The large-$N$ limit of 4D $SU(N)$ gauge theories can be investigated
numerically, by performing Monte Carlo simulations of their lattice
formulation, for various values of $N$, and checking the convergence
to $N=\infty$ with  the expected $1/N^2$ approach.  This idea has been recently
exploited in several papers, see
Refs.~\cite{LT-01,DPRV-02,DPV-02,LTW-04,LTW-04b,Teper-04,DPV-04,MT-04,LTW-05,
  LTW-05b,BT-05,NN-05,NN-06,BT-06,DMPSV-06,NN-07,DPV-07,DLPP-07}.  

The topological properties of 4D $SU(N)$ gauge theories in the large-$N$ limit
have also been examined using this approach in several numerical
investigations, providing interesting checks of the predictions obtained using
large-$N$ arguments, discussed in Sec.~\ref{largeNsec}.  The main limitation
of these numerical studies is essentially related to the fact that the correct
sampling of the topological charge becomes more and more difficult at large
$N$, due to a peculiar phenomenon of critical slowing down, which affects the
topological modes and which significantly worsens with increasing
$N$~\cite{DPV-02,LTW-05}.  We shall return to this point in
Sec.~\ref{qsampling}.

Results for $N>3$ can be found in Refs.~\cite{LT-01,DPV-02,CTW-02,LTW-05}, up
to $N=8$.  They are listed in Table~\ref{sunchi}.  They were essentially
obtained by the cooling method. The only exception is Ref.~\cite{CTW-02} which
presents a comparison between measurements using the cooling and overlap
methods on the same sample of configurations, up to $N=5$.  Some results are
shown in Fig.~\ref{overcoolcomp}, which presents various averages involving
lattice topological charges $Q_f$ and $Q_g$ obtained using the overlap and
cooling methods respectively, from MC simulations at fixed
$a\sqrt{\sigma}=0.261$ and for various values of $N$, i.e.  $N=2,3,4,5$.
Although the sample size in this comparison is relatively small, it clearly
shows that there is a strong correlation between the cooling and overlap
determinations, which increases with increasing $N$.

\begin{table*}
  \caption{
    MC results for the topological susceptibility of the 4D
    $SU(N)$ gauge theory with $N>3$. All of them have been obtained
    using the Wilson action and the cooling method,
    with the only exception of the results of Ref.~\cite{CTW-02}
    which were obtained using the overlap definition.
    Only the results of Refs.~\cite{LT-01,DPV-02} were obtained
    by extrapolating the data to the continuum limit. The data shown from
    Ref.~\cite{LTW-05} were obtained using values of $Q$ rounded to
    the appropriate neighboring integer value; they correspond to a
    fixed value of the lattice spacing $a=1/5T_c$\,. 
    Results from Ref.~\cite{CTW-02} were taken at one value of $\beta$ only.
  }
\label{sunchi}
\hspace*{-1cm}    
\tabcolsep 4pt        
\begin{center}
\begin{tabular}{crl}
\hline
\multicolumn{1}{c}{$N$}& 
\multicolumn{1}{c}{Ref.}& 
\multicolumn{1}{c}{$C\equiv \chi/\sigma^2$}\\
\hline  
4  & \cite{LT-01}  & 0.0224(39) \\
   & \cite{DPV-02} & 0.0257(10) \\
   & \cite{CTW-02} & 0.0213(47) \\
   & \cite{LTW-05} & 0.0301(10) \\

5 & \cite{LT-01}  & 0.0224(49) \\
  & \cite{CTW-02}  & 0.0234(51) \\

6 & \cite{DPV-02} & 0.0236(10) \\
  & \cite{LTW-05} & 0.0265(14) \\

8 & \cite{LTW-05} & 0.0236(31) \\
\hline
\end{tabular}
\end{center}
\end{table*}

\begin{figure}
\centerline{\psfig{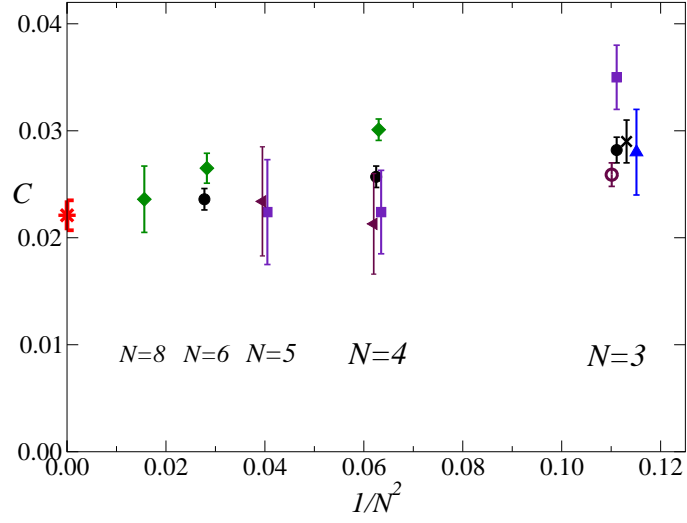}}
\caption{
  Results for $C\equiv \chi/\sigma^2$ versus $1/N^2$, for
  $N=3,4,5,6,8$. We show MC data taken from 
  Refs.~\cite{DPV-02} (filled circle, by cooling), 
  \cite{LT-01} (square, by cooling), 
  \cite{LTW-05} (diamonds, by cooling), 
  \cite{CTW-02} (left triangle, by overlap), 
  \cite{DGP-05} (cross, by overlap), 
  \cite{ADD-97} (triangle, by heating), 
  \cite{DFHK-07} (open circle, by HYP smoothing).  The result on the
  $y$-axis, indicated by an asterisk, shows the extrapolations to
  $N=\infty$ obtained in Ref.~\cite{DPV-02}.  
  Some data have been slightly shifted along the $x$-axis to make them
 more visible.
}
\label{largenres}
\end{figure}

\begin  {figure}[p]
\begin  {center}
\leavevmode
\setlength{\unitlength}{0.1606pt}
\ifx\plotpoint\undefined\newsavebox{\plotpoint}\fi
\begin{picture}(1500,1800)(0,0)
\font\gnuplot=cmr10 at 12pt
\gnuplot
\sbox{\plotpoint}{\rule[-0.200pt]{0.400pt}{0.400pt}}%
\put(300.0,250.0){\rule[-0.200pt]{3.218pt}{0.400pt}}
\put(275,250){\makebox(0,0)[r]{\ \ {$0$}}}
\put(1405.0,250.0){\rule[-0.200pt]{3.218pt}{0.400pt}}
\put(300.0,625.0){\rule[-0.200pt]{3.218pt}{0.400pt}}
\put(275,625){\makebox(0,0)[r]{\ \ {$1$}}}
\put(1405.0,625.0){\rule[-0.200pt]{3.218pt}{0.400pt}}
\put(300.0,1000.0){\rule[-0.200pt]{3.218pt}{0.400pt}}
\put(275,1000){\makebox(0,0)[r]{\ \ {$2$}}}
\put(1405.0,1000.0){\rule[-0.200pt]{3.218pt}{0.400pt}}
\put(300.0,1375.0){\rule[-0.200pt]{3.218pt}{0.400pt}}
\put(275,1375){\makebox(0,0)[r]{\ \ {$3$}}}
\put(1405.0,1375.0){\rule[-0.200pt]{3.218pt}{0.400pt}}
\put(300.0,1750.0){\rule[-0.200pt]{3.218pt}{0.400pt}}
\put(275,1750){\makebox(0,0)[r]{\ \ {$4$}}}
\put(1405.0,1750.0){\rule[-0.200pt]{3.218pt}{0.400pt}}
\put(300.0,250.0){\rule[-0.200pt]{0.400pt}{3.218pt}}
\put(300,200){\makebox(0,0){\ {$0$}}}
\put(300.0,1730.0){\rule[-0.200pt]{0.400pt}{3.218pt}}
\put(394.0,250.0){\rule[-0.200pt]{0.400pt}{3.218pt}}
\put(394,200){\makebox(0,0){\ {$1$}}}
\put(394.0,1730.0){\rule[-0.200pt]{0.400pt}{3.218pt}}
\put(488.0,250.0){\rule[-0.200pt]{0.400pt}{3.218pt}}
\put(488,200){\makebox(0,0){\ {$2$}}}
\put(488.0,1730.0){\rule[-0.200pt]{0.400pt}{3.218pt}}
\put(581.0,250.0){\rule[-0.200pt]{0.400pt}{3.218pt}}
\put(581,200){\makebox(0,0){\ {$3$}}}
\put(581.0,1730.0){\rule[-0.200pt]{0.400pt}{3.218pt}}
\put(675.0,250.0){\rule[-0.200pt]{0.400pt}{3.218pt}}
\put(675,200){\makebox(0,0){\ {$4$}}}
\put(675.0,1730.0){\rule[-0.200pt]{0.400pt}{3.218pt}}
\put(769.0,250.0){\rule[-0.200pt]{0.400pt}{3.218pt}}
\put(769,200){\makebox(0,0){\ {$5$}}}
\put(769.0,1730.0){\rule[-0.200pt]{0.400pt}{3.218pt}}
\put(863.0,250.0){\rule[-0.200pt]{0.400pt}{3.218pt}}
\put(863,200){\makebox(0,0){\ {$6$}}}
\put(863.0,1730.0){\rule[-0.200pt]{0.400pt}{3.218pt}}
\put(956.0,250.0){\rule[-0.200pt]{0.400pt}{3.218pt}}
\put(956,200){\makebox(0,0){\ {$7$}}}
\put(956.0,1730.0){\rule[-0.200pt]{0.400pt}{3.218pt}}
\put(1050.0,250.0){\rule[-0.200pt]{0.400pt}{3.218pt}}
\put(1050,200){\makebox(0,0){\ {$8$}}}
\put(1050.0,1730.0){\rule[-0.200pt]{0.400pt}{3.218pt}}
\put(1144.0,250.0){\rule[-0.200pt]{0.400pt}{3.218pt}}
\put(1144,200){\makebox(0,0){\ {$9$}}}
\put(1144.0,1730.0){\rule[-0.200pt]{0.400pt}{3.218pt}}
\put(1238.0,250.0){\rule[-0.200pt]{0.400pt}{3.218pt}}
\put(1238,200){\makebox(0,0){\ {$10$}}}
\put(1238.0,1730.0){\rule[-0.200pt]{0.400pt}{3.218pt}}
\put(1331.0,250.0){\rule[-0.200pt]{0.400pt}{3.218pt}}
\put(1331,200){\makebox(0,0){\ {$11$}}}
\put(1331.0,1730.0){\rule[-0.200pt]{0.400pt}{3.218pt}}
\put(1425.0,250.0){\rule[-0.200pt]{0.400pt}{3.218pt}}
\put(1425,200){\makebox(0,0){\ {$12$}}}
\put(1425.0,1730.0){\rule[-0.200pt]{0.400pt}{3.218pt}}
\put(300.0,250.0){\rule[-0.200pt]{179pt}{0.400pt}}
\put(1425.0,250.0){\rule[-0.200pt]{0.400pt}{239pt}}
\put(300.0,1750.0){\rule[-0.200pt]{179pt}{0.400pt}}
\put(100,1300){\makebox(0,0){\Large{$D(\rho)$}}}
\put(837,75){\makebox(0,0){\large{$\rho$}}}
\put(300.0,250.0){\rule[-0.200pt]{0.400pt}{239pt}}
\put(640,252){\circle*{12}}
\put(668,252){\circle*{12}}
\put(687,254){\circle*{12}}
\put(713,258){\circle*{12}}
\put(732,268){\circle*{12}}
\put(758,309){\circle*{12}}
\put(782,353){\circle*{12}}
\put(805,416){\circle*{12}}
\put(828,566){\circle*{12}}
\put(851,780){\circle*{12}}
\put(875,1010){\circle*{12}}
\put(898,1226){\circle*{12}}
\put(922,1347){\circle*{12}}
\put(945,1607){\circle*{12}}
\put(968,1591){\circle*{12}}
\put(991,1668){\circle*{12}}
\put(1014,1402){\circle*{12}}
\put(1038,1473){\circle*{12}}
\put(1062,1199){\circle*{12}}
\put(1086,1108){\circle*{12}}
\put(1110,934){\circle*{12}}
\put(1132,744){\circle*{12}}
\put(1152,705){\circle*{12}}
\put(1178,683){\circle*{12}}
\put(1201,503){\circle*{12}}
\put(1220,418){\circle*{12}}
\put(501,251){\circle{18}}
\put(522,251){\circle{18}}
\put(548,253){\circle{18}}
\put(569,259){\circle{18}}
\put(594,258){\circle{18}}
\put(618,267){\circle{18}}
\put(640,274){\circle{18}}
\put(664,296){\circle{18}}
\put(688,319){\circle{18}}
\put(711,339){\circle{18}}
\put(734,389){\circle{18}}
\put(758,430){\circle{18}}
\put(781,503){\circle{18}}
\put(804,555){\circle{18}}
\put(828,630){\circle{18}}
\put(850,698){\circle{18}}
\put(875,784){\circle{18}}
\put(898,815){\circle{18}}
\put(922,927){\circle{18}}
\put(945,986){\circle{18}}
\put(968,954){\circle{18}}
\put(992,1015){\circle{18}}
\put(1014,968){\circle{18}}
\put(1039,1106){\circle{18}}
\put(1062,941){\circle{18}}
\put(1086,1017){\circle{18}}
\put(1110,954){\circle{18}}
\put(1132,775){\circle{18}}
\put(1153,833){\circle{18}}
\put(1178,887){\circle{18}}
\put(1202,665){\circle{18}}
\put(1220,590){\circle{18}}
\put(462,250){\makebox(0,0){$+$}}
\put(480,254){\makebox(0,0){$+$}}
\put(502,269){\makebox(0,0){$+$}}
\put(525,278){\makebox(0,0){$+$}}
\put(547,289){\makebox(0,0){$+$}}
\put(571,299){\makebox(0,0){$+$}}
\put(593,312){\makebox(0,0){$+$}}
\put(617,327){\makebox(0,0){$+$}}
\put(640,349){\makebox(0,0){$+$}}
\put(664,368){\makebox(0,0){$+$}}
\put(687,400){\makebox(0,0){$+$}}
\put(710,421){\makebox(0,0){$+$}}
\put(734,451){\makebox(0,0){$+$}}
\put(758,480){\makebox(0,0){$+$}}
\put(781,492){\makebox(0,0){$+$}}
\put(804,508){\makebox(0,0){$+$}}
\put(828,524){\makebox(0,0){$+$}}
\put(850,523){\makebox(0,0){$+$}}
\put(874,562){\makebox(0,0){$+$}}
\put(898,562){\makebox(0,0){$+$}}
\put(921,554){\makebox(0,0){$+$}}
\put(945,563){\makebox(0,0){$+$}}
\put(968,515){\makebox(0,0){$+$}}
\put(991,528){\makebox(0,0){$+$}}
\put(1014,498){\makebox(0,0){$+$}}
\put(1038,518){\makebox(0,0){$+$}}
\put(1062,463){\makebox(0,0){$+$}}
\put(1086,478){\makebox(0,0){$+$}}
\put(1110,457){\makebox(0,0){$+$}}
\put(1132,394){\makebox(0,0){$+$}}
\put(1153,414){\makebox(0,0){$+$}}
\put(1178,430){\makebox(0,0){$+$}}
\put(1202,360){\makebox(0,0){$+$}}
\put(1220,343){\makebox(0,0){$+$}}
\put(300,250){\makebox(0,0){$\star$}}
\put(459,250){\makebox(0,0){$\star$}}
\put(482,259){\makebox(0,0){$\star$}}
\put(501,308){\makebox(0,0){$\star$}}
\put(522,329){\makebox(0,0){$\star$}}
\put(547,342){\makebox(0,0){$\star$}}
\put(570,350){\makebox(0,0){$\star$}}
\put(593,358){\makebox(0,0){$\star$}}
\put(617,364){\makebox(0,0){$\star$}}
\put(640,373){\makebox(0,0){$\star$}}
\put(664,382){\makebox(0,0){$\star$}}
\put(686,384){\makebox(0,0){$\star$}}
\put(711,394){\makebox(0,0){$\star$}}
\put(734,397){\makebox(0,0){$\star$}}
\put(757,392){\makebox(0,0){$\star$}}
\put(780,395){\makebox(0,0){$\star$}}
\put(804,401){\makebox(0,0){$\star$}}
\put(828,402){\makebox(0,0){$\star$}}
\put(851,405){\makebox(0,0){$\star$}}
\put(874,411){\makebox(0,0){$\star$}}
\put(898,404){\makebox(0,0){$\star$}}
\put(921,397){\makebox(0,0){$\star$}}
\put(945,412){\makebox(0,0){$\star$}}
\put(968,396){\makebox(0,0){$\star$}}
\put(992,393){\makebox(0,0){$\star$}}
\put(1014,385){\makebox(0,0){$\star$}}
\put(1039,401){\makebox(0,0){$\star$}}
\put(1062,379){\makebox(0,0){$\star$}}
\put(1086,386){\makebox(0,0){$\star$}}
\put(1110,379){\makebox(0,0){$\star$}}
\put(1132,346){\makebox(0,0){$\star$}}
\put(1153,361){\makebox(0,0){$\star$}}
\put(1178,372){\makebox(0,0){$\star$}}
\put(1202,331){\makebox(0,0){$\star$}}
\put(1220,320){\makebox(0,0){$\star$}}
\end{picture}

\phantom{$a$}
\end    {center}
\vskip -0.15in
\caption{The instanton size density, $D(\rho)$, for
$N=2(\star),3(+),4(\circ),8(\bullet)$
on $16^4$ lattices with $a \simeq 1/8T_c$.}
From Ref.~\cite{LTW-05}.
\label{fig_drho16}
\end    {figure}
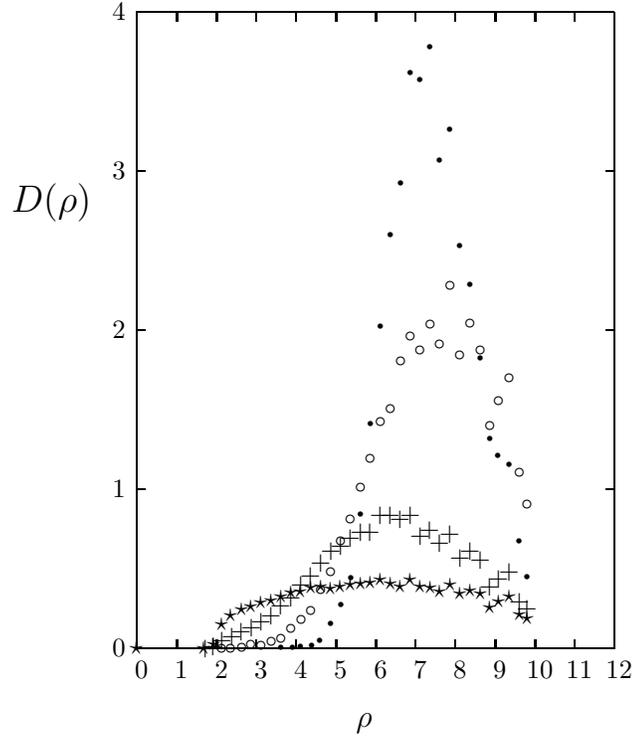

MC data for $N=3,4,5,6,8$ are shown in Fig.~\ref{largenres}.  They fit well
the expected large-$N$ behavior:
\begin{equation}
C=C_\infty+{c_2\over N^2}, 
\label{lnfit}
\end{equation}
thus providing an
estimate of $C_\infty$, and therefore of the topological susceptibility in the
large-$N$ limit: 
\begin{eqnarray}
&&C_\infty=0.0221(14),\nonumber \\
&&C_\infty=0.0200(43),\\
&&C_\infty=0.0248(18),\nonumber
\end{eqnarray}
respectively from Ref.~\cite{DPV-02}, Ref.~\cite{LT-01}, and
Ref.~\cite{LTW-05}; the latter was obtained using $N\le 8$ and keeping
$a=1/5T_c$ fixed, where $T_c$ is the critical temperature at the deconfinement
transition.  The coefficient $c_2$ in Eq.~(\ref{lnfit}) turns out to be quite
small,~\cite{DPV-02}
\begin{equation}
c_2\approx 0.06.  
\label{c2est}
\end{equation}
We stress that the good agreement for $N=3$ of the cooling method with the
more rigorous overlap result make us quite confident on the reliability of
cooling results for higher values of $N$, since there are no arguments to
suggest that this agreement could be spoiled with increasing $N$; actually,
there are arguments and evidence in favor of improved agreement
\cite{CTW-02,RRV-97}, as shown by the data presented in Fig.~\ref{overcoolcomp}.

These results are in substantial agreement with the large-$N$ relation
(\ref{wittenformula}).  In order to compare them to the Witten-Veneziano
formulae, one may translate them in physical units assuming the standard value
$\sqrt{\sigma}=440 \, {\rm MeV}$, thus obtaining $\chi_\infty^{1/4}=170(3) \,
{\rm MeV}$ from Ref.~\cite{DPV-02}, $\chi_\infty^{1/4}=165(9) \, {\rm MeV}$
from Ref.~\cite{LT-01}, and $\chi_\infty^{1/4}=175(3) \, {\rm MeV}$ from
Ref.~\cite{LTW-05}.

We finally mention some interesting comparisons of instanton size
distributions for various values of $N$ which have been presented in
Refs.~\cite{CTW-02,LTW-05}.  They provide significant evidence for the
suppression of small instantons with increasing $N$. Fig.~\ref{fig_drho16}
shows the distribution of instanton size as obtained after a number of cooling
steps at $a=1/8T_c$ fixed and for various values of $N$, up to $N=8$.  Besides
the suppression of small instantons, one also observes a rapid narrowing of
the distribution as $N$ grows.  The data suggest that at large $N$ the
distribution is peaked around a scale $\rho\sim 1/T_c$\,, where $T_c$ is the
critical temperature of the deconfinement transition~\cite{CTW-02}.  The
suppression of small topological structures can be also explained by
semiclassical instanton calculations, which shows that the instanton size
distribution is suppressed for small size $\rho$ as~\cite{GPY-81}
\begin{equation}
D(\rho) \sim {1\over\rho^5}\,\exp\left({-8\pi^2 
\over g^2(\rho)}\right) \sim \rho^{11N/3-5},
\label{insuppr}
\end{equation}
suggesting that small instantons are heavily suppressed with increasing $N$.
Instanton-size distributions qualitatively similar to those shown in
Fig.~\ref{fig_drho16} have been also obtained within the instanton liquid
model~\cite{Schafer-04}.

\subsection{The $O(\theta^4)$ term in the expansion
  of the ground-state energy}
\label{b2res}

\begin{table*}
\caption{ 
Results for the coefficient $b_2$ of the $O(\theta^4)$ term in
the expansion (\ref{stheta})
of the ground-state energy around $\theta=0$.
}
\label{b2}
\begin{center}
\begin{tabular}{cccl}
\hline
\multicolumn{1}{c}{$N$}&
\multicolumn{1}{c}{Ref.}&
\multicolumn{1}{c}{method}&
\multicolumn{1}{c}{$b_2$}\\
\hline 
3  &  \cite{DPV-02} & cooling & $-$0.023(7) \\  
   &  \cite{Delia-03} & heating & $-$0.024(6) \\  
   &  \cite{GPT-07} & overlap & $-$0.025(9) \\  
4  &  \cite{DPV-02} & cooling & $-$0.013(7) \\  
6  &  \cite{DPV-02} & cooling & $-$0.01(2) \\  
\hline
\end{tabular}
\end{center}
\end{table*}

Higher moments of the topological charge distribution provide estimates of the
coefficients $b_{2n}$ in the expansion of the scaling energy density
$f(\theta)$, cf. Eqs.~(\ref{ftheta}) and (\ref{stheta}).  In particular $b_2$
can be estimated using formulae (\ref{b2chi4}), (\ref{chi4}).  Note that a
reasonably accurate computation of $\chi_4$ in a MC simulation, which is
necessary to compute $b_2$, requires very high statistics due to a large
cancellation between the two terms in its definition (\ref{chi4}). Of course,
things worsen for the determination of higher-order coefficients $b_{2n}$.

There are a number of results for $b_2$ at $N=3$, obtained by different
approaches: Ref.~\cite{DPV-02} used the cooling method, Ref.~\cite{Delia-03}
used the heating technique to estimate additive and multiplicative
renormalizations in zero-momentum correlations of lattice discretizations of
$q(x)$, and finally Ref.~\cite{GPT-07} used the overlap method.  The results
reported in Table~\ref{b2} are in good agreement, suggesting that the
systematic errors of the various methods are sufficiently small.  The fourth
moment of the topological charge distribution was investigated also in
Ref.~\cite{DFHK-07}, obtaining substantially consistent results, although the
authors left open the possibility that the ratio $\chi_4/\chi$, and therefore
$b_2$, may vanish in the large-volume limit.

The results of Table~\ref{b2} provide robust evidence that $b_2$ is nonzero,
and therefore that there are deviations from a Gaussian distribution of the
topological charge. However, $b_2$ turns out to be quite small, indeed
$|b_2|\ll 1$. Thus deviations from a simple Gaussian behavior are already
small at $N=3$.

\begin{figure}
\centerline{\psfig{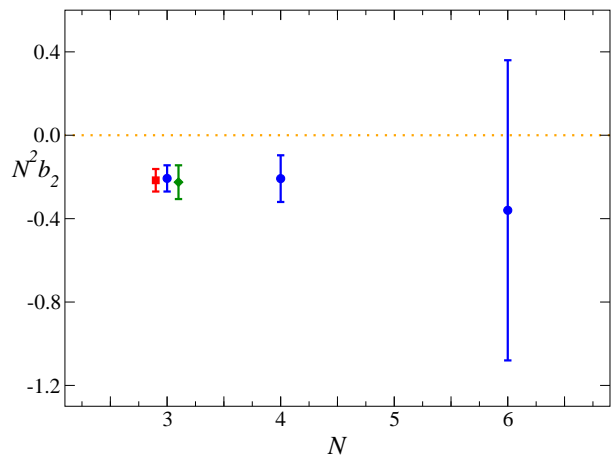}}
\caption{ Plot of $N^2 b_2$ versus $N$, for $N=3,4,6$.  We show MC data taken
  from Ref.~\cite{DPV-02} (circles, by cooling), from
  Ref.~\cite{Delia-03} (square, by heating), from
  Ref.~\cite{GPT-07} (diamond, by overlap).  Some data for $N=3$ have been slightly
  shifted along the $x$-axis to make them more visible.  }
\label{b2fig}
\end{figure}

There are also estimates for larger values of $N$, see Table~\ref{b2}, but
only using the cooling method. Again, given the agreement found at $N=3$,
higher $N$ results should be sufficiently reliable.  They appear to decrease
consistently with the expectation from the large-$N$ scaling arguments, i.e.
\begin{equation}
b_2\approx {\bar{b}_2\over N^2} \qquad {\rm with} \qquad  \bar{b}_2\approx -0.2.
\label{b2lnest}
\end{equation}
Fig.~\ref{b2fig} shows a plot of the available results for $b_2$,
actually $N^2 b_2$, versus $N$.

Overall, these results support the scenario obtained by general large-$N$
scaling arguments, which indicate $\bar{\theta}\equiv \theta/N$ as the
relevant Lagrangian parameter in the large-$N$ expansion.  They also show that
$N=3$ is already in the regime of the large-$N$ behavior.  For $N\ge 3$ the
simple quadratic form
\begin{equation}
F(\theta) - F(0) \approx {1\over 2} \chi \theta^2
\label{gauform}
\end{equation}
provides a good approximation of the dependence on $\theta$ for a relatively
large range of values of $\theta$ around $\theta=0$.

\subsection{$\theta$ dependence at finite temperature and across the phase
  transition}
\label{fintthe}

Another interesting issue concerns the behavior of topological properties at
finite temperature, and in particular their change at the finite-temperature
deconfining transition, which is first order for $N\ge 3$, and second order
for $N=2$; see e.g. Refs.~\cite{LTW-02,LTW-04,LTW-05b} and references therein.

At high temperature, $T\gg T_c$ where $T_c$ is the transition temperature, one
can compute the $\theta$ dependence semiclassically.  At zero temperature
semiclassical calculations fail because of the absence of any large-distance
cutoff on the instanton length scale~\cite{CDG-78}.  However, at finite
temperature, the temperature $T$, which is related to the size of the
Euclidean time dimension, is expected to act as a natural infrared
cutoff~\cite{GPY-81}.  Consequently, at sufficiently high temperature, say
$T\gg T_c$\,, one may compute the $\theta$ dependence from the one-loop
contribution of instantons to the functional integral, obtaining~\cite{GPY-81}
\begin{equation}
F(\theta) - F(0) \sim (1-{\rm cos} \theta) T^4 \exp[-8\pi^2/g^2(T)],
\label{largeTins}
\end{equation}
where $g(T)$ is the running coupling constant at the scale $T$.
In the case of a pure gauge theory, 
\begin{equation}
{8 \pi^2\over g^2(T)} \approx  {11\over 3} N \ln T 
\label{gscalt}
\end{equation}
asymptotically at large $T$.  This shows that the $\theta$ dependence gets
suppressed at high temperature.

The finite-temperature behavior of the topological susceptibility, and in
particular its behavior across the transition, has been investigated in
several numerical MC works, see
Refs.~\cite{Teper-88,DMP-92,ADD-97,FIMT-98,DGHS-98,EHKN-00-2,GGRSS-01,GHS-02,
GGL-02,DPV-04,LTW-05,BLMPIM-07}, 
using different methods to determine the topological susceptibility.

Results for the $SU(3)$ gauge theories can be found in
Refs.~\cite{ADD-97,EHKN-00-2,GHS-02,LTW-05}. They have been obtained by the
heating method~\cite{ADD-97}, the cooling method~\cite{LTW-05}, and the
overlap method~\cite{EHKN-00-2,GHS-02}, and give substantially consistent
results.  They show that the topological properties, and in particular the
topological susceptibility $\chi$, vary very little up to $T\lesssim T_c$.
They change across the transition, where $\chi$ shows a significant decrease.
For example~\cite{GHS-02}, $\chi$ decreases by approximately a factor 13 as
the temperature is increased from $0.88\, T_c$ to $1.31\, T_c$.  Then, at high
temperature, $T\gg T_c$, where the instanton calculus (\ref{largeTins}) should
become reliable~\cite{GPY-81},
\begin{equation}
\chi \sim T^4 \exp[-8\pi^2/g^2(T)].
\label{chilate}
\end{equation}

Concerning the large-$N$ behavior (investigated by performing simulations at
various values of $N\ge 3$ \cite{DPV-04,LTW-05}), the results indicate that
$\chi$ has a nonvanishing large-$N$ limit for $T<T_c$, as at $T=0$, and that
the topological properties, and therefore the ground state energy $F(\theta)$,
remain substantially unchanged in the low-temperature phase, up to $T_c$.  On
the other hand, above the deconfinement phase transition, $T>T_c$, $\chi$
shows a large suppression, hinting at a vanishing large-$N$ limit for $T>T_c$.

\begin{figure}
\centerline{\psfig{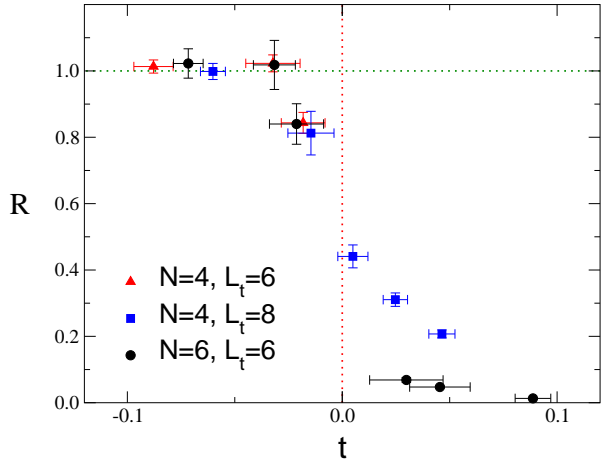}}
\caption{ The ratio $R(T)\equiv \chi(T)/\chi(T=0)$ as a function of the
  reduced temperature $t\equiv T/T_c-1$.  For each $SU(N)$ gauge theory, these
  curves are expected to converge to a continuum curve when $L_t\to\infty$,
  where $L_t$ is the length, in units of the lattice spacing, of the lattice
  along the Euclidean time direction.  From Ref.~\cite{DPV-04}.  }
\label{chit}
\end{figure}

Fig.~\ref{chit} shows results for the scaling ratio 
\begin{equation}
R(T)\equiv {\chi(T)\over \chi(T=0)}
\label{ratior}
\end{equation}
versus the reduced temperature $t\equiv T/T_c - 1$ for 4D $SU(N)$ gauge
theories with $N=4,6$~\cite{DPV-04}, and Euclidean temporal size $L_t=6,8$ (we
recall that the temperature is related to $L_t$ by $1/T=a L_t$).  One can
immediately observe that its behavior is drastically different in the low- and
high-temperature phases.  In the low-temperature phase, all data for $N=4$,
$L_t=6,8$ and $N=6$, $L_t=6$ appear to lie on the same curve, showing that
scaling corrections are small and also that the large-$N$ limit is quickly
approached.  The ratio $R$ remains constant and compatible with the value
$R=1$. As shown in Ref.~\cite{LTW-05}, also the instanton size distribution
appears substantially unchanged for $T\lesssim T_c$.
Only close to $T_c$, i.e. for $T > 0.97 \, T_c$, does this ratio
appear to decrease. These results show that in the confined phase the
topological properties remain substantially unchanged up to $T_c$. On the
other hand, above the deconfinement phase transition, $\chi$ shows a
significant decrease. The comparison between the $N=4$ and $N=6$ data shows
that the ratio $R$ decreases much faster for $N=6$, hinting at a vanishing
large-$N$ limit of $R$ for $T>T_c$.  A comparison with the results for $N=3$
of Refs.~\cite{ADD-97,GHS-02} suggests that the suppression of topological
fluctuations is faster in $SU(4)$ than it is in $SU(3)$.

\begin{figure}
\centerline{\psfig{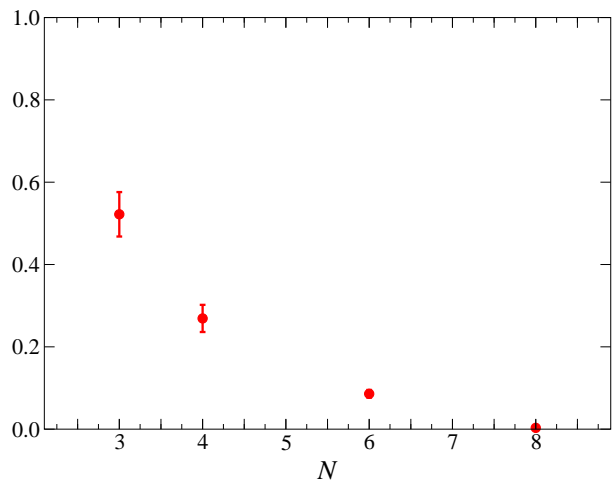}}
\caption{ 
The ratio of the topological susceptibility in the deconfined and confined
phases at $T_c$. Data from Ref.~\cite{LTW-04}.
}
\label{LTW05}
\end{figure}

Numerical results supporting the same picture have also been reported in
Ref.~\cite{LTW-04}. The numerical evidence of the topological suppression
across the transition was inferred from simulations at $T_c$, by monitoring
the correlation of the topological charge with the Polyakov line, whose value
is used to infer the actual phase of the configurations generated along the
given Monte Carlo run for $N\ge 3$ at the first-order transition.  This has
led to estimates of the topological susceptibility in the deconfined and
confined phase at $T_c$, whose ratio is shown in Fig.~\ref{LTW05} for various
$N$. 

Ref.~\cite{DPV-04} presented also some results for the
finite-temperature behavior of the coefficient $b_2$ of the $\theta$ expansion
around $\theta=0$: it remains small and substantially unchanged up to
$T\approx T_c$, then it appears to slightly increase  at $T\gtrsim T_c$.

Summarizing, the lattice results suggest that physical properties
determined by topological effects remain substantially unchanged in the
low-temperature confined phase.  On the other hand, in the high-temperature
phase there is a sharp change of regime where the topological susceptibility is
largely suppressed.  Such suppression becomes larger with increasing $N$,
suggesting that the topological susceptibility vanishes above the critical
temperature. Monte Carlo results seem to support the scenario presented
in~\cite{KPT-98}: at large $N$ the topological properties in the
high-temperature phase are essentially determined by instantons,
from very high temperature down to $T_c$; the exponential suppression of
instantons, as $e^{-N}$, induces the rapid decrease of the topological
activity observed in the large-$N$ limit.

\subsection{Results for the $\theta$ dependence of the spectrum}
\label{thspres}

Another interesting issue concerns the $\theta$ dependence of the spectrum of
the theory.  A numerical MC study for 4D $SU(N)$ gauge theories was reported
in Ref.~\cite{DMPSV-06}. Again numerical simulations of the Wilson lattice
formulation were employed to investigate the $\theta$ dependence of the string
tension $\sigma(\theta)$ and the lowest glueball mass $M(\theta)$. They can be
expanded around $\theta=0$ as in Eqs.~(\ref{sigmaex}) and (\ref{gmex}).  Then
the coefficients $s_i$ and $g_i$ of these expansions can be computed from
appropriate correlators at $\theta=0$.  In particular, $s_2$ can be
determined~\cite{DMPSV-06} from the large-$t$ behavior of connected
correlation functions of two Polyakov lines at distance $t$ and the square
topological charge, such as
\begin{eqnarray}
\langle A_P(t) Q^2 \rangle_{\theta=0} -
\langle A_P(t) \rangle_{\theta=0} \langle Q^2 \rangle_{\theta=0}
\label{g2}
\end{eqnarray}
where
\begin{equation}
A_P(t) = \sum_{x_1,x_2}  {\rm Tr}\,P^\dagger(0;0) \; {\rm Tr}\,P(x_1,x_2;t),  
\label{apdef}
\end{equation}
$P(x_1,x_2;t)$ is the Polyakov line of size $L$ along the $x_3$ direction, and
$Q$ is the topological charge.  Analogously, the $O(\theta^2)$ term of the
glueball mass can be obtained from appropriate connected correlation functions
of plaquette operators and $Q^2$. The $O(\theta^2)$ coefficients $s_2$ and
$g_2$, and in general all coefficients of those $\theta$ expansions, 
 are dimensionless scaling quantities, which should approach a constant
in the continuum limit, with $O(a^2)$ scaling corrections.

Ref.~\cite{DMPSV-06} obtained the first estimates of $s_2$ and $g_2$,
cf. Eqs.~(\ref{sigmaex}) and (\ref{gmex}), using the cooling method to
determine the topological charge, and for $N=3,4,6$ to also check their
large-$N$ behavior.  The $O(\theta^2)$ terms in the expansion around
$\theta=0$ of the spectrum of $SU(N)$ gauge theories are small for all $N\ge
3$. For example we mention the estimates 
\begin{equation}
s_2=-0.08(1),\quad g_2=-0.06(2)\qquad {\rm for} \;\; N=3.
\end{equation}
The $\theta$ dependence appears even smaller when dimensionless combinations
are considered, such as $M/\sqrt{\sigma}$ and, for $N>3$, the ratio of
tensions corresponding to strings in different representations. For
example, in the case of the scaling ratio 
\begin{equation}
{M(\theta)\over \sqrt{\sigma(\theta)}} =
{M\over \sqrt{\sigma}} ( 1 +  r_2 \theta^2 + ... ),
\label{ratioex}
\end{equation}
we find $r_2=g_2-s_2/2$, thus $r_2=-0.02(2)$ for $N=3$.

The $O(\theta^2)$ corrections appear to decrease with increasing $N$, and the
coefficients do not show evidence of convergence to a nonzero value.  This is
suggestive of a scenario in which the $\theta$ dependence of the spectrum
disappears in the large-$N$ limit, at least for sufficiently small values of
$\theta$ around $\theta=0$.  In the case of the spectrum, the general
large-$N$ scaling arguments of Sec.~\ref{largeNsec}, which indicate
$\bar{\theta}\equiv \theta/N$ as the relevant Lagrangian parameter in the
large-$N$ limit, imply that $O(\theta^2)$ coefficients should decrease as
$1/N^{2}$.  The results of Ref.~\cite{DMPSV-06} appear substantially
consistent: In the case of the string tension they suggest
\begin{equation}
\sigma(\theta) = \sigma \left( 1 + s_2 \theta^2 + ... \right),
\qquad s_2\approx -0.9/N^2.
\label{sigmaex2}
\end{equation}

Further investigation would be useful to put this scenario on a firmer
ground, using for example other definitions of topological charge.

\section{Results in full QCD}
\label{fullQCD}

In this section we discuss a number of issues related to the $\theta$
dependence and $U(1)_A$ symmetry breaking in QCD.  We recall that in the
presence of $N_f$ fermions, the general QCD Lagrangian with a $\theta$ term
reads
\begin{equation}
{\cal L_\theta}= {1\over 4} F_{\mu\nu}^a(x)F_{\mu\nu}^a(x)
+ \bar\psi\, D \psi
+ \bar\psi_L \,M\,\psi_R + \bar\psi_R \,M^\dagger\,\psi_L
- i \theta_q q(x),
\end{equation}
where $\psi$ is an $N_f$ component vector in flavor space, and $M$ is the
quark mass matrix.  As already discussed in Sec.~\ref{ua1pro}, the parameter
$\theta_q$ and the imaginary part of the quark mass are related, so that the
relevant $\theta$ parameter is given by
\begin{equation}
\theta = \theta_q + {\rm arg}\,{\rm det}\, M.
\label{thetafe}
\end{equation}

\subsection{The neutron electric dipole moment}
\label{neutron}

In strong interactions, one of the most stringent constraints on possible
violations of parity and time reversal symmetries is inferred from
measurements of the neutron electric dipole moment (NEDM).  The experimental
analyses of Refs.~\cite{Harris-etal-99} and \cite{Baker} lead to the upper
bounds
\begin{equation}
|d_n| < 6.3 \times 10^{-26}\,e\,{\rm cm},\qquad
|d_n| < 2.9 \times 10^{-26}\,e\,{\rm cm},
\label{upperbound}
\end{equation}
respectively.  In the standard model, CP violation in the
Cabibbo-Kobayashi-Maskawa mixing matrix gives also rise to a nonzero NEDM, but
only beyond the one-loop approximation in the weak interaction, thus leading
to a very small value~\cite{EGN-76,NYC-80}, i.e. $|d_n| \lesssim
10^{-30}e\,{\rm cm}$, much smaller than the actual experimental bound
(\ref{upperbound}).

\begin{table*}
  \caption{
  Estimates of the NEDM in the presence of a $\theta$ term.
  We report the quantity $c_n$ defined by  
  writing $d_n = c_n \, \theta \,e\,{\rm fm}$.
Results from lattice QCD are reported in the text.  
}
\label{nedm}
\hspace*{-1cm}    
\tabcolsep 4pt        
\begin{center}
\begin{tabular}{rll}
\hline
\multicolumn{1}{c}{Ref. $_{\rm year}$}& 
\multicolumn{1}{c}{approach/model}& 
\multicolumn{1}{c}{$c_n$} \\
\hline  
\cite{Baluni-79} $_{1979}$ & bag model & \phantom{$-$}0.0027 \\

\cite{CDVW-79} $_{1980}$ & ChPT & \phantom{$-$}0.0036 \\

\cite{KO-81} $_{1981}$ & ChPT & \phantom{$-$}0.001 \\

\cite{KK-81} $_{1981}$ & ChPT & \phantom{$-$}0.0055 \\

\cite{Shabalin-82} $_{1982}$ & ChPT & \phantom{$-$}0.02 \\

\cite{MI-84} $_{1984}$ & chiral bag model & \phantom{$-$}0.0030 \\

\cite{Schnitzer-84} $_{1984}$ & soft pion Skyrmion model & \phantom{$-$}0.0012 \\

\cite{CN-84} $_{1984}$ & single nucleon contribution & \phantom{$-$}0.011 \\

\cite{PR-91} $_{1991}$ & ChPT & \phantom{$-$}0.0033(18) \\

\cite{Cheng-91} $_{1991}$ & ChPT & \phantom{$-$}0.0048 \\

\cite{AH-92} $_{1992}$ & ChPT & $-$0.0072,$\,-$0.0039 \\

\cite{PR-99} $_{1999}$ & sum rules & \phantom{$-$}0.0024(10) \\

\cite{Borasoy-00} $_{2000}$ & heavy baryon ChPT & $-$0.0075(32) \\

\cite{FGS-04} $_{2004}$ & instanton liquid & \phantom{$-$}0.010(4) \\

\cite{HKS-07} $_{2007}$ & holographic QCD & \phantom{$-$}0.00108 \\

\hline
\end{tabular}
\end{center}
\end{table*}

In order to translate the above experimental bound into a constraint on
$\theta$, one needs to compute the NEDM in QCD in the presence of a $\theta$
term, i.e. to compute the quantity $c_n$ defined as
\begin{equation}
d_n \approx c_n \,\theta \,e\,{\rm fm}
\label{cndef}
\end{equation}
in the presence of a small $\theta$ term.  Calculations have been done within
models, such as the bag models~\cite{Baluni-79,MI-84,MM-86}, the instanton
liquid model~\cite{FGS-04}, holographic QCD \cite{HKS-07}, or by approaches
based on a CP violating effective chiral Lagrangian and chiral perturbation
theory
(ChPT)~\cite{CDVW-79,DiVecchia-80,KO-81,KK-81,PR-91,Cheng-91,AH-92,Borasoy-00},
on QCD sum rules~\cite{SVZ-80,PR-99}, etc...  Table~\ref{nedm} presents a list
of results for $c_n$.

The present theoretical estimations of the NEDM in the presence of a $\theta$
term significantly vary among the different models, approaches and
approximations which are considered. It is hard to get an accurate estimate
from the results reported in Table~\ref{nedm}.  There is a substantial global
agreement on the size of the quantity $c_n$, indeed they suggest
\begin{equation}
0.001 \lesssim |c_n|  \lesssim 0.01,
\label{bounddn}
\end{equation}
however, there is not a complete agreement on the sign of $c_n$\,.  Anyway,
these crude estimates of $c_n$ strongly suggest that $\theta$ is very small:
By taking as lower bound $|c_n|\gtrsim 0.001$, the experimental upper bounds
(\ref{upperbound}) lead to
\begin{equation}
|\theta| \ltapprox O(10^{-10}).
\label{thetabound}
\end{equation}
Clearly, an accurate estimate of $c_n$ is essential, to translate the
experimental bound on the NEDM into a reliable stringent bound on $\theta$.

The computation of the ratio $c_n\equiv d_n/\theta$ can also be addressed by
calculations within the lattice formulation of QCD, which represents the most
fundamental source of theoretical nonperturbative information on the strong
interactions.  The dipole moment of nucleons is obtained from the form factors
that parametrize the electromagnetic current between nucleon states in the
$\theta$ vacuum,
\begin{eqnarray}
\langle p',s'| J_{\rm em}^\mu | p,s \rangle = 
\bar{u}_{s'}(p') \Gamma^\mu(q^2) u_s(p), \label{eleform}
\end{eqnarray}
where $q=p'-p$ and $\Gamma^\mu(q^2)$ has the most general four-vector
structure consistent with the gauge, Lorentz, and CPT invariance of QCD, see
e.g. \cite{BBOS-05},
\begin{equation}
\Gamma^\mu(q^2) = F_1(q^2) \gamma^\mu  + 
{1\over 2m_n} F_2(q^2) i\sigma^{\mu\nu} q_\nu 
 + F_A(q^2) (\gamma^\mu \gamma^5 q^2 - 2 m_n \gamma^5 q^\mu) +
{1\over 2 m_n} F_3(q^2) \sigma^{\mu\nu}\gamma_5 q_\nu.
\end{equation}
$F_3(0)/2m_n$ gives the electric dipole moment, which vanishes when $\theta\to
0$.  In lattice calculations matrix elements can be extracted from the
large-distance behavior of appropriate correlation functions in Euclidean
spacetime. In the case at hand one considers three-point correlations such as
\begin{equation}
\langle N(t_2,\vec{p}_2) J_{\rm em}^\mu(t,\vec{q}) 
N^\dagger(t_1,\vec{p}_1) \rangle,
\end{equation}
where $N$ is the interpolating operator which creates the given nucleon.

The NEDM can be also extracted by computing the variation of the energy of
neutron states in the presence of an external electric field~\cite{AG-89}. In
a CP-violating theory, a static constant electric field $\vec{E}$ gives rise
to a change in the energy ${\cal E}$ of the nucleon state:
\begin{equation} 
\delta {\cal E} \approx d_n \vec{S}\cdot \vec{E},
\label{deltaE} 
\end{equation}
where $\vec{S}$ is the nucleon spin, and $d_n \sim \theta$.  An estimate of
the NEDM is obtained by measuring the energy difference of nucleon states with
opposite spin along the electric field, i.e.
\begin{equation}
{\cal E}_+ - {\cal E}_- = 2 d_n \vec{S}\cdot \vec{E} + O(|E|^3).
\label{enediff}
\end{equation}

Although the action with $\theta\ne 0$ cannot be directly studied by numerical
Monte Carlo methods, in the small $\theta$ limit one can obtain the desired
result for $c_n$, Eq.~(\ref{cndef}), by expanding around $\theta=0$ and taking
only the linear term, i.e.  for a generic expectation value of product of
operators $P$,
\begin{equation}
\langle  P \rangle_\theta \approx 
\langle  P \rangle_{\theta=0} + i \theta \langle  P Q \rangle_{\theta=0}
\end{equation}
where $Q$ is the topological charge. Alternatively, the nucleon electric
dipole at finite $\theta$ can be determined using reweighting techniques with
the complex weight factor $e^{i\theta Q}$.  Another approach, pursued in
Ref.~\cite{Izubuchi-etal-07}, utilizes MC simulations at
imaginary $\theta$ values.

Lattice calculations have been reported and discussed in
Refs.~\cite{AG-89,AGMS-90,GLMS-03,Shintani-etal-05,BBOS-05,
  Shintani-etal-07,Izubuchi-etal-07,SAK-08}.  Several
investigations~\cite{AG-89,AGMS-90,GLMS-03,Shintani-etal-05,Shintani-etal-07}
have been performed within the quenched approximation, i.e. neglecting fermion
loops, leading to the estimate~\cite{Shintani-etal-07,SAK-08} $c_n\approx
0.02$-$0.04$ for the linear coefficient defined in Eq.~(\ref{cndef}), using
both the form factor and external electric field methods.  However,
calculations within this approximation are not expected to give accurate
results because they miss the fact that the effect of the $\theta$ term must
vanish in the chiral limit, and therefore they may overestimate the ratio
$d_n/\theta$ at the physical small values of the $u$ and $d$ quark masses.
First results for full QCD have been reported only
recently~\cite{BBOS-05,Shintani-etal-07,Izubuchi-etal-07,SAK-08}.
Ref.~\cite{BBOS-05} reported a calculation in full QCD with two dynamical
domain-wall fermions, by measuring the electromagnetic form factor.  However,
the precision was not sufficient to estimate the ratio $d_n/\theta$, but
provided only the bound 
\begin{equation}
|c_n| \lesssim 0.02, 
\label{lattbound}
\end{equation}
which is larger than the calculations reported in Table~\ref{nedm}, see
also Eq.~(\ref{bounddn}).  In Ref.~\cite{SAK-08} the NEDM has been estimated
using the external electric field method, with two dynamical clover quarks,
obtaining the estimate $c_n \approx 0.04$ from the smallest quark mass values
considered in the MC simulations.  Although this estimate comes apparently
with a relatively large uncertainty, it is significantly larger than the
above-mentioned estimates by other approaches, summarized by
Eq.~(\ref{thetabound}).  In particular, it confirms the results obtained
within the quenched approximation, without showing the expected decrease when
approaching the chiral limit; the original paper contains a discussion of the
possible reasons, which, for example, may be related to the fact that the
small quark-mass regime (chiral limit) has not yet been reached by numerical
simulations. Therefore, further work is called for, in
order to accurately determine the NEDM in the presence of a $\theta$ term from
lattice QCD at sufficiently small values of the quark masses.

\subsection{The topological susceptibility in the chiral limit}
\label{topchiral}

The behavior of $\chi$ as a function of the quark masses can be understood in
the framework of the chiral effective Lagrangian of QCD, see e.g.
Refs.~\cite{Crewther-77,DV-80,LS-92,Weinberg,Leutwyler-00}.  
In the case of $N_f$ degenerate flavors, the topological
susceptibility is expected to vanish at small quark masses $m$ as
\begin{equation}
\chi = {m \Sigma \over N_f} + O(m^2) = {f_\pi^2 M_\pi^2\over 2 N_f} + O(M_\pi^4),
\label{chimq}
\end{equation}
where $\Sigma=-\langle \bar\psi \psi \rangle/N_f$.  Eq.~(\ref{chimq}) does not
apply to one-flavor QCD, $N_f=1$, where there is no spontaneous breaking of
chiral symmetry (the anomaly breaks the whole axial symmetry), thus no
Goldstone bosons such as pions are expected.
Refs.~\cite{Creutz-04-2,Creutz-07-2} discuss the case of one-flavor QCD, and
in particular the meaning of its massless quark limit.  Note that while $m$
and $\Sigma$ are not renormalization-group invariant, since each one depends
on the renormalization-group scheme and on the corresponding scale, their
product $m\Sigma$ is a renormalization-group invariant quantity.  When $m\to
0$, $\chi$ vanishes, consistently with the fact that in the presence of a
massless flavor the $\theta$ dependence disappears.  On the other hand, in the
limit of large quark masses, the topological susceptibility is expected to
approach the nonzero value $\chi_\infty$ of the pure gauge theory.  In the
large-$N$ limit, it has been argued that~\cite{DV-80,LS-92}
\begin{equation}
{1\over \chi} = {2 N_f\over f_{\pi,\infty}^2 M_{\pi,\infty}^2} + 
{1\over \chi_\infty}\,,
\label{chimq2}
\end{equation}
which describes the crossover behavior of $\chi$ from the small-$M_\pi$ and
large-$M_\pi$ asymptotic regions.  We recall that a relation analogous to
(\ref{chimq}) holds also on the lattice, at finite lattice spacing, when
fermions satisfying the Ginsparg-Wilson relation are considered, see
Sec.~\ref{gwferm}. In the case of other discretizations of the Dirac operator,
such as Wilson and staggered fermions, the relation (\ref{chimq}) is expected
to hold only in the continuum limit.

\begin{figure}
\centerline{\psfig{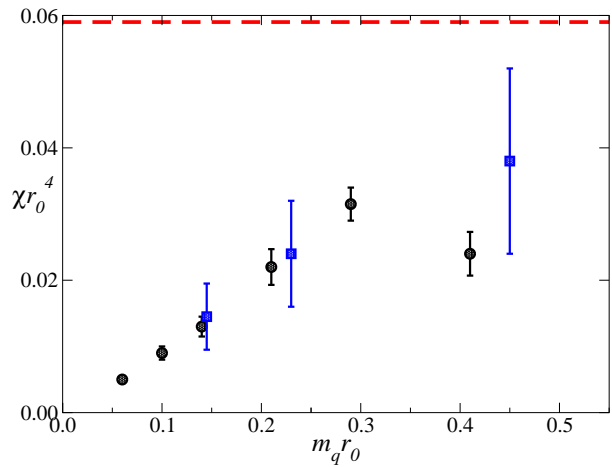}}
\caption{
  $\chi r_0^4$ vs $m_q r_0$ where $m_q$ is the $\overline{\rm MS}$ quark mass
  at the scale $\mu=2$ GeV.  We show data from Ref.~\cite{DS-07} (squares) and
  from Ref.~\cite{Aoki-etal-07} (circles), 
  obtained by Monte Carlo simulations of two flavors of dynamical
  overlap fermions (the results of Ref.~\cite{Aoki-etal-07} were
  obtained at fixed topological charge $Q=0$,
  as discussed at the end of Sec.~\ref{fermdefGW}). 
  The dashed line indicates the quenched result, which
  corresponds to the limit $m_q r_0\to \infty$.  }
\label{chivsm}
\end{figure}

Several numerical simulations have been devoted to the study of the behavior
of the topological susceptibility in the presence of dynamical quarks, see
Refs.~\cite{Bitar-etal-91,KFMOU-93,DGHS-98-2,Hasenfratz-00,ADD-00,Hasenfratz-01,
HT-01,CPPACS-01,BELNOSSV-01,Durr-01,Bernard-etal-03,BDO-04,Hart-04,Allton-etal-04,
Aoki-etal-05,DS-05,BBOS-05,EFKS-06,LMO-06,Aoki-etal-07,DS-07,Bernard-etal-07},
and in particular to a check of the asymptotic behavior (\ref{chimq})
in the regime of small masses.  The results obtained using
Wilson~\cite{HT-01,CPPACS-01,BELNOSSV-01,Durr-01} and
staggered~\cite{Bitar-etal-91,KFMOU-93,DGHS-98-2,ADD-00,Durr-01,Bernard-etal-03,
Bernard-etal-07}
fermions are substantially consistent with the expected behavior (\ref{chimq})
and (\ref{chimq2}). However, a conclusive quantitative check of
Eq.~(\ref{chimq}) has not been achieved yet.  One should not forget that the
numerical check of Eq.~(\ref{chimq}) is subject to scaling corrections (i.e.
it is expected to hold only in the continuum limit), to possible systematic
errors in the measurement of the topological susceptibility depending on the
method employed, and also to the fact that the topological modes are very
slowly sampled in Monte Carlo simulations (we shall return to this point in
Sec.~\ref{qsampling}).  Results obtained using chirally improved fermions,
such as domain wall fermions, can be found in
Refs.~\cite{Aoki-etal-05,BBOS-05,LMO-06}, and are substantially consistent
with a suppression of the topological susceptibility in the chiral limit.  The
use of overlap fermions should simplify the check of Eq.~(\ref{chimq}),
because it suppresses some sources of systematic errors.  Although overlap
fermions require a much larger numerical effort, the check of
Eq.~(\ref{chimq}) should benefit from the fact that it holds also at finite
lattice spacing if one employs the fermionic definition of topological charge
based on the index theorem. Therefore, it can be done on relatively small
lattice sizes, without the need of carefully studying the continuum limit.
The results obtained so far are quite
promising~\cite{EFKS-06,Aoki-etal-07,DS-07}. They show clear evidence of the
suppression of $\chi$ when decreasing the quark mass $m$, consistently with
the prediction of chiral perturbation theory.  In Fig.~\ref{chivsm} we show
some recent results from Refs.~\cite{DS-07,Aoki-etal-07}, obtained by Monte
Carlo simulations of two flavors of dynamical overlap fermions; $\chi r_0^4$
is plotted versus $m_q r_0$\,, where $r_0$ is the Sommer
scale~\cite{Sommer-94} and $m_q$ is the $\overline{\rm MS}$ quark mass at the
scale $\mu=2$ GeV.  The data are nicely consistent with the expected linear
behavior $\chi\sim m_q$ at small quark masses.

The behavior of $\chi$ has also been investigated at
finite density, in Refs.~\cite{MZ-05,ADL-06}.

\subsection{The $U(1)_A$ axial anomaly at finite temperature}
\label{ftqcd}

Another interesting issue concerns the temperature dependence of the
$U(1)_A$ axial symmetry breaking, and, in particular, the possibility of its
effective restoration at finite temperature, for example at the chiral
transition where the chiral $SU(N_f)_A$ symmetry is restored in the massless
limit. We recall that the anomaly equation holds also at finite temperature
because of its ultraviolet origin.

As already discussed in Sec.~\ref{ua1pro}, at $T=0$ the $U(1)_A$ axial anomaly
gives rise to a quite large splitting in mass of pseudoscalar flavor singlet
and nonsinglet mesons, such as $\eta$ and $\eta'$, which have the same quark
content.  This issue has been investigated on the lattice in
Refs.~\cite{HI-08,Aoki-etal-07-2,GIRN-07,SNL-05,CPPACS-03,MMS-02,
Struckmann-etal-01,MM-00,FKOU-95-2}.
The results for the $\eta$ and $\eta'$ masses are in substantial agreement
with the experimental values.  

The thermodynamics of strong interactions is characterized by a transition at
$T\approx 200$ MeV from a low-$T$ hadronic phase, in which chiral symmetry is
broken, to a high-$T$ phase with deconfined quarks and gluons (quark-gluon
plasma), in which chiral symmetry is restored~\cite{Wilczek-00,Karsch-02}.
The nature of the phase transition is also sensitive to the symmetry breaking
pattern, and therefore to whether the $U(1)_A$ symmetry is effectively restored
at $T_c$~\cite{PW-84,BPV-05-2,v-07}.

A complete suppression of the anomaly effects at $T_c$ is however unlikely in
QCD.  Semiclassical calculations in the high-temperature phase~\cite{GPY-81}
show that the instanton contribution is suppressed for $T\gg T_c$, implying a
suppression of the anomaly effects in the high-temperature limit, but not
their complete vanishing at finite $T$.  In Sec.~\ref{fintthe} we have
discussed the case of pure $SU(N)$ gauge theories, where the topological
susceptibility appears to suffer a significant decrease across the transition,
but remains nonzero above $T_c$ for finite $N$.

The effects of the $U(1)_A$ anomaly at finite temperature in full QCD
can be investigated by comparing the behavior of fermionic
correlators in different meson channels which are related by the $U(1)_A$
symmetry, see e.g.
Refs.~\cite{Shuryak-94,Cohen-96,LH-96,EHS-96,BCM-96}.  For example, in
the case of two flavors, we have the isoscalar $I=0$ scalar channel $\sigma$
associated with the operator $O_\sigma=\bar\psi\psi$, the $I=1$ scalar channel
$\delta$ associated with $\vec{O}_\delta=\bar\psi \vec{\tau} \psi$, and the
corresponding pseudoscalar $I=0$ $\eta'$ and $I=1$ $\pi$ channels associated
with the operators $O_{\eta'}=\bar\psi\gamma_5 \psi$ and
$\vec{O}_{\pi}=\bar\psi\vec{\tau} \gamma_5 \psi$, respectively.  When the
$SU(2)_L\otimes SU(2)_R$ chiral symmetry with two massless flavors is realized
explicitly, rather than spontaneously broken, it implies degeneracies between
mesonic correlators.  $SU(2)_A$ transformations mix the $\sigma$ and $\pi$
channels, and the $\delta$ and $\eta'$ channels. Therefore, above $T_c$, where
the $SU(2)_A$ symmetry is restored, the correlations in the corresponding
channels are expected to become identical (of course in the chiral limit
only), apart from trivial factors, see e.g. Ref.~\cite{KLS-98}.  On the other
hand, $U(1)_A$ transformations relate correlators in the $\sigma$ and $\eta'$
channels, and correlators in the $\delta$ and $\pi$ channels.  Therefore an
effective further restoring of the $U(1)_A$ symmetry at $T\ge T_c$ would lead
to equal correlators in all these channels, and in particular, to equal
susceptibilities and screening masses.

\begin{figure}
\centerline{\psfig{width=8truecm,angle=0,file=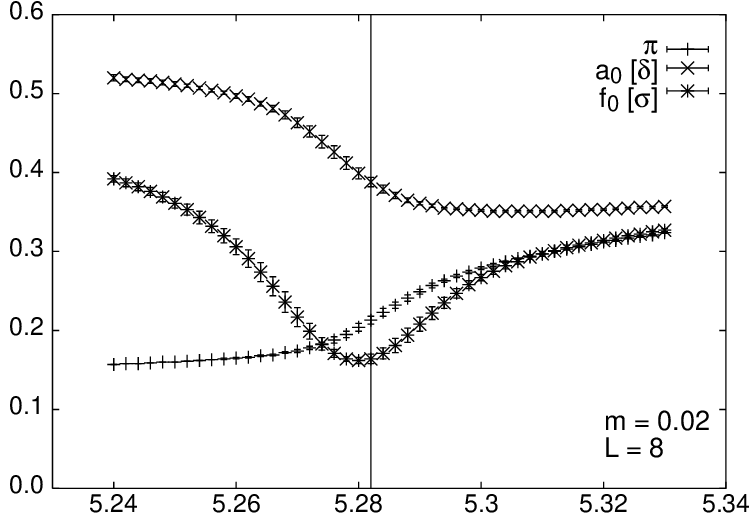}}
\caption{
  Susceptibilities in different channels for 2-flavor QCD with staggered
  fermions on lattices of size $8^3\times 4$, versus $\beta$.  
  The vertical line indicates the value of $\beta$ at the
  transition. Taken from Ref.~\cite{Karsch-00}.  }
\label{karsch01}
\end{figure}

The issue has been investigated by lattice Monte Carlo studies, see
Refs.~\cite{Karsch-00,Vranas-99,CCCLMV-99,KLS-98,Bernard-etal-97}.  Although
lattice results have been obtained for relatively small lattice sizes and
convincing checks of scaling in the continuum limit have not been reported,
they favor the scenario in which the $U(1)_A$ symmetry is not restored at
$T_c$.  Fig.~\ref{karsch01}, taken from Ref.~\cite{Karsch-00}, shows results
for the susceptibilities of different channels.  Similar results have been
obtained using domain wall fermions in Ref.~\cite{Vranas-99}.  The screening
masses in the $\sigma$ and $\pi$ appear equal above $T_c$ signaling that the
chiral symmetry $SU(2)_L\otimes SU(2)_R$ is restored above $T_c$.  The
effective breaking of the axial $U(1)_A$ symmetry appears substantially
reduced above $T_c$, but the difference in the $\delta$ and $\pi$ channels,
and in particular in their screening masses, although small above $T_c$,
appears significantly different from zero at least up to $T\approx 1.2 T_c$.
At even higher temperatures, we expect that this difference decrease smoothly
to zero, as indicated by semiclassical instanton calculations~\cite{GPY-81}.

The $U(1)_A$ symmetry breaking at finite temperature has also been discussed
within the instanton liquid model~\cite{SS-98,SS-96}, leading to analogous
conclusions. Its effects have been also investigated within the AdS/CFT
framework in Refs.~\cite{Schaefer-07,BL-06}.  

The phase structure of the theory above $T_c$, and in particular a possible
effective restoration of the $U(1)_A$ symmetry above $T_c$, has been
investigated in Refs.~\cite{Meggiolaro-94,MM-03,Meggiolaro-04}, by assuming
the existence of quark condensates which break the $U(1)_A$ axial symmetry,
but are invariant under the symmetry $SU(N_f)_V \otimes SU(N_f)_A \otimes
U(1)_V$, such as the vacuum expectation value of the $2N_f$-quark operator
${\rm det}\, (\bar{\psi}_{R,i} \psi_{L,j})+{\rm det}\,(\bar{\psi}_{L,i}
\psi_{R,j})$ (where $i,j$ are flavour indices) and appropriate extensions,
which may play a relevant role above $T_c$ where the bilinear quark condensate
$\langle \bar{\psi} \psi\rangle$ vanishes.

\subsection{$U(1)_A$ breaking and the nature of the finite-temperature chiral transition}
\label{u1atrans}

Our understanding of the nature of the finite-$T$ phase transition in the
chiral limit is essentially based on the relevant symmetry and
symmetry-breaking pattern. In the presence of $N_f$ light quarks, the relevant
symmetry is the chiral symmetry $SU(N_f)_L\otimes SU(N_f)_R$. At $T=0$ this
symmetry is spontaneously broken to $SU(N_f)_V$ with a nonzero quark
condensate $\langle {\bar \psi} \psi \rangle$.  The finite-$T$ transition is
related to the restoring of the chiral symmetry.  It is therefore
characterized by the symmetry breaking
\begin{equation}
SU(N_f)_L\otimes SU(N_f)_R   \rightarrow SU(N_f)_V . 
\label{qcdsb}
\end{equation}
An effective restoring of the $U(1)_A$ axial symmetry at the
finite-temperature transition can have important consequences for the nature
of the transition in the chiral limit~\cite{PW-84}.  Indeed, if the effects of
the axial anomaly are effectively suppressed at the transition, instead of the
symmetry breaking pattern (\ref{qcdsb}), we would have
\begin{eqnarray}
U(N_f)_L\otimes U(N_f)_R  \rightarrow U(N_f)_V .
\label{qcdsbwa}
\end{eqnarray}
Table~\ref{summarytqcd} lists the possible transitions as predicted by
renormalization-group analyses based on universality
arguments~\cite{BPV-05-2,BPV-05,BPV-03,PW-84,v-07}, for various values of
$N_f$, and for the two symmetry breaking patterns (\ref{qcdsb}) and
(\ref{qcdsbwa}).  In those cases where a continuous transition is possible,
the corresponding universality class is reported.

\begin{table*}
  \caption{
    This Table summarizes the predictions by renormalization-group 
    analyses based on 
    universality arguments on the nature of the finite-temperature transition 
    of QCD, for different numbers   
    $N_f$ of light flavors, and in the case the effects of the $U(1)_A$
    axial anomaly are relevant or effectively suppressed. 
    In those cases where a continuous transition is possible, the corresponding
    universality class is reported, and indicated by the
    corresponding symmetry breaking pattern.
    See Refs.~\cite{BPV-05-2,v-07} for more details and references.
  }
\label{summarytqcd}
\begin{center}
\begin{tabular}{ccc}
\hline\hline
\multicolumn{1}{c}{}& 
\multicolumn{1}{c}{$U(1)_A$ anomaly}& 
\multicolumn{1}{c}{suppressed anomaly at $T_c$}\\
\hline
\multicolumn{1}{c}{$N_f$}& 
\multicolumn{1}{c}{$SU(N_f)_L\otimes SU(N_f)_R\rightarrow SU(N_f)_V$}& 
\multicolumn{1}{c}{$U(N_f)_L\otimes U(N_f)_R\rightarrow U(N_f)_V$}\\ 
\hline
$N_f=1$ & crossover or first order & $O(2)/Z_2$ or first order  \\
$N_f=2$ & $O(4)/O(3)$ or first order & 
$[\,U(2)_L\otimes U(2)_R\,]\,/\,U(2)_V$  or first order  \\
$N_f\ge 3$ & first order & first order \\
\hline\hline
\end{tabular}
\end{center}
\end{table*}

In the physically relevant case of $N_f=2$ light quarks, renormalization-group
arguments show that the chiral transition can be continuous, and, in this
case, its critical behavior belongs to the 3D $O(4)$ vector universality class
(essentially because the expected symmetry breaking $SU(2)_L\otimes SU(2)_R
\rightarrow SU(2)_V$ is equivalent to $O(4)\rightarrow O(3)$, as in the $O(4)$
vector model), characterized by the critical exponents $\nu=0.749(2)$ and
$\eta=0.0365(10)$ \cite{Hasenbusch-01-2,PV-02}.  The transition can be
continuous also in the case the $U(1)_A$ symmetry is effectively restored at
$T_c$, but in a different $[\,U(2)_L\otimes U(2)_R\,]\,/\,U(2)_V$ universality
class~\cite{BPV-05,BPV-05-2}, characterized by the critical exponents
$\nu\approx 0.7 $ and $\eta \approx 0.1 $.  As we have discussed above, this
should not occur in three-color QCD, but it may be relevant in the large-$N_c$
limit, where the anomaly gets suppressed.  We also recall that the existence
of a corresponding universality class does not exclude that the transition may
be first order, when the system is outside the attraction domain of the stable
fixed point of the RG flow which describes the universality class.

The nature of the transition in QCD can be investigated within the lattice
formulation of QCD. Many studies based on MC simulations of different lattice
formulations of QCD have been performed, see e.g.
Refs.~\cite{CP-PACS-01,MILC-00,KLP-01,KS-01,EHMS-01,DDP-05,KS-06}, but this
issue is still controversial. Some MC results favor a continuous transition.
However, the results have not been sufficient to settle its O(4) nature yet.
There are also claims~\cite{DDP-05} in favor of a first-order transition.
Unlike $N_f=2$ QCD, the transition scenario appears settled for $N_f\ge 3$: MC
simulations \cite{KLP-01,IKSY-96,FP-03,Ch-etal-07,FP-07} show a first-order
transition, in agreement with the RG predictions.

The effects of the $U(1)_A$ anomaly on the chiral transition have been
numerically investigated in Ref.~\cite{CM-07} using strongly coupled lattice
QED with two flavors as a model for the chiral transition in two-flavor QCD.
In this model the transition appears first order in the absence of terms
breaking the $U(1)$ symmetry corresponding to the $U(1)_A$ symmetry of QCD.
Then, the explicit breaking of this $U(1)$ symmetry weakens the transition,
and turns it to an $O(4)$ continuous transition for a sufficiently large
strength of the $U(1)$ symmetry breaking. These features are consistent with
one of the different scenarios allowed by RG arguments, which are listed in
Table~\ref{summarytqcd}. However, these results cannot be extended to QCD,
because the actual realization of one of the scenarios allowed by the RG
arguments, i.e. the fact that the transition is first order or continuous
(when the universality class exists), depends on the details of the model
under investigation, not only on the global features which characterize the
given universality class.

In nature quarks are not massless, although some of them, the quarks $u$ and
$d$, are light. The physically interesting case is QCD with $N_f=2$ light
quarks and four heavier quarks (in particular the quark $s$ with $m_s\approx
100$ MeV).  Therefore, it is important to consider the effects of the quark
masses in the above transition scenarios.  Renormalization-group arguments
show that, if the transition is continuous in the chiral limit, then an
analytic crossover is expected for nonzero values of the quark masses $m_f$,
because the quark masses act as external fields coupled to the order
parameter. On the other hand, a first-order transition is generally robust
against perturbations, and therefore it is expected to persist for $m_f>0$, up
to an Ising end point.  Actually, the presence of the massive strange quark
$s$ makes the above scenario more complicated, because the nature of the
transition may be sensitive to its mass $m_s$.  Since the transition is
expected to be first order in the chiral limit of $N_f=3$ degenerate quarks,
see Table~\ref{summarytqcd}, we also expect that the first-order transition
persists for sufficiently small values of $m_s$.  On the other hand, if the
transition is continuous in the limit $m_s\rightarrow \infty$ corresponding to
$N_f=2$ degenerate quarks, then there must be a tricritical point at
$m_s=m_s^*$ (where the critical behavior is as described by mean field theory,
apart from logarithms) separating the first-order transition line from the
O(4) critical line.

For quark masses around their physical values, the results of MC simulations,
see Refs.~\cite{Be-etal-05,Ch-etal-06,AFKS-06,FP-07}, support a crossover
scenario, i.e. the low-$T$ hadronic and high-$T$ quark-gluon plasma regimes
are not separated by a phase transition, but rather by an analytic crossover,
where the thermodynamic quantities change rapidly in a relatively narrow
temperature interval.  Nevertheless, the nature of the transition in the
chiral limit is still of interest. Since the physical masses of the lightest
quarks $u$ and $d$ are very small, some scaling relations, valid in the chiral
limit, may still hold at the physical values of the quark masses.

\section{$\theta$ dependence in the two-dimensional $CP^{N-1}$ models}
\label{cpn}

\subsection{The two-dimensional $CP^{N-1}$ model 
as a theoretical laboratory}
\label{cpngen}

Issues concerning the $\theta$ dependence can also be discussed in
two-dimensional CP$^{N-1}$ models~\cite{DDL-79,Witten-79b}, whose
Lagrangian is given by
\begin{equation}
{\cal L} = {N\over 2g} \overline{D_\mu z}\, D_\mu z ,
\label{lagrangiancpn}
\end{equation}
where $z$ is a complex $N$-component scalar field subject to the constraint
$\bar{z}z=1$,
\begin{equation}
A_\mu=i\bar{z}\partial_\mu z
\label{adef}
\end{equation}
is a composite gauge field, and $D_\mu =\partial_\mu +iA_\mu$ is a covariant
derivative.  They provide an interesting theoretical laboratory. Indeed they
present several features that hold in QCD: asymptotic freedom, gauge
invariance, existence of a confining potential between non gauge invariant
states (that is eventually screened by the dynamical constituents), and
nontrivial topological structure (semiclassical instanton solutions, $\theta$
vacua). We recall that the $CP^1$ model is equivalent to the O(3) $\sigma$
model, as can be easily seen by a change of variables. Indeed, in the case
$N=2$, the action can be written in a O(3)-symmetric form by reexpressing the
$z$ field in terms of $\vec{s}=\bar{z}\vec{\sigma} z$ where $\sigma_i$ are the
Pauli matrices.

An appealing feature of 2D $CP^{N-1}$ models is the possibility of perfoming a
systematic $1/N$ expansion, keeping $g$ fixed, around the large-$N$ saddle-point
solution~\cite{DDL-79,Witten-79b,CR-92,CR-93}, unlike 4D $SU(N)$ gauge theories.
This makes these models particularly interesting, because they allow us to
also check general nonperturbative scenarios by analytic calculations, without
necessarily resorting to numerical Monte Carlo methods of their lattice
formulation.

Analogously to 4D $SU(N)$ gauge theories, one may add a $\theta$ term to the
Lagrangian, writing
\begin{equation}
{\cal L}_\theta  = {N\over 2g} \overline{D_\mu z}\, D_\mu z -
i \theta {1\over2\pi}\,\epsilon_{\mu\nu}\, \partial_\mu A_\nu,
\label{lagrangiancpntheta}
\end{equation}
where 
\begin{equation}
q(x) ={1\over2\pi}\,\epsilon_{\mu\nu}\, \partial_\mu A_\nu
= {i\over2\pi}\,\epsilon_{\mu\nu}\, \overline{D_\mu z} D_\nu z 
\label{qcpn}
\end{equation}
is the topological charge density.  Then one may study the $\theta$ dependence
of the ground state and other observables. As in 4D $SU(N)$ gauge
theory, $q(x)$ is a renormalization-group invariant operator.

\subsection{The ground-state energy}
\label{groundstate energy}

By analogy with $SU(N)$ gauge theories, the ground state energy $F(\theta)$,
defined as in Eq.~(\ref{vftheta}), depends on $\theta$.  One may define a
dimensionless function $f(\theta)$ related to the ground state energy,
\begin{equation}
f(\theta) \equiv \xi^{2} [F(\theta)-F(0)] ,
\label{fthetacpn} 
\end{equation}
where $\xi \equiv \xi (\theta=0)$ is a length scale defined at $\theta=0$.
$\xi(\theta)$ can be defined from the second moment of the two-point
correlation function $G_P(x-y) = \langle {\rm Tr}\, P(x) P(0) \rangle$ of the
operator $P_{ij}(x) \equiv \bar{z}_i(x) z_j(x)$, i.e.  \footnote{The
  second-moment correlation length $\xi$ of $G_P(x)$ turns out to be more
  suitable for a $1/N$-expansion than the length scale $\xi_w$ (inverse mass
  gap) determined from the large-distance exponential decay of $G_P(x)$, i.e.
  $G_P(x)\sim e^{-|x|/\xi_w}$ at large $|x|$, due to its analytical properties
  in $1/N$~\cite{CR-91,CR-92}.  While using $\xi$ leads to an expansion in
  powers of $1/N$, the use of $\xi_w$ gives rise to nontrivial power
  corrections: Indeed, in the large-$N$ limit $\xi/\xi_w=\sqrt{2/3} +
  O(N^{-2/3})$.  }
\begin{equation}
\xi(\theta)^2 \equiv { \int d^2 x \;{1\over 4} x^2 G_P(x)\over  \int d^2 x  \;G_P(x) }.
\label{xidefcpn}
\end{equation}
The scaling function $f(\theta)$ can be expanded around $\theta=0$ as
\begin{equation}
f(\theta) = {1\over 2} C \theta^2 \left( 1 + \sum_{n=1} b_{2n} \theta^{2n} \right) ,
\label{fthetacpnexp} 
\end{equation}
where $C$ is the dimensionless quantity $\chi\xi^2$ at $\theta=0$, and $\chi$
is the topological susceptibility
\begin{equation}
\chi = \int d^2 x \langle q(0) q(x) \rangle .
\label{chicpn}
\end{equation}

\subsection{$\theta$ dependence in the large-$N$ limit}
\label{thetadepcpnln}

The $\theta$ dependence of the theory can be investigated within the $1/N$
expansion.  In the large-$N$ limit we
have~\cite{luescher-78,DDL-79,Witten-79b,CR-91}
\begin{equation}
f(\theta) = {1\over 4\pi N}  \theta^2  + O(1/N^2)
\label{thetadepcpn}
\end{equation}
for $|\theta|<\pi$, thus the topological susceptibility is $O(1/N)$,
\begin{equation}
C = {\partial^2 f(\theta)\over d \theta^2}\Bigg|_{\theta{=}0} 
= \chi\xi^2 = {1\over 2\pi N} + O(1/N^2).
\label{chim2largen}
\end{equation} 
As with 4D $SU(N)$ gauge theories (see Sec.~\ref{largeNsec} and in particular
Eqs.~(\ref{conj1}) and~(\ref{conj2})), the $2\pi$ periodicity of the $\theta$
dependence must give rise to a cusp at $\theta=\pi$, and therefore to a
singular behavior, which will be further discussed in Sec.~\ref{thetapi}.

The coefficients $b_{2n}$ of the expansion (\ref{fthetacpnexp}) are obtained
from appropriate $2n$-point correlation functions of the topological charge
density operators at $\theta=0$.  The analysis of the $1/N$-expansion Feynman
diagrams~\cite{CR-92} of the connected correlations necessary to compute
$b_{2n}$ shows that they are suppressed in the large-$N$ limit,
as~\cite{DMPSV-06}
\begin{equation}
b_{2n} = O(1/N^{2n}).
\end{equation}
Rather cumbersome calculations lead to the results~\cite{DMPSV-06} 
\begin{eqnarray}
b_2= -{27\over 5 N^2}+ O(1/N^3), \qquad
b_4= -{1830\over 7 N^4}+ O(1/N^5). 
\end{eqnarray} 

The above results are consistent with large-$N$ scaling arguments applied to
the Lagrangian (\ref{lagrangiancpn}), which indicate that the relevant
$\theta$ parameter in the large-$N$ limit should be $\bar{\theta}\equiv
\theta/N$.  This implies that the ground-state energy can be rewritten as
\begin{eqnarray} 
&&f(\theta) = N \bar{f}(\bar{\theta}\equiv \theta/N), 
\label{fthetabarcpn}\\ 
&& \bar{f}(\bar{\theta}) = 
{1\over 2} \bar{C} \bar{\theta}^2 
( 1 + \sum_{n=1} \bar{b}_{2n} \bar{\theta}^{2n} ), 
\nonumber
\end{eqnarray}
where $\bar{C}\equiv N C$ and $\bar{b}_{2n}=N^{2n}b_{2n}$ are $O(N^0)$.
Note the analogy with the expected $\theta$ dependence of the ground-state
energy in 4D $SU(N)$ gauge theories, cf.  Eq.~(\ref{lnexp}).

Within the $1/N$ expansion one may also study the dependence of the mass
$M(\theta)\equiv \xi(\theta)^{-1}$
on the parameter $\theta$. We write
\begin{equation}
M(\theta) \equiv \xi(\theta)^{-1} = M\left( 1 + m_2 \theta^2 + ... \right).
\label{gmexcpn}
\end{equation} 
The analysis of the corresponding diagrams in the $1/N$ expansion indicates
that $m_2$ is suppressed as
\begin{equation}
m_2 = O(1/N^2).
\end{equation}
Once again, the relevant parameter is seen to be $\bar{\theta}\equiv
\theta/N$.

Concerning the relevance of semiclassical instanton solutions in the context
of the $1/N$ expansion in $CP^{N-1}$ models, it was shown in
Ref.~\cite{Jevicki-79} that, at the quantum level, instantons appear in the
form of poles of the effective action, instead of stationary points, and that
the $1/N$ expansion and the semiclassical method correspond to two alternative
contour integrations of the path integral. The problem of the summability of
the instanton contributions~\cite{FFS-79} has been discussed in
Ref.~\cite{David-84}.  The compatibility of instanton models and large-$N$
expansion has been also addressed in Refs.~\cite{DM-00,KT-08}.

\subsection{Lattice calculations at $\theta=0$}

2D $CP^{N-1}$ models have also been studied by exploiting lattice techniques,
based on the straighforward discretization of the continuum
action~\cite{DHMNP-81} (in the following we set the lattice spacing $a=1$) 
\begin{equation}
S_L = -N\beta\sum_{n,\mu}\left|\bar z_{n+\mu}z_n\right|^2
\label{standard}
\end{equation}
or the lattice action~\cite{RS-81,DHMNP-81,BL-81}:
\begin{equation}
S_\lambda = -N\beta\sum_{n,\mu}\left( 
   \bar z_{n+\mu}z_n\lambda_{n,\mu} +
   \bar z_nz_{n+\mu}\bar\lambda_{n,\mu} - 2\right),
\label{basic}
\end{equation}
where $\beta=1/g$, $z_n$ is a complex $N$-component vector, constrained by the
condition $\bar z_nz_n = 1$, and $\lambda_{n,\mu}$ is an auxiliary $U(1)$ variable.
The lattice $CP^{N-1}$ model represents a very useful theoretical laboratory for
testing the various numerical methods to determine topological
properties, and in particular the topological susceptibility, also because the
results can be compared with the analytic large-$N$ calculations.

A lattice topological charge density operator $q_L$, having the correct
classical continuum limit $q^L(x) = a^2 q(x) + O(a^4)$, can be defined as
\begin{equation}
q_L(x) = -{i\over 2\pi}\sum_{\mu\nu} \epsilon_{\mu\nu}
   {\rm Tr}\left[ P(x)\Delta_\mu^{\rm s} P(x) 
   \Delta_\nu^{\rm s} P(x) \right],
\label{localq}
\end{equation}
where $\Delta^{\rm s}$ is a symmetrized version of the finite derivative:
$\Delta^{\rm s}_\mu P(x) = {1\over 2} [P(x{+}\mu) - P(x{-}\mu)]$,
and $P_{ij}(x)\equiv \bar{z}_i(x) z_j(x)$. Then, one may define the
corresponding lattice susceptibility $\chi_L=\langle \sum_x q_L(0) q_L(x)
\rangle$.  Analogously to the 4D $SU(N)$ gauge theories, see
Sec.~\ref{chilren}, the relation with the continuum topological susceptibility
is given by
\begin{equation}
\chi_{L} = a^2 Z_L^2\chi+B,
\label{chileqcpn}
\end{equation}
i.e. it involves a multiplicative renormalization and a background term which
eventually becomes dominant in the continuum limit.  
Some perturbative calculations can be found in
Refs.~\cite{FP-93,DFPV-92}.

\begin{figure}
\centerline{\psfig{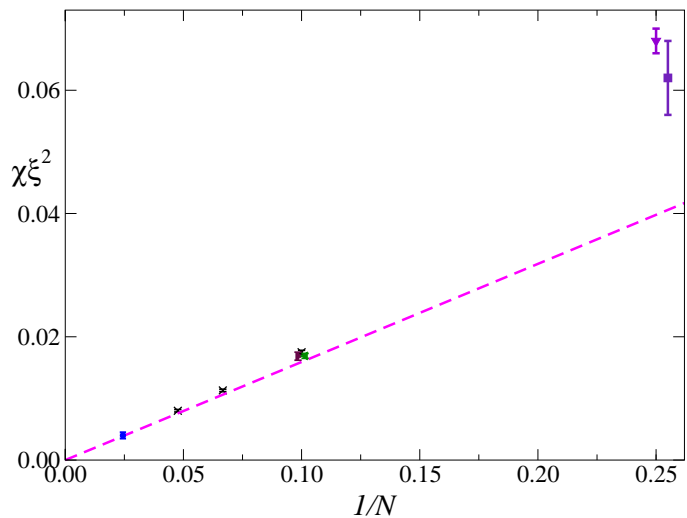}}
\caption{ MC results for the scaling ratio $\chi\xi^2$ versus $1/N$.  Results
  indicated by crosses are taken from Ref.~\cite{DMV-04} ($N=10,15,21$,
  obtained by the geometrical method), by a circle from
  Ref.~\cite{v-93} 
  ($N=41$, geometrical method), by a box from Ref.~\cite{CRV-92-2} ($N=4$,
  heating), by a downward pointing triangle from Ref.~\cite{Burkhalter-96} ($N=4$, obtained
  using also~\cite{CRV-92-2} $\xi/\xi_w=0.985(4)$ where $\xi_w$ is the length
  scale extracted from the exponential long-distance behavior of the $P_{ij}$
  correlation function), by a right triangle from Ref.~\cite{ACDP-00} ($N=10$,
  heating) and by a left triangle from Ref.~\cite{ACDP-00} ($N=10$, cooling).
  The line shows the asymptotic large-$N$ behavior $\chi\xi^2 = 1/(2\pi N)$.
  In some cases the data are slightly shifted along the $x$-axis to make them
  visible.}
\label{chicpnfig}
\end{figure}

One can also define the topological charge by a geometrical approach.  The
original geometrical construction for the topological charge proposed in
Ref.~\cite{BL-81} is
\begin{eqnarray}
Q_g&=&\sum_n q_{n},\nonumber \\
q_{n}&=& 
{1\over 2 \pi} 
{\rm Im}\left[ \ln{\rm Tr}(P_{n+\mu+\nu}
P_{n+\mu}P_n)+ \ln{\rm Tr}
(P_{n+\nu}P_{n+\mu+\nu}P_n)\right],
\qquad \mu \neq \nu, 
\label{qgeomP}
\end{eqnarray}
where the imaginary part of the logarithm is to be taken in the interval $(-\pi, \pi)$.
For the lattice formulation (\ref{basic}) an alternative geometrical
definition $Q_{g,\lambda}$ can be given in terms of the ``gauge'' field
$\lambda_{n,\mu}$~\cite{CRV-92}. Introducing the plaquette operator
\begin{equation}
 u_{\lambda,n} =  \lambda_{n,\hat{1}} \, \lambda_{n+\hat{1},\hat{2}} 
\bar{\lambda}_{n+\hat{2},\hat{1}} \bar{\lambda}_{n,\hat{2}}, 
\label{uqt}
\end{equation}
one defines $Q_{g,\lambda}=\sum_n q_{\lambda,n}$ where $u_{\lambda,n} = \exp(i
2 \pi q_{\lambda,n})$ and $q_{\lambda,n}\in (-1/2,1/2)$.  On a finite volume
and for periodic boundary conditions $q_{n}$ and $q_{\lambda,n}$ generate
integer values of the total topological charge for each configuration.  They
are not analytical functions of the lattice fields $z$ and $\lambda$, and fail
to be defined on certain ``exceptional'' configurations, which constitute a
set of measure zero.

Fermionic constructions of the lattice topological charge density, based on
the overlap Dirac operator, have been considered in Refs.~\cite{ALT-05,LT-07}.

The topological properties of 2D $CP^{N-1}$ models have been investigated by a
large number of studies exploiting lattice techniques, see
Refs.~\cite{Berg-81,MPV-82,Seiberg-84,BP-85,Wolff-92,IM-92,JW-92,CRV-92,CRV-92-2,
HM-93,v-93,DFP-95,BBHN-96,Burkhalter-96,DFP-97,RRV-97,PS-97-2,PS-97,v-99,ACDP-00,
BISY-01,ADGL-03,DMV-04,BPRW-05,LT-07,KT-08}. 
A wide range of values of $N$ has been considered, both small and large, in
order to test large-$N$ calculations.  The present state of art is briefly
summarized in the following.

Many numerical works, see
Refs.~\cite{Berg-81,MPV-82,BP-85,DFPV-92,CRV-92,FP-94,DFP-95,BBHN-96,DFP-97},
have been dedicated to the $N=2$ case, which also corresponds to the $O(3)$
nonlinear $\sigma$ model. The most recent simulations using the so-called
classical perfect action~\cite{BBHN-96,DFP-97} favor what is suggested by
semiclassical arguments, that is that $\chi$ would not be a physical quantity
for this model, in that a nonremovable ultraviolet divergence affects the
instanton size distribution. The manifestation of this property on the lattice
would be that lattice estimators of $\chi$ do not properly scale approaching
the continuum limit.

This problem should not affect the topological properties of $CP^{N-1}$ models
for $N>2$.  Monte Carlo results can be found in
Refs.~\cite{PL-83,JW-92,Wolff-92,HM-92,CRV-92,CRV-92-2,HM-93,v-93,Burkhalter-96,
RRV-97,ACDP-00,DMV-04,ALT-05}.
They have been obtained using the various methods discussed in
Sec.~\ref{bosmeth}.  In particular, at large $N$, say $N\gtrsim 10$, the
geometrical estimator (\ref{qgeomP}) shows scaling already at reasonable
values of the correlation length~\cite{CRV-92-2,v-93,DMV-04}.  Some results
for the dimensionless quantity $\chi\xi^2$ at $N\ge 4$ are shown in
Fig.~\ref{chicpnfig}. For $N=10$, there are accurate estimates of the
continuum limit of $\chi\xi^2$ by different
methods: $\chi\xi^2=0.0175(3)$ using the geometrical method~\cite{DMV-04},
$\chi\xi^2=0.0169(7)$ and $\chi\xi^2=0.0169(4)$ using respectively heating and
cooling~\cite{ACDP-00}, which are in good agreement.  The results clearly
approach the large-$N$ asymptotic behavior (\ref{chim2largen}).  Deviations
from this behavior are seen to be quite small, and clearly
suppressed at large $N$.  Writing 
\begin{equation}
\chi\xi^2 = {1\over 2\pi N} + {e_2\over N^2} + O(1/N^3),
\label{o1n2chi}
\end{equation}
the MC results suggest a small $O(1/N^2)$ coefficient:  $e_2\approx
0.15$.  This number can be compared with the rather cumbersome $O(1/N^2)$
calculation in the framework of the large-$N$ expansion~\cite{CR-91}, which
confirms that the $O(1/N^2)$ correction is small, but gives the number
$e_2\approx -0.060$.  The origin of this apparent discrepancy remains unclear.

Numerical MC investigations of topological charge structure in 2D $CP^{N-1}$
models are reported in Refs.~\cite{ALT-05,LT-07,KT-08}, finding evidence of
extended coherent structures, such as instantons and domain walls, which led
to a qualitative picture of the relevant vacua, and of the
changes with increasing $N$; they point to a transition from an
instanton dominated vacuum, at small values 
of $N$, to a domain wall dominated vacuum at large $N$.

\subsection{Results around $\theta=\pi$}
\label{thetapi}

The $\theta$ dependence of $CP^{N-1}$ models, and in particular around $\theta=\pi$,
is physically relevant in condensed matter physics.  The $CP^1$ or O(3)
$\sigma$ model at $\theta=\pi$ should describe the antiferromagnetic spin 1/2
chain~\cite{Haldane-83,Affleck-89}, which is gapless.  Thus the mass gap in
the 2D O(3) $\sigma$ model is expected to vanish when $\theta\to
\pi$~\cite{AH-87,AGSZ-89,Affleck-91}.  The spectrum around $\theta=\pi$ has
been studied in Ref.~\cite{CM-04}.  The behavior of $CP^{N-1}$ models around
$\theta=\pi$ has also been discussed in connection with the theory of the
quantum Hall effect, see e.g. Ref.~\cite{PB-05} and references therein. It has
been conjectured~\cite{Affleck-88,Affleck-91} that the ground-state energy
$F(\theta)$ of $CP^{N-1}$ models with $N>2$ presents a discontinuity at
$\theta=\pi$, like the large-$N$ limit~\cite{Seiberg-84}.  We also mention the
quite general scenario argued in Ref.~\cite{AGL-03}: assuming a nontrivial
$\theta$ dependence and quantization of the topological charge, the theory
either breaks spontaneously CP at $\theta=\pi$ or shows a singular behavior at
some critical $\theta_c$ between 0 and $\pi$.

Several lattice studies have also investigated the behavior of the theory at
finite $\theta$ and not only around $\theta=0$, see
e.g. Refs.~\cite{Seiberg-84,OS-94,BPW-95,PS-97,PS-97-2,
BISY-01,ADGL-02,ADGL-04,BPRW-05,AGL-07,AP-07,KT-08}.
Like 4D $SU(N)$ gauge theories, the Euclidean action with the $\theta$ term is
complex, which impedes a Monte Carlo simulation of the theory using its
straightforward discretization, i.e. by adding the $\theta$ term to the lattice
formulations (\ref{standard}) or (\ref{basic}).  Essentially two approaches
have been pursued.  In one of them the topological charge distribution is
determined at $\theta=0$ by Monte Carlo simulations, and then it is used to
estimate quantities at finite $\theta$.  In an alternative approach one
gets information of the finite-$\theta$ behavior by performing Monte Carlo
simulations at complex values of $\theta$ where the action is real, and then
somehow extending the results to real $\theta$ values.  The numerical
studies~\cite{BPW-95,AGL-07,AP-07} of the critical behavior of the $CP^1$ or O(3)
$\sigma$ model have confirmed the theoretical expectation of a gapless theory
at $\theta=\pi$.  Results for $CP^{N-1}$ models with larger values of $N$, see
Refs.~\cite{BPRW-05,ADGL-04,BISY-01,PS-97,Seiberg-84}, confirm a smooth
$\theta$ dependence up to $\theta=\pi$, where there is a singularity, like
the one found in the large-$N$ limit.  In particular, in Ref.~\cite{BPRW-05} the
sign problem was overcome by simulating an appropriate $SU(N)$ quantum spin
ladder equivalent to a $CP^{N-1}$ model with $\theta=\pi$; the results provided
evidence of a first-order transition at $\theta=\pi$ (discontinuity in
the ground-state energy $F(\theta)$) with spontaneous breaking of charge
conjugation symmetry for $CP^{N-1}$ models with $N>2$.
Ref.~\cite{KT-08} used fractionally charged Wilson loops 
to infer properties related to the $\theta$ dependence, 
providing further support to the first-order transition scenario
at $\theta=\pi$ for sufficiently large values of $N$.

\section{The two-point correlation function of the topological charge density}
\label{qqcorr}

The two-point correlation function
\begin{equation}
G(x-y)= \langle q(x) q(y) \rangle
\label{equation}
\end{equation}
of the topological charge density presents a peculiar behavior, especially if
contrasted with the physical renormalization-group invariant meaning of its
positive integral, i.e. the topological susceptibility.  Reflection positivity
implies $G(x)\leq 0$ for $|x|>0$.  On the other hand, the susceptibility must
be positive, trivially from its definition.  These facts indicate that there
is a positive contact term at $x=0$, that contributes to the determination of
the physical quantity $\chi$. In this section we discuss the main features of
the correlation function $G(x)$.

\subsection{General features of the two-point function and reflection positivity}

In order that Euclidean correlation functions can be continued back to
Minkowski space, they have to obey a positivity condition: the so-called
reflection positivity~\cite{OS-73,OS-78}.  The general statement concerning
reflection positivity is that
\begin{equation}
\langle ( \Theta F) F \rangle \geq 0,
\label{rfst}
\end{equation}
where $\Theta$ is the antilinear reflection operator consisting in a Euclidean
time reflection and a complex conjugation, and $F$ is an arbitrary gauge
invariant function of the fields having support only at positive Euclidean
times (see also Ref.~\cite{MM-94}).  As a consequence of the intrinsic odd
parity of $q(x)$ under reflection,
\begin{equation}
\Theta q(x_1,x_2,x_3,x_4)  = - q(x_1,x_2,x_3,-x_4), 
\label{qtr}
\end{equation}
reflection positivity states that~\cite{SS-87,v-99,Seiler-02}
\begin{equation}
G(x) \leq  0 \quad {\rm for} \quad |x|>0.
\label{cqxd}
\end{equation}
This fact holds for any operator that is intrisically odd with respect to
reflection symmetry in the Euclidean space.

The asymptotic large- and small-distance behavior of $G(x)$ can be inferred by
general arguments.  At large distance $G(x)$ should decay exponentially as
$e^{-m |x|}$ where $m$ is the lowest mass in the corresponding $0^{-+}$
channel.  In the presence of fermions we expect $G(x) \sim e^{-m_{\eta'} |x|}$
apart from negative powers of $|x|$.  Dimensional, perturbative and
renormalization group arguments~\cite{v-99} tell us that for $r
\rightarrow 0$
\begin{equation}
G(x) = {c\over r^{8}(\ln r )^2 }
\left[ 1 + O\left({1\over \ln r}\right)\right], 
\label{r0beh}
\end{equation}
where $c$ is a negative constant.  The logarithms can be related to a running
coupling constant; indeed, in perturbation theory $G(x)$ is $O(g^2)$.  Since
the topological susceptibility $\chi$ is positive ($\chi=0$ in the presence of
a massless fermion) and $G(x)<0$ for $x\neq 0 $, $G(x)$ should develop a
positive diverging term at $x=0$, that compensates the negative contribution
of its integral for $x\neq 0$ and makes $\chi$ positive.  Inspite of this
singular short-distance behavior, the low-momentum behavior and in particular
the moments of $G(x)$,
\begin{equation}
\mu_{2n}\equiv \int d^d x\, (x^{2})^n G(x)
\label{moments}
\end{equation}
($\mu_0\equiv \chi$), are expected to be well defined and finite.  Below we
will see in a solvable model, the large-$N$ limit of the two-dimensional
$CP^{N-1}$ model, how these features can coexist.

It is worth mentioning that, besides the topological susceptibility, also the
zero-momentum slope $\chi'$ is of phenomenological relevance, in connection
with the spin content of the proton~\cite{SV-90,NSV-95,NSV-99}.  In the
Minkowskian space-time the Fourier transformed correlation function of the
topological charge density is given by
\begin{equation}
\chi(p^2) = - i \int d^d x e^{ipx} \langle 0 | T q(x) q(0) | 0\rangle ,
\label{chip2}
\end{equation}
$\chi\equiv\chi(0)$ is the zero-momentum 
topological susceptibility, while the so-called
zero-momentum slope is defined as
\begin{equation}
\chi' \equiv -{d \over d p^2}\chi(p^2)|_{p=0} \,.
\label{chip}
\end{equation}
In Euclidean space it corresponds to
\begin{equation}
\chi' = {1\over 2d} \, \mu_2\,,
\end{equation}
where $\mu_{2}$ is the second moment of the Euclidean correlation function
$G(x)$, cf. Eq.~(\ref{moments}).  $\chi'$ has been determined within chiral
perturbation theory, obtaining~\cite{Leutwyler-00}
\begin{equation}
\chi' \approx 
{1\over 4} f_\pi^2 \Bigl( \sum_f {1\over m_f^2} \Bigr)
\Bigl( \sum_f {1\over m_f} \Bigr)^{-2}.
\label{chipest}
\end{equation}
For $N_f=3$, and inserting the phenomenological values of the mass
ratios~\cite{Leutwyler-96}, Eq.~(\ref{chipest}) gives $\chi'\approx (47 \,{\rm
  MeV})^2$.  Consistent results have also been obtained by QCD sum
rules~\cite{IO-98}. $\chi'$ has been also estimated in the massless
limit~\cite{NSV-95,NSV-99,Narison-06} obtaining $\chi'=[26.5(3.1) \,{\rm
  MeV}]^2$.  There are also results for the theory without quarks, i.e. the
pure gauge theory. An analysis~\cite{Narison-91,Narison-06} based on the QCD
sum rules gave $\chi'\approx - [7(3) \,{\rm MeV}]^2$.  Lattice studies of this
quantity have been reported in Refs.~\cite{BDP-91,DMP-92-2,BADD-97}.

\subsection{The two-point function on the lattice}

\begin{figure}
\centerline{\psfig{width=8truecm,angle=0,file=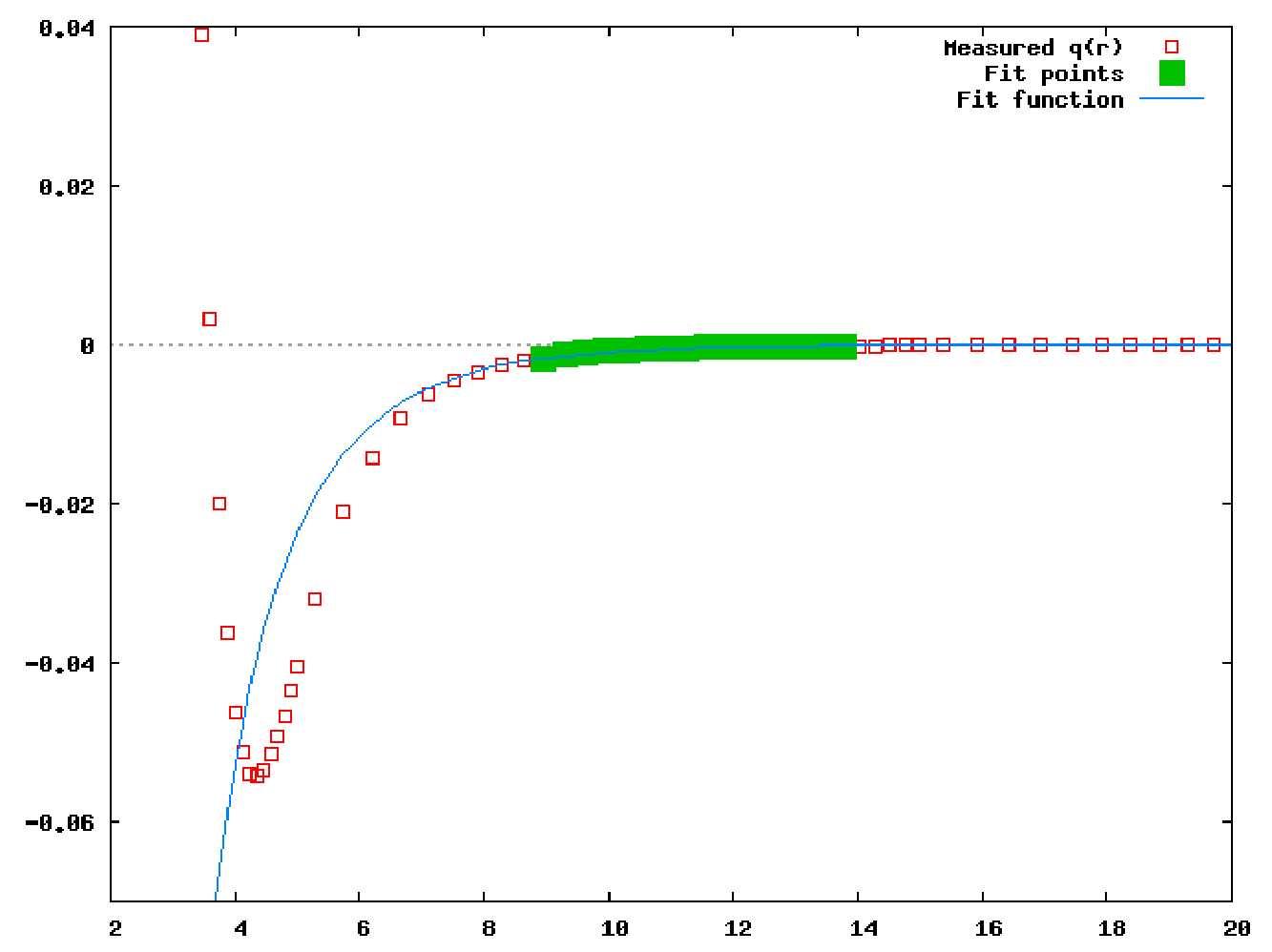}}
\caption{
  Results (indicated by boxes) for the two-point function $\langle
  q_L(0)q_L(r)\rangle$ of lattice topological charge densities versus the
  distance $r$, determined using HYP smearing techniques, computed from 2+1
  flavor dynamical configurations using staggered fermions.  From
  Ref.~\cite{Bernard-etal-07}.  }
\label{qqstag}
\end{figure}

In the lattice formulation of a theory one may define two versions of
reflection symmetry that are equivalent in the continuum limit: site- and
link-reflection symmetry.  We recall that reflection positivity is essential
on the lattice for the existence of a self-adjoint Hamiltonian at finite
lattice spacing defined from the transfer matrix.  Site- and link-reflection
symmetries are both satisfied by the Wilson lattice action of $SU(N)$ gauge
theories~\cite{OS-78}, and by Wilson fermions with Wilson parameter
$r=1$~\cite{MP-87}.  So Eq.~(\ref{rfst}) must hold also on the lattice for
finite lattice spacing.  A lattice discretization of the topological charge
density operator is for example that given in Eq.~(\ref{qL}).  One may easily
verify that $q_L(x)$ transforms as $q(x)$, cf. Eq.(\ref{qtr}), under
site reflection.  Again, reflection positivity tells us that
\begin{equation}
G_L(x)\equiv\langle q_L(x) q_L(0) \rangle < 0, 
\qquad {\rm for}\ x\neq 0
\label{GLx}
\end{equation}
(at least for $x$ lying along the
lattice axes and when there is no overlap among the link
variables of the two operators).  In the continuum limit the lattice
correlation $G_L(x)$ should reproduce the continuum correlation function
$G(x)$.

The two-point function of the lattice version of the topological charge
density has been numerically investigated by Monte Carlo simulations in
Refs.~\cite{ADDK-98,Hasenfratz-00,CTW-02,Ho-etal-03,HAZCDDLMTT-05,
  Thacker-06,ILMKSW-08,Bernard-etal-07,BG-06}.  Their results show the expected
behavior: they are negative at sufficiently large distances, i.e.  when the
effect of the overlap between the links involved in the definition of the
lattice density disappears.  This is clearly observed in Fig.~\ref{qqstag}
obtained in Ref.~\cite{Bernard-etal-07} from Monte Carlo simulations with
staggered fermions, and the HYP smoothing technique to determine the
topological charge density~\cite{DHK-97,HK-01}.  Similar results have been
obtained by employing other lattice definitions of 
the topological charge 
density~\cite{ADDK-98,Hasenfratz-00,CTW-02,Ho-etal-03,HAZCDDLMTT-05,
  Thacker-06,ILMKSW-08},
such as the overlap definition in pure $SU(3)$ gauge theory.

\subsection{Analytic results in the large-$N$ limit of the 2D $CP^{N-1}$ model}

The arguments of the previous section can be applied to the
two-dimensional $CP^{N-1}$ models as well.  The topological charge density
$q(x)$ has been defined in Eq.~\ref{qcpn}. 
Like the topological charge density of QCD, $q(x)$ transforms as
\begin{equation}
\Theta q(x_1,x_2)  = - q(x_1,-x_2)
\end{equation}
under reflection symmetry.  Therefore, as a consequence of reflection
positivity, 
\begin{equation}
G(x-y) \equiv \langle q(x) q(y) \rangle <0 \qquad{\rm for} \qquad  x-y\neq 0.
\end{equation}
The two-point function $G(x)$ of $q(x)$ can be
computed in the large-$N$ limit, in the continuum and on the lattice.  These
analytical results show explicitly that $G(x)$ develops a singular behavior
at the origin consistently with the reflection positivity requirement
$G(x)<0$ for $x\neq 0$, and the positivity of the corresponding topological
susceptibility, i.e. of its space integral.  Nevertheless, the low-momentum
behavior of $G(x)$ turns out to be well defined without the need of
special subtractions.  This provides an explicit example where the conjectured
main features of the two-point function $G(x)$ of QCD are analytically
verified.

\subsubsection{The large-$N$ limit in the continuum}

Straightforward calculations in the large-$N$ limit lead to the following
expression for the Fourier transform of $G(x)$~\cite{v-99}:
\begin{eqnarray}
N \widetilde{G}(p) &=& {1\over 2\pi} p^2 \left[ u(p) 
\ln {u(p)+1\over u(p)-1} - 2\right]^{-1},
\label{tG}\\
u(p) &=& \sqrt{ 1 + {2\over 3 p^2\xi^2} },\nonumber 
\end{eqnarray}
where $\xi$ is the length scale defined in Eq.~(\ref{xidefcpn}).
The scaling function 
\begin{equation}
B(k) \equiv  \xi^2 N \widetilde{G}(p=k/\xi),
\label{bk}
\end{equation}
has the following asymptotic behavior
\begin{eqnarray}
k\to\infty\,:\quad B(k)&=& {k^2\over 2\pi (\ln (6 k^2)-2)} + O\left( {1\over \ln k}\right),
\label{yinf}\\
k\to 0\,:\quad B(k) &=& {1\over 2\pi} + {3\over 10\pi} k^2  - {27\over 350\pi} k^4 + O(k^6).
\label{y0}
\end{eqnarray}
The singular behavior of $G(x)$ at
small distance is already apparent from the asymptotic behavior (\ref{yinf})
of its Fourier transform.

\begin{figure}
\centerline{\psfig{width=8truecm,angle=-90,file=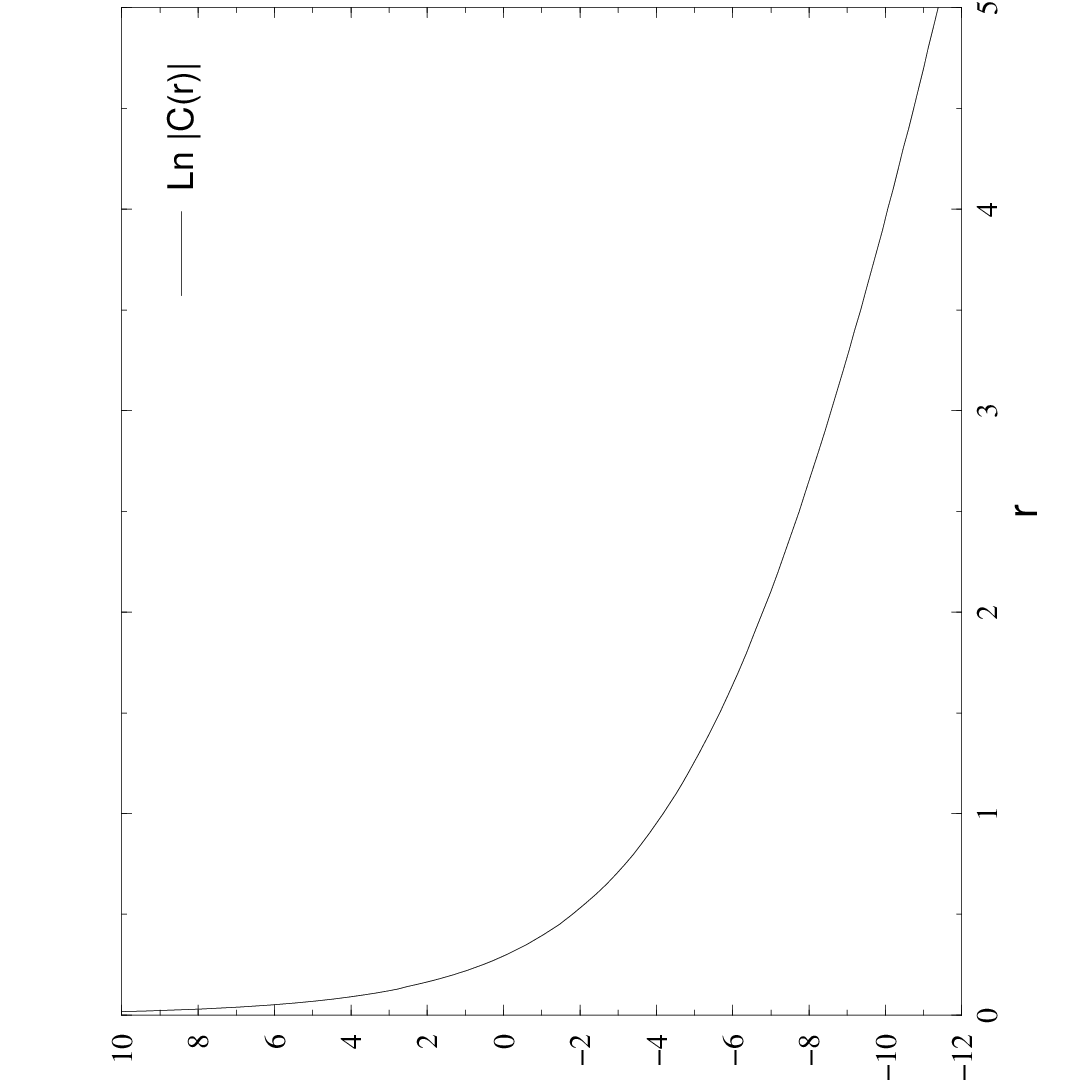}}
\caption{
Plot of $\ln \left[ - C(\tilde x)\right]$ versus $r\equiv |\tilde x|$.
}
\label{cpntwop}
\end{figure}

The calculation of the large-$N$ limit of $G(x)$ requires performing the
Fourier transform of the expression (\ref{tG}).  As before, let us introduce
the dimensionless quantities $\tilde x \equiv x/\xi$ and:
\begin{equation}
C(\tilde x) \equiv \xi^4 G(\tilde x\xi) = 
\int {d^2 k\over (2\pi)^2} e^{ik\cdot \tilde x} B(k).
\label{cz}
\end{equation}
The moments of $C(\tilde x)$ 
\begin{equation}
\bar\mu_{2n} \equiv \int d^2\tilde x\;(\tilde x^2)^n C(\tilde x), 
\label{mubar}
\end{equation}
and therefore of
$G(\tilde x)$, can be easily obtained from the expansion of $B(k)$ in powers of
$k^2$, cf. Eq.~(\ref{y0}).  In the large-$N$ limit one finds
\begin{eqnarray}
&&\bar{\mu}_0=\chi \xi^2={1\over 2\pi N} + O\left( {1\over N^2}\right),\\
&&\bar{\mu}_{2} = -{6\over 5\pi N } + O\left( {1\over N^2}\right),
\end{eqnarray}
and so on. By rotational invariance, $C(\tilde x)$ depends only on $r\equiv
|\tilde x|$.  For $r>0$
\begin{equation}
C(\tilde x) = - {1\over 2\pi^2} \int_{\sqrt{2\over 3}}^\infty 
dt K_0(tr) t^3 v(t) \left[ \left( v(t)\ln{1+v(t)\over 1-v(t)} 
- 2\right)^2 + \pi^2 v(t)^2\right]^{-1},
\label{Kr}
\end{equation}
where $v(t) = \sqrt{ 1 - {2/(3t^2)}}$. Since the Bessel function $K_0(x)$
satisfies $K_0(x)>0$, Eq.~(\ref{Kr}) shows that $C(\tilde x)<0$ for $r>0$ as
expected.  In Fig.~\ref{cpntwop} we show the function $C(\tilde x)$.  The
integral representation (\ref{Kr}) for $C(\tilde x)$ holds only for $r>0$.
For $r=0$ the contour rotation leading to the integral representation
(\ref{Kr}) misses the contribution from the path at infinite distance, which
is not suppressed anymore and should represent the positive contact term (see
Ref.~\cite{v-99} for details).  The integral representation (\ref{Kr}) allows
us to derive the asymptotic behavior of $C(\tilde x)$.  At large distance
$C(\tilde x)$ decays exponentially:
\begin{equation}
C(\tilde x) =  -{1\over 24\pi} {e^{-\sqrt{2\over 3}r}\over r^2} 
\left[ 1 + O\left({1\over r}\right)\right].
\end{equation}
For $r\rightarrow 0$, $C(\tilde x)$ diverges as
\begin{equation}
C(\tilde x) = - {1\over 2\pi^2} \,{1\over r^4 \left(\ln r\right)^2 } 
\left[ 1 + O\left( {1\over \ln r}\right)\right].
\label{leadterm}
\end{equation}
Note that one could have obtained this short-distance behavior by calculating
the leading order of perturbation theory, which is given by
\begin{equation}
C(\tilde x)\approx -{g^2\over 2\pi^4 r^4}.
\end{equation} 
Then, using renormalization group arguments, one replaces the coupling $g$
with a running coupling constant 
\begin{equation}
g(r)\approx {\pi\over \ln (1/r\Lambda)}, 
\end{equation}
thus recovering Eq.~(\ref{leadterm}).

The diverging negative integral of $C(\tilde x)$ for $r>0$ must be compensated
by a diverging positive contribution of the contact term at $r=0$, so that
\begin{equation}
\int d^2 \tilde x\; C(\tilde x) = B(0)={1\over 2\pi}.  
\end{equation}
Notice that, as a consequence of the $r^{-4}$ short-distance behavior and the
positivity of $\chi$, we have formally
\begin{equation}
\lim_{\delta\to 0} \int_{|\tilde x|<\delta} d^2 \tilde x\;C(\tilde x) = \infty.
\label{limc}
\end{equation}
Thus $\delta$-like distributions cannot represent a contact term with such
singular properties.  The behavior at $x=0$ should be described by more
complicated distributions, which act on a finite interval around
$x=0$~\cite{v-99}, see e.g. Ref.~\cite{Vladimirov-book} for a standard
mathematical text.  An example of such distributions may be
\begin{equation}
{\rm lim}_{\varepsilon\rightarrow 0^+} 
\left[ P_\varepsilon(\partial) \delta(\vec{x}) - 
\lambda_\varepsilon {1\over |x|^4} f\left( \ln |x| \right)
\theta(x-\varepsilon)\right],
\end{equation}
where the polynomial $P_\varepsilon(\partial)$ and $\lambda_\varepsilon$ are
appropriate functions of $\varepsilon$ and the limit $\varepsilon\rightarrow
0$ must be considered in a weak sense, i.e. after performing the integral with
the test function. Analogous considerations apply to QCD.

In Refs.~\cite{BN-97,AS-05,AS-05-2} the behavior of the two-point correlation
function of the topological charge density is discussed within the
two-dimensional $CP^1$ / O(3) $\sigma$ model.  In particular, it has been
shown~\cite{AS-05-2} that one-loop perturbative calculations are consistent
with the positivity requirements for such a correlation function.

\subsubsection{The large-$N$ limit on the lattice}

It is also interesting to understand how the peculiar behavior of $G(x)$ is
recovered in the continuum limit of lattice models, and, in particular, how
the short-distance diverging behavior for $x\to 0$ and the $x=0$
distribution-like behavior is obtained from the lattice calculations where
everything is finite. We consider the lattice action (\ref{basic}) which is
more convenient for a large-$N$
expansion~\cite{DHMNP-81,DMNPR-84,CR-93}. One can easily prove
that site- and link-reflection positivity holds for the lattice action
(\ref{basic}).  

\begin{table}
\caption{
The large-$N$ limit of $N G_L(0)$, 
$C_L(\tilde x,0)\equiv \xi^4 N G_L(x_1=\tilde x\xi,\,x_2=0)$ and $\xi^2 N \chi_L$ 
for various values of $\xi$.
For $\xi=\infty$ the continuum results are recovered.
}
\label{qqcorrl}
\tabcolsep 4pt        
\begin{center}
\begin{tabular}{rlllllll} 
\hline
\multicolumn{1}{c}{$\xi$}&
\multicolumn{1}{c}{$\beta$}&
\multicolumn{1}{c}{$N G_L(0)$}&
\multicolumn{1}{c}{$C_L(1/4,0)$}&
\multicolumn{1}{c}{$C_L(1/2,0)$}&
\multicolumn{1}{c}{$C_L(1,0)$}&
\multicolumn{1}{c}{$C_L(2,0)$}&
\multicolumn{1}{c}{$\xi^2 N \chi_L$}\\
\hline
1 & 0.41829... & 0.5371 &  &  &  $-$0.07888 & $-$0.001376 & 0.199791 \\
2 & 0.52868... & 0.2751 &  &  $-$0.8148 & $-$0.01634 & $-$0.001257 & 0.171593 \\
4 & 0.63901... & 0.1759 &  $-$9.314 &  $-$0.1713 & $-$0.01615 & $-$0.001160 & 0.162993 \\
8 & 0.74933... & 0.1277 &  $-$1.790 &  $-$0.1728 & $-$0.01547 & $-$0.001136 & 0.160316 \\
16& 0.85965... & 0.09995&  $-$1.808 &  $-$0.1662 & $-$0.01535 & $-$0.001131 & 0.159498 \\
$\infty$ & $\infty$ & 0 &  $-$1.718 &  $-$0.1651 & $-$0.01532 & $-$0.001129 & 0.159155 \\
\hline
\end{tabular}
\end{center}
\end{table}

In an infinite lattice (free boundary conditions are assumed) one may consider
the following discretization of the topological charge density operator
\begin{equation}
q_L(n) = {1\over4\pi}\epsilon_{\mu\nu} (\theta_{n,\mu} + 
   \theta_{n+\mu,\nu} - \theta_{n+\nu,\mu} - \theta_{n,\nu}),
\label{qgeomtheta2}
\end{equation}
where $\theta_{n,\mu}$ is the phase of the field $\lambda_{n,\mu}$, i.e.
$\lambda_{n,\mu} \equiv e^{i\theta_{n,\mu}}$, introduced in the lattice
formulation (\ref{basic}).  Using the property of $q_L(n)$ under
site reflection, one can prove Eq. (\ref{GLx}).
At large $N$ one can explicitly show that $q_L(n)$ has the correct continuum
limit, and no lattice renormalizations are necessary~\cite{DMNPR-84,RRV-97}.
Thus in the continuum limit and for $|x|>0$ one expects
\begin{equation}
C_L(\tilde x)\equiv \xi^4 N G_L(\tilde x\xi) = C(\tilde x) + O(\xi^{-2}),
\label{clx}
\end{equation}
where $G_L(x-y) = \langle q_L(x) q_L(y) \rangle$, and $C(\tilde x)$ is the
continuum function defined in Eq.~(\ref{cz}); $\xi$ the second-moment
correlation length associated with the lattice correlation function $G_P(x)
\equiv \langle {\rm Tr}\,P(x) P(0)\rangle$, defined
as in Eq.~(\ref{xidefcpn}).

$G_L(x)$ can be calculated in the large-$N$ limit~\cite{v-99}. In
Table~\ref{qqcorrl} we report some results. In the following
we list some interesting features arising from these lattice calculations.

\noindent 
(i) The continuum limit of $\xi^2 \widetilde{G}_L(k)$ at $k\xi$ fixed is
$B(k\xi)$, cf. Eq.~(\ref{bk}).  Indeed at large $\xi$
\begin{equation}
\xi^2 \widetilde{G}_L(k) = B(k\xi)+O(\xi^{-2}). 
\label{xigl}
\end{equation}
This may be proved by performing an asymptotic expansion of
$\widetilde{G}_L(k)$ (at fixed $k\xi$) in powers of $\xi^{-2}$, following
the procedure outlined in Ref.~\cite{CR-93}.

\noindent 
(ii) $G_L(x)$ is negative everywhere for $x\neq 0$, consistently with
reflection positivity.

\noindent 
(iii) At fixed physical distance $r=x/\xi >0$ the continuum limit exists and
it is given by $C(r)$, in agreement with Eq.~(\ref{clx}).  Notice that the
convergence is not uniform in $r$.  Moreover, from Eq.~(\ref{xigl}) it follows
that
\begin{equation}
\xi^{2(1-j)} \mu_{L,2j} = \bar{\mu}_{2j} + O(\xi^{-2}),
\end{equation}
where $\mu_{L,2j}$ are the lattice moments of $G_L(x)$, defined as in
Eq.~(\ref{moments}), and $\bar{\mu}_{2j}$ are the moments of $C(\tilde x)$, cf.
Eq.~(\ref{mubar}).

\noindent 
(iv) $G_L(0)$ compensates the negative sum $\sum_{x\neq 0} G_L(x)$ and makes
$\chi$ positive. Moreover, at large $\xi$, where $\xi\sim \exp (2\pi\beta)$,
\begin{equation}
G_L(0) \sim {1\over (\ln\xi )^2}\sim {1\over  \beta^2},
\end{equation}
giving rise to a positive contact term in the continuum limit.

The continuum limit of the correlation function $G_L(x)$ is regular at fixed
physical distances and it is given by $G(x)$.  Also its moments have
a regular continuum limit. On the other hand, a singular behavior is found at
$x=0$ consistently with reflection positivity and positivity of $\chi$.  These
features should also characterize the Euclidean correlation function of the
topological charge density in the continuum limit of lattice QCD.

\section{Slow dynamics of  topological modes in Monte Carlo simulations}
\label{qsampling}

Monte Carlo simulations of statistical systems at the critical point and of
quantum field theories, such as QCD, in the continuum limit are hampered by
the problem of critical slowing down (CSD).  For a general introduction to
critical slowing down in Monte Carlo simulations, see, e.g.,
Ref.~\cite{Sokal-92}.  The autocorrelation time $\tau$, which is related to
the number of iterations needed to generate a new independent configuration,
grows with increasing length scale $\xi$.  In simulations of lattice QCD and
$CP^{N-1}$ models, where the upgrading methods are essentially local, it has
been observed, see e.g. Refs.
\cite{CRV-92,DGHS-98-2,DPRV-02,DPV-02,DMV-04,LTW-04b}, that the topological
modes show autocorrelation times that are typically much larger than those of
other observables not related to topology, such as Wilson loops and their
correlators.  Actually, the heating method~\cite{DV-92}, used to estimate the
topological susceptibility, essentially relies on this phenomenon,
as discussed in Sec.~\ref{heatingmeth}.  The slow
sampling of the topological charge has been also discussed within MC simulations
of full QCD, see e.g.
Refs.~\cite{ABDDV-96,Alles-etal-98,Stuben-99,BJNNSS-06,BBOS-05,DS-07}.

\begin{figure}
\centerline{\psfig{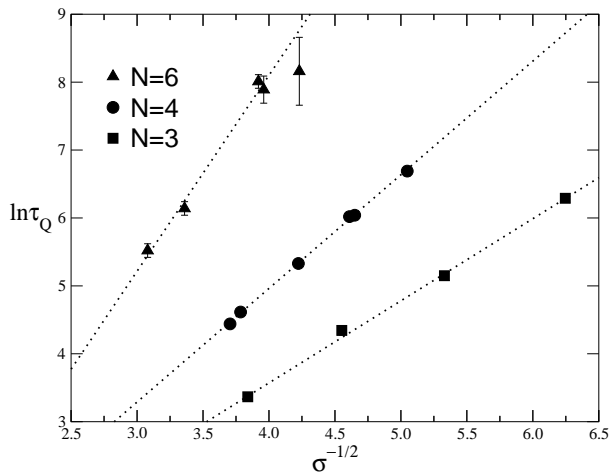}}
\caption{
Autocorrelation times of the topological charge from MC simulations 
of 4D $SU(N)$ lattice gauge theories for $N=3,4,6$,
versus $\sigma^{-1/2}$ where $\sigma$ is the string tension.
From Ref.~\cite{DPV-02}.
}
\label{tausun}
\end{figure}

Recent Monte Carlo simulations \cite{DPV-02,DPRV-02} of 4D $SU(N)$ lattice
gauge theories have provided evidence of severe CSD for the topological
modes, using a rather standard local overrelaxed upgrading algorithm.  Indeed,
the autocorrelation time $\tau_{\rm top}$ of the topological charge grows very
rapidly with the length scale $\xi\equiv \sigma^{-1/2}$, where $\sigma$ is the
string tension, exhibiting an apparent exponential behavior $\tau_{\rm
  top}\sim \exp (c\,\xi)$ in the range of values of $\xi$ where data are
available, as shown in Fig.~\ref{tausun}.  Such a phenomenon worsens with
increasing $N$, indeed the constant $c$ appears to increase as $c\propto N$.
The worsening of the CSD with increasing $N$ may be also related to the
suppression of small instantons discussed in Refs.~\cite{CTW-02,LTW-05}, see
also Sec.~\ref{largentop}.  A possible mechanism for the change of the
topological charge in a MC simulation is that an instanton-like structure
gradually shrinks until it disappears within the single lattice cell, or
appears from it. This process requires a significant probability of having
small instantons. But this probability is strongly suppressed in the
large-$N$ limit, cf. Eq.~(\ref{insuppr}).  Thus, at fixed length scale, the
change of topological charge becomes more amd more difficult with increasing
$N$~\cite{LTW-05}, giving rise to a substantial worsening of the corresponding
CSD. 

Of course, this behavior does not depend on the particular estimator of the
topological charge.  This peculiar effect has not been observed in
plaquette-plaquette or Polyakov line correlations, indicating a small
effective coupling between topological modes and nontopological ones, such as
those determining the confining properties.

These results suggest that the dynamics of the topological modes in Monte
Carlo simulations is rather different from that of quasi-Gaussian modes.  CSD
of quasi-Gaussian modes for traditional local algorithms, such as those based
on standard Metropolis, heat bath and molecular dynamics updatings, is related
to an approximate random-walk spread of information around the lattice.  Thus,
the corresponding autocorrelation time $\tau$ is expected to behave as
$\tau\sim\xi^2$ (an independent configuration is obtained when the information
travels a distance of the order of the correlation length $\xi$, and the
information is transmitted from a given site/link to the nearest neighbors).
This guess is correct for Gaussian (free field) models; in general one expects
that $\tau\sim \xi^z$ where $z$ is a dynamical critical exponent, and
$z\approx 2$ for quasi-Gaussian modes. On the other hand, in the presence of
relevant topological modes, the random-walk picture may fail, and therefore we
may have qualitatively different types of CSD.  These modes may give rise to
sizeable free-energy barriers separating different regions of configuration
space.  The evolution in configuration space may then present a long-time
relaxation due to transitions between different topological charge sectors,
and the corresponding autocorrelation time should behave as $\tau_{\rm
  top}\sim \exp F_b$ where $F_b$ is the typical free-energy barrier among
different topological sectors.  However, this picture remains rather
qualitative, because it does not tell us how the typical free-energy barriers
scale with the correlation length. For example, we may still have a power-law
behavior if $F_b \sim \ln \xi$, or an exponential behavior if $F_b\sim \xi^s$.
It is worth mentioning that in physical systems, such as random-field Ising
systems \cite{Fisher-86} and glass models \cite{Parisi-92}, the presence of
significant free-energy barriers in the configuration space causes a very slow
dynamics, and an effective separation of short-time relaxation within the
free-energy basins from long-time relaxation related to the transitions
between basins.  In the case of three-dimensional random-field Ising systems
the free-energy barrier picture supplemented with scaling arguments leads to
the prediction that $\tau\sim \exp (c \,\xi^s)$, where $s$ is a universal
critical exponent \cite{Fisher-86}.

\begin{figure}
\centerline{\psfig{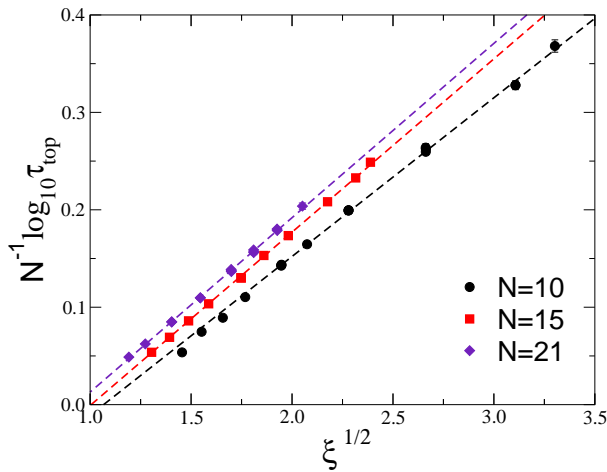}}
\caption{
Autocorrelation times of the topological charge from MC simulations 
of 2D $CP^{N-1}$ models.  From Ref.~\cite{DMV-04}.
}
\label{taucpn}
\end{figure}

The severe CSD experienced by the topological modes under local updating
algorithms is expected to be a general feature of Monte Carlo simulations of
lattice models with nontrivial topological properties, since the mechanism
behind this phenomenon should be similar.  This has been also observed in
two-dimensional CP$^{N-1}$ models \cite{DMV-04,CRV-92}.  The numerical study
of Ref.~\cite{DMV-04} for various values of $N$ shows that an exponential
Ansatz, i.e. $\tau_{\rm top}\sim \exp (c\,\xi^s)$ with $s\approx
1/2$, and $c\propto N$, provides a good effective description in the range of
the correlation length $\xi$ where data are available (however, the
statistical analysis of the data did not allow one to exclude an asymptotic
power-law behavior $\tau\sim \xi^z$ with $z\gtapprox N/2$ setting in at
relatively large $\xi$). Some data for the autocorrelation time of the
topological charge are shown in Fig.~\ref{taucpn}.

Two remarks are in order.  The first one is that, although the effects of the
topological CSD have not been directly observed in plaquette-plaquette or
Polyakov line correlations, the CSD of the topological modes will eventually
affect those observables as well.  The point is that the results of
Ref.~\cite{DMPSV-06}, summarized in Sec.~\ref{thspres}, show that the
correlators of plaquette operators and topological charge do not vanish at
finite $N$, although they are quite small, and therefore there is not a
complete decoupling between topological and nontopological modes.  The strong
critical slowing down that is clearly observed in the topological sector will
eventually affect also the measurements of nontopological quantities, such as
those related to the string and glueball spectrum. The second remark is that
the above CSD is not restricted only to the total topological charge of the
lattice configuration, which is the strictly zero-momentum topological mode,
but it should affect general low-momentum topological modes.  For example, it
would also affect the topological charge as measured on a part of the lattice,
a half say.  Therefore, it may not be possible to reliably estimate quantities
which are sensitive to the topological modes when the simulations are not able
to sample them, and in particular when simulations remain trapped at a fixed
topological charge sector.

The issue of CSD of topological modes has serious implications for Monte Carlo
simulations of full QCD. It may pose a serious limitation for numerical
studies of physical issues related to topological properties, such as the mass
and the matrix elements of the $\eta'$ meson, and in general the physics
related to the broken $U(1)_A$ symmetry. Indeed, it is expected to
substantially worsen the cost estimates of the dynamical fermion simulations
for lattice QCD using the actual optimal hybrid Monte Carlo algorithms, see
e.g.  Refs.~\cite{DKPR-87,Hasenbusch-01,Peardon-02,Luscher-03,Jansen-03}.
Note that standard cost estimates of these algorithms, see e.g.
Refs.~\cite{Ukawa-02,Bernard-etal-02,Jansen-03,DGLPT-07}, which are already
quite elevated, have been obtained without taking into account the CSD of
topological modes. In this respect, it would be important to devise algorithms
which sample topological sectors more efficiently.

The correct sampling of topology becomes increasingly more difficult towards
realistic simulations with light quarks and sufficiently fine lattices.
Standard local simulations, which change the gauge field configuration by
small steps, may get trapped in a fixed topological charge sector, without being
able to change it during the whole simulation. The question arises whether one
can obtain the interesting observables averaged over the full $\theta$ vacuum.
Of course, physical issues that are particularly sensitive to the topological
sampling cannot be addressed.  However, one should also consider that a fixed
finite global topological charge $Q$ is equivalent to a boundary condition,
thus the effects get suppressed in the thermodynamic $V\to \infty$ limit,
although such finite size effects may be much larger than those observed when
$Q$ is correctly sampled, which are expected to be suppressed as $e^{-m_\pi
  L}$.  However, the quenched study reported in Ref.~\cite{Gallety-06} indicates
quite a strong dependence of the pseudoscalar correlation function on $Q$.

Knowing the $\theta$ dependence of a given observable allows one to estimate
the finite-size effects when averaging at fixed topological
charge~\cite{AFHO-07,BCNW-03}.  For example, let us consider an observable $P$
whose $\theta$ dependence around $\theta=0$ is given by
\begin{equation}
P_\theta = P_{\theta=0}\left( 1 + p_2 \theta^2 + ... \right).
\label{gmex2}
\end{equation} 
$P$ may be the string tension or the glueball mass in pure gauge theories or
a hadron mass in full QCD.  The average $P_Q$ of $P$ at fixed $Q$ is given
by~\cite{BCNW-03}
\begin{equation}
P_Q = P_{\theta=0} \left[ 1 + p_2 {1\over V\chi} 
\left( 1 - {Q^2\over V\chi}\right) + ...\right].
\label{fssq}
\end{equation}
Therefore, if the calculation over a spacetime volume $V$ is performed at
fixed finite $Q$, the error is order $1/V$.  This formula may be used to
correct averages obtained at a fixed topological charge.

The variations of topological sectors are particularly delicate in Monte Carlo
simulations of overlap fermions, and in general of Ginsparg-Wilson fermions,
because they concide with the nonanalyticity points of the Dirac operator, and
require a particular numerical effort to be
handled~\cite{FKS-04,Schaefer-06,CKLS-08}.  In this respect, the simulation
performance is expected to improve by suppressing the occurrence of plaquette
values far away from the identity, which are essentially responsible for the
change of topological sector.  Of course, as a consequence, the simulation is
performed at a fixed topological sector.  This can be achieved by
appropriately choosing the action.  Numerical studies of these actions can be
found in Refs.~\cite{BJNNSS-06,FHHOO-06}.

Simulations at fixed topological charge turn out to be particularly useful in
the so-called $\epsilon$ regime, where $\xi\gg L$, and the finite-size effects
can be analytically computed by an appropriate
expansion~\cite{GL-87,Neuberger-88}.  In this regime observables are strongly
sensitive to the topological sector and predictions exist for each
sector~\cite{LS-92,DDHJ-02,DHJLL-03}. This is important because the low-energy
physical quantities of the infinite volume limit also appear, and therefore
may be estimated by numerical simulations in this regime, see e.g.
Ref.~\cite{GLWW-03}.

\bigskip
\bigskip

{\bf Acknowledgements}

\bigskip

We thank Luigi Del Debbio for recent collaboration on some of the issues
considered in this review.  We also thank Ken Konishi and Mihail Mintchev for
useful discussions, as well as Constantia Alexandrou and Nick Toumbas for
their comments on our manuscript. Finally, we thank all colleagues who sent us
their interesting comments on the first draft of this review.

\footnotesize
\newcommand{\arXiv}[1]{\href{http://xxx.lanl.gov/abs/#1}{\tt arXiv:#1}}

\end{document}